# Radio-Frequency Method for Detecting Superconductivity Under High Pressure

Dmitrii V. Semenok[1,†,*], Di Zhou[1,†,*], Jianbo Zhang[1,†],Toni Helm[2,3], Wuhao Chen[4], Yang Ding[1] , Ho-kwang Mao[5,6,] Ivan A. Troyan[6] and Viktor V. Struzhkin[5,6,*]

*[1] Center for High Pressure Science and Technology Advanced Research (HPSTAR), Beijing 100193, P. R. China*

*[2] Dresden High Magnetic Field Laboratory (HLD-EMFL) and Würzburg-Dresden Cluster of Excellencect.qmat, Helmholtz-Zentrum Dresden-Rossendorf, 01328 Dresden, Germany*

*[3] Max Planck Institute for Chemical Physics of Solids, 01187 Dresden, Germany*

*[4] Quantum Science Center of Guangdong–Hong Kong–Macao Greater Bay Area (Guangdong), Shenzhen 518000, China*

*[5] Center for High Pressure Science and Technology Advanced Research (HPSTAR), Shanghai 201203, P. R. China*

*[6] Shanghai Key Laboratory of Material Frontiers Research in Extreme Environments (MFree), Shanghai Advanced Research in Physical Sciences (SHARPS), Pudong, Shanghai 201203, China*

[*]Corresponding authors, emails: dmitrii.semenok@hpstar.ac.cn (Dmitrii V. Semenok), di.zhou@hpstar.ac.cn (Di Zhou), viktor.struzhkin@hpstar.ac.cn (Viktor V. Struzhkin).

[†]These authors contributed equally to this work

## Abstract

We introduce a contactless technique for probing superconductivity, metal-insulator transitions, and magnetic ordering in micron-sized samples under extreme pressure. Utilizing a multistage Lenz lens system, directly sputtered onto diamond anvils, we realize a radio-frequency (RF, 50 kHz – 200 MHz) transformer with a sample, of 50-100 μm in diameter, as its core. This configuration enables efficient transfer and focusing of an electromagnetic field within the diamond anvil cell's chamber. Consequently, the transmitted RF signal exhibits high sensitivity to variations in the sample's surface conductivity and magnetic permeability. We validate this method by determining the critical temperatures ($T_c$) of known superconductors, including NbTi, $MgB_2$, Hg-1223, Bi-2212, YBCO and REBCO in various magnetic fields, as well as the magnetic ordering temperatures of Gd and Tb. Notably, we apply this technique to the $LaH_{10-x}$, $CeH_{9-10}$, and (La,Ce)$H_{10-12}$ superhydrides at a pressure of about 1.5 Mbar. The observed superconducting transitions in La and Ce superhydrides at 90-110 K, and 215-242 K, respectively, correlate with the $T_c$'s determined via traditional electrical-resistance measurements. Moreover, we show how multiple repetitions of the RF experiment with the La-Ce superhydride make it possible to detect the increase in $T_c$ over time up to ≈ 260-270 K. This finding indicates possibility of reaching a critical $T_c$ around 0°C in the La-based superhydrides.

## Introduction

Superhydrides – are compounds of hydrogen with various elements that exhibit superconductivity (SC)[1]. As a rule, superconductivity in hydrides manifests itself at high pressures of about 100 GPa, when chemical compounds with high hydrogen content are formed. Examples include sulfur hydrides $H_2S$ and $H_3S$[2], yttrium hydrides $YH_4$[3], $YH_6$[3]and $YH_9$[4], lanthanum hydrides $LaH_4$[5], $La_4H_{23}$[6,7], $LaH_{10}$[8,9], $LaH_{12}$[10], and many other compounds. There are exceptions to this rule. For example, although molecular bismuth hydride $BiH_2$ contains only two hydrogen atoms per unit cell, it nevertheless demonstrates superconductivity at temperatures below 64-69 K[11,12]. Notably, all these different superhydrides share a certain degree of heterogeneity[5,13,14], i.e., the samples usually contain more than one phase with varying superconducting properties. This heterogeneity means that the standard method of studying their superconducting properties using four or more electrical contacts on diamond anvils is not always effective. This also means that observing the true Meissner-Ochsenfeld [15] effect in hydrides is challenging [16], since the magnetic field cannot be completely expelled from the sample, being trapped by pinning centers at grain boundaries. Nevertheless, heterogeneity does not interfere with their potential for practical applications, for example, as a superconducting quantum interference device (SQUID) or magnetic field sensors[17].

Even though, resistance is considered bulk-sensitive, the current usually probes only the path of least resistance. Consequently, some high-pressure phases that are obtained only in small amounts cannot be detected using electrical transport methods (e.g., Supplementary Fig. S3). In this case, it is necessary to use complementary volumetric methods for the detection of superconductivity. Such a method is AC[18,19] or DC magnetometry based on sensitive SQUID sensors [16]. A key limitation of these methods is that the signal strength scales with the sample volume, thus becoming very weak at pressures above 100 GPa compared to the overwhelming background signal from the diamond anvil cell (DAC).

We report here, that the detection of superconducting transitions based on radio-frequency (RF) transmission could provide a solution to many problems mentioned above. Moreover, counter to volumetric methods, the sensitivity of RF transmission is proportional to the surface area of the sample. For finely dispersed nanograined samples of hydrides[14], the developed surface area and the corresponding surface impedance are of greater importance than the volume and can lead to a much better noise/signal ratio than in the case of magnetometry. While superhydride samples are rather thin (1-2 μm at 150-200 GPa), they possess a significant (~50-100 μm diameter) surface area.

The radio-frequency transmission method for detecting superconductivity has been known since the 1980s, when Sakakibara et al. used it to detect the superconducting transition in YBCO in a strong pulsed magnetic field[20]. This method is sensitive to changes in the magnetic permeability and surface impedance of superconducting microsamples. It is based on a high-frequency transformer, whose core is the sample being measured (Fig. 1). In their experiments, Sakakibara et al. demonstrated that a 160-micron-thick disk-shaped sample of YBCO effectively shielded radio waves at a frequency of 20 MHz below the superconducting transition temperature, reducing the gain of the high-frequency transformer to almost zero[20,21]. An improved method of RF detection of superconductivity was proposed in 2002 by Timofeev *et al.*[22]. In this method, in addition to the RF transmission, a low-frequency modulation by an external magnetic field of the order of 1 mT is used, which periodically transfers part of the bulk superconductor to the normal state by exceeding $H_{c1}$. As a result, the high-frequency transmission signal experiences a modulation with twice the frequency of the applied magnetic field. This second-harmonic signal serves as an additional indicator of the superconducting nature of the transition in the studied specimen.

This approach has a lot in common with the method of detecting changes in the resonance frequency of a micro-resonator with a superconductor placed inside[23,24]. A high-pressure DAC with deposited Lenz lenses or micro loops, including the connecting wires, is, in principle, a resonator of this kind. However, inhomogeneous superconducting samples may induce rather complex outputs, as each local superconducting grain acts as its own resonator with different diamagnetic permeability. Nevertheless, as we show in this work, it can be utilized to reliably detect phase transitions at extreme pressures.

Our work consists of two parts. In the first part, we validate the proposed RF transmission method, using various high-temperature superconductors: (i) BSCCO and REBCO ($T_c$ = 92-95 K) (ii) Hg-1223 ($T_c$ = 128 K) (iii) NbTi ($T_c \approx 9$ K) and (iv) $MgB_2$ ($T_c \approx 40$ K). In the second part, we investigate the transmittance of magnetic metals, gadolinium and terbium, in high-pressure DACs. Finally, in the third part, we apply our RF technique to the lanthanum-based superhydrides and $CeH_{9-10}$ using DACs with Lenz lenses deposited on the diamond anvils. We demonstrate the feasibility of the RF approach under 1.5 megabar pressures and provide further evidence for high-temperature superconductivity in lanthanum-based superhydrides, also indicated in our previous publications on the La-Sc-H [25] and $LaH_{12}$ [10] phases.

## Results

### Experimental setup and computer modeling

Figure 1 shows how the ideas for contactless measurements of microsamples in high-pressure DACs have evolved since 2002. Initially, compensated multi-turn coils, wrapped around the anvils (Fig. 1a), i.e., separated from the sample by the gasket, were used for the contactless detection of changes in the magnetic susceptibility of pressurized samples. This method, however, suffers from a poor filling factor, as the sample space of DACs is limited to only few tens of microns, resulting in a very low signal-to-noise (S/N) ratio and the need to separately measure and subtract the background signal[26].

To solve this problem, there were attempts with both, the coil and the sample, placed inside the sample space of the DAC (Fig. 1b)[27]. Even though this improves the filling factor, such deposited microcoils have a high electrical resistance of several tens of Ohms or even kiloohms, which leads to strong thermal noise. A significant part of the resistance in such circuits stems from the leads and contacts that connect to the coil.

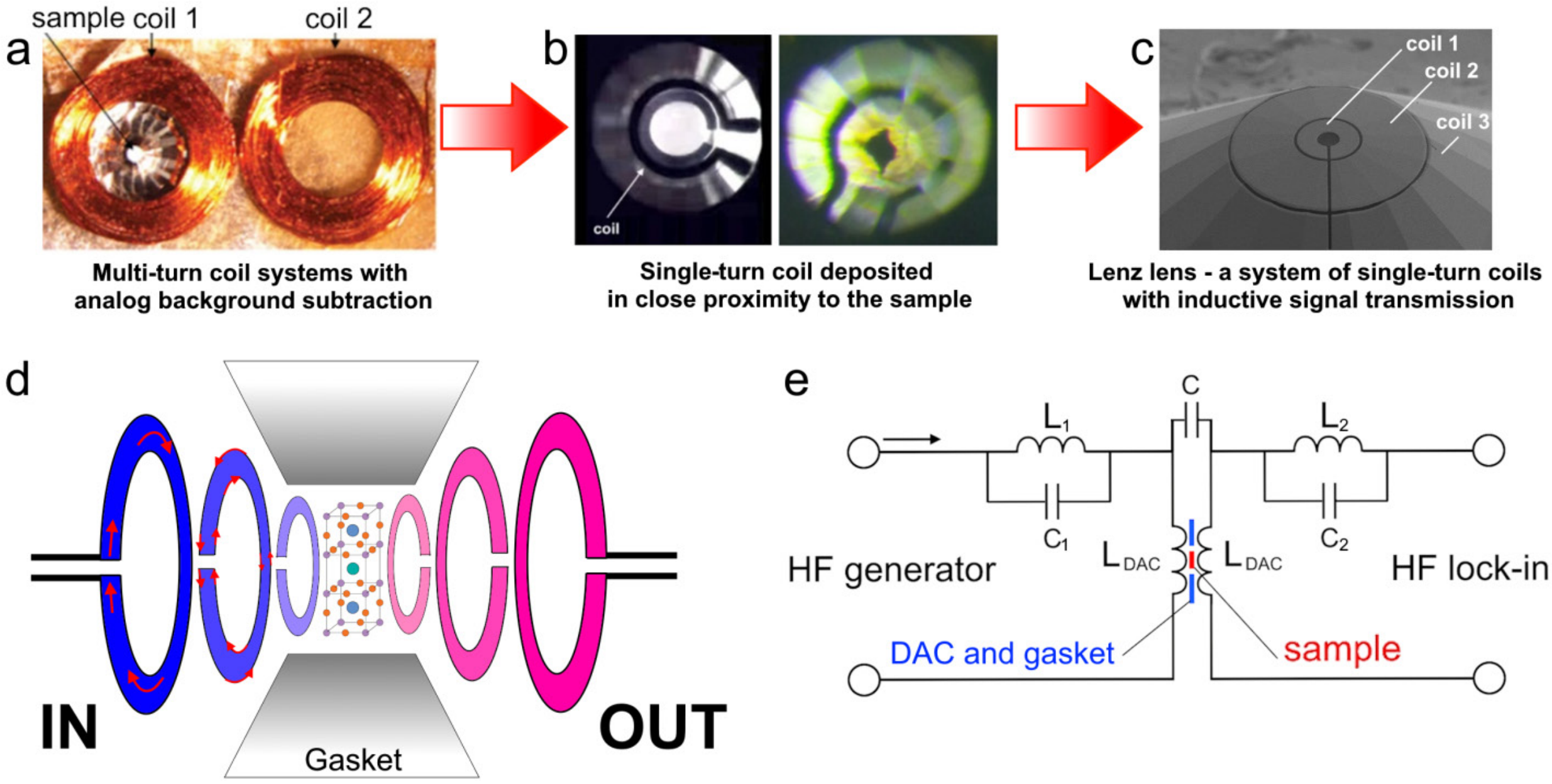

**Fig. 1 Development of a contactless RF detection method in diamond anvil pressure cells. a** Alternative current magnetometry setup utilizing multi-turn coils with a reference coil. **b** Single-turn coils deposited in close proximity to the sample for improved sensitivity under pressure. **c** A Lenz lens system, comprising multiple single-turn coils, enabling inductive signal transmission and enhanced electromagnetic field focusing. **d** Schematic representation of the Lenz lens system in a diamond anvil cell, illustrating the inductive coupling between the input (IN) and output (OUT) coils, with the sample positioned on a diamond culet. **e** Simplified electrical circuit of the measurement setup, showing the HF (high-frequency) generator, Lenz lens coils (L - inductance), tuning capacitors (C - capacitance), sample (red) in-between the gasket (blue), and HF lock-in amplifier.

In recent years, a design with inductive energy transfer called Lenz lens (Figs. 1c, d) has been applied successfully for studies of high-pressure nuclear magnetic resonance (NMR)[28-32]. Hydrogen NMR ($^1$H NMR) is a rather delicate technique that requires very specialized high-frequency equipment. It requires a lot of time for the relaxation-time measurements and is very sensitive to any hydrogen impurities in the sample of the measurement setup. We noticed that a simple change (i.e., independent use of two Lenz lenses as in/out instead of their sequential connection) in the measurement scheme (Fig. 1e) makes it possible to implement Timofeev's idea [22] of a high-frequency transformer with an additional modulation of the external magnetic field (Fig. 2) in combination with wide single-turn low-resistance coils and inductive current transfer mostly through a sample space. DACs prepared for NMR studies can be tested for superconducting transitions by Timofeev's technique in a few days using the simplest cryostat and two lock-in amplifiers (Fig. 2b). As we have recently shown using Bi-2212 and Cu-1234 superconductors as examples, the Lenz lens can be transferred to an insulating gasket [33].

As illustrated in Fig. 2a, the modulation of the high-frequency RF signal arises from the sensitivity of superconductivity to an external magnetic field, with the required field strength decreasing as the temperature approaches $T_c$(onset). Near $T_c$(onset), all characteristic magnetic fields ($H_{c1}$, $H_p$, $H_{c2}$) become very low. Since the magnetic field suppresses superconductivity regardless of its direction, the modulation occurs at twice the applied frequency, $f_{mod}$ of the alternating magnetic field ($2f_{mod}$). Furthermore, due to the presence of numerous grains with varying $T_c$ values, SC signals can be observed at different temperatures for various RF frequencies within the superconducting transition interval. This interval may be broader than that determined by temperature-dependent electrical-resistance measurements (Figs. 2c, d), as not all grains are located along the path of the electrical current through the sample (see Supplementary Note 1 for more details).

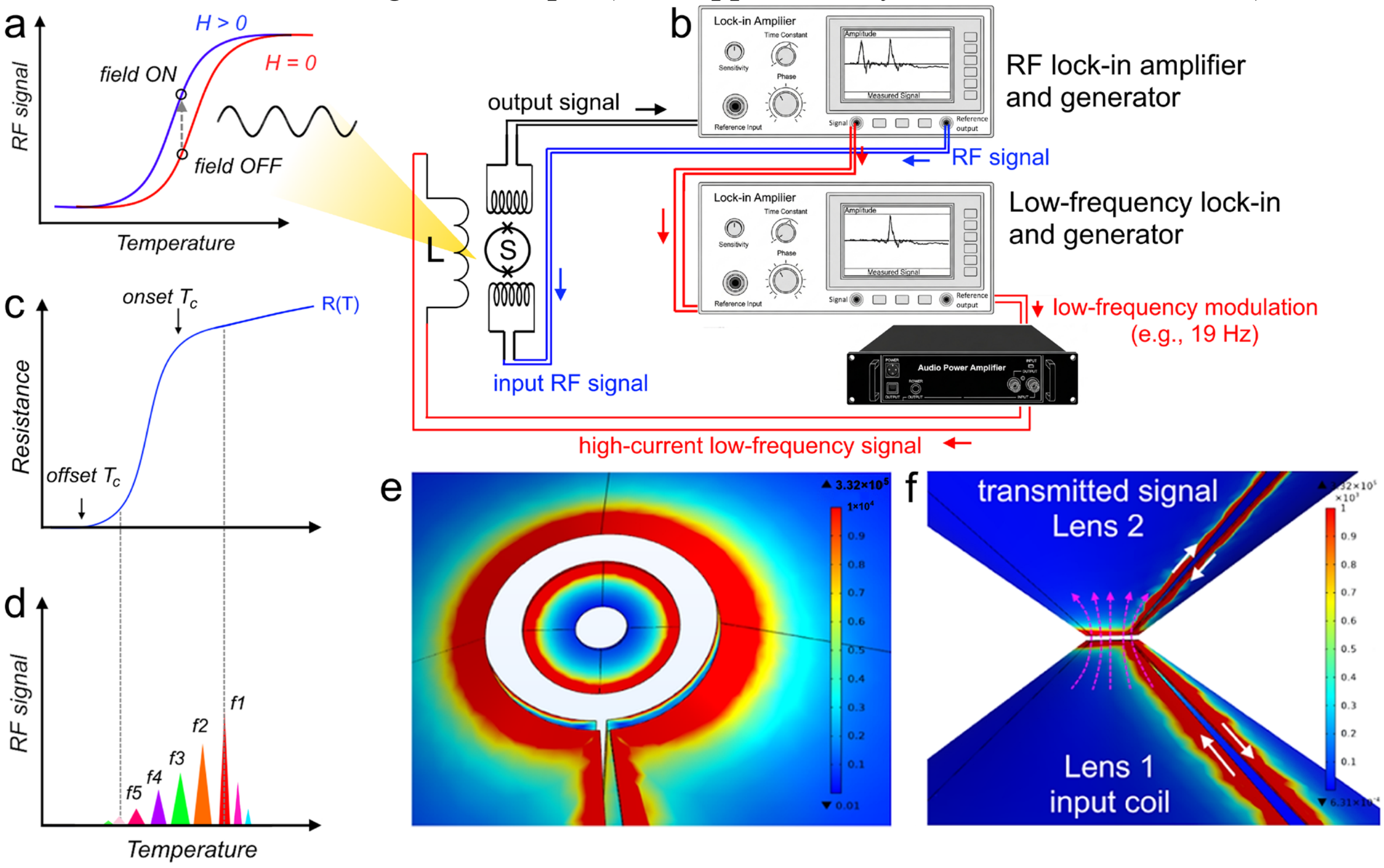

**Fig. 2 Experimental setup and principles of radio-frequency superconductivity detection.**
**a** Schematic representation of the second harmonic modulation of the RF signal induced by an applied AC magnetic field. The graph depicts how the RF signal in the vicinity of a SC phase transition changes with time as the AC magnetic field penetrates the sample (S). **b** Experimental setup. It includes two lock-in amplifiers and an audio-power amplifier. The RF lock-in generates the RF signal, while the low-frequency lock-in provides the low-frequency modulating field. The audio amplifier boosts the low-frequency signal current. **c** Schematic representation of a superconducting transition in the resistance of metals. The plot shows a rather broad transition from the normal to SC state with a gradual decrease to zero over a finite temperature range. **d** Schematic representation of the observation of RF transition signals at different temperatures for various RF frequencies (*f1, f2, f3...*). This can occur in heterogeneous samples containing individual grains transitioning into the SC state at slightly different temperatures between $T_c$(offset)and $T_c$(onset). **e, f** Finite-element simulations [34] of the surface loss density (SLD) distribution in a Lenz lens and a disk-shaped sample. Red regions indicate areas of high inductive current localization (max value is $3.32\times10^5$ W/m$^2$, SI units). **f** A scheme depicting the operation of a RF transformer based on Lenz lenses within a diamond anvil cell. The red dashed arrows illustrate the transmission of the RF signal through the lenses and the sample.

Computer simulations in the COMSOL[34] show (Figs. 2e, f) that the induction currents and their surface loss density (SLD) are mainly distributed at the rims of the Lenz lenses and at the edges of the sample, which is especially noticeable for high frequencies. This indicates, on one hand, that the Lenz lenses exhibit greater resistance to mechanical damage away from their rims. On the other hand, any alterations in the edge shape, e.g., burning of some grains due to excessive inductive current density or changes in the inductive current path, can lead to abrupt variations in the RF signal characteristics.

The radio-frequency detection method is more sensitive than the resistive and AC susceptibility measurements with macroscopic coils, which is its positive distinctive feature. Considering the high sensitivity and, accordingly, the noise level caused by random micro movements of individual parts of the wires, the Lenz lenses, or elements of the high-pressure DAC, analyses need to be done with great caution. We will, therefore, consider as a "significant signal" only features that, first are reproducible in independent warming/cooling cycles at several different carrier frequencies *f*, second, consist of many data points (i.e., not isolated artifacts), and third, are accompanied by a feature in the higher harmonics at low or high frequencies. A sign of the superconducting nature of the transition could also be the displacement of the target signal in a strong external magnetic field (see Fig. 4).

Finally, transport measurements of the sample's resistance – performed either simultaneously or separately – allow focusing on a narrow temperature range of interest (±30 K), eliminating most artifacts inherent to measurements over hundreds of degrees.

## Details of the experiment

The radio-frequency transmission method is usually implemented as follows. About 10-12 different frequencies of the SR844 lock-in (time constant 100 μs) generator from 50 kHz to 200 MHz, or Hantek 2102B (in the case of low frequency up to 50 MHz) are used as the carrier frequency, $f_{carr}$. The signal voltage, $U$, of the receiving coil varies from 0.5 mV to 50-100 mV. It is also possible to select a grid of carrier frequencies with a step of $\Delta f_{carr}$ = 5-10 MHz. The induction and detection coils were connected with rigid shielded coaxial wires fixed with glue to prevent vibrations. The coaxial wires were spaced 10 cm apart along the entire length of the cryostat insert to reduce mutual induction. Work at a high carrier frequency is usually accompanied by a large amount of noise and artifacts, but has an advantage in sensitivity ($\propto f$). Operation at frequencies below 0.1 MHz is complicated by the low signal level of the receiving coil. Typical cooling/warming rates were about 3-6 K/min or less, the warming rate is usually higher due to inductive heating. Comparison of measurement results in warming and cooling cycles, as well as warming cycles at different rates (3-5.5 K/min) shows that temperature hysteresis does not exceed 5 K (Fig. 6d, e, Supplementary Figs. S6, S22).

Target signal in the RF approach is a step-like or peak-like change in the output voltage, $U$, of the receiving high-frequency coil, and the simultaneous occurrence of a modulation of the RF signal amplitude with a frequency double ($2f_{mod}$, the second harmonic) that of the AC magnetic field of the external solenoid. An empty standard high-pressure DAC BX-90 mini (BeCu) for superhydride synthesis with an $W/Ta/Ta_2O_5$[35] insulating gasket, ammonia borane (pressure-transmitting media), a sputtered Ta/Au Lenz lens (culet: 100 μm), and emitting micro coil was investigated separately (Supplementary Fig. S5). The RF transmission signal as well as the modulation on the 2nd harmonic in this empty DAC do not show any pronounced anomalies in the region of superconducting transitions in considered polyhydrides (160-300 K, (Supplementary Fig. S5).

## Feasibility tests I: Detection of SC in $HgBa_2Ca_2Cu_3O_{8+\delta}$ (Hg-1223), BSCCO and REBCO tape at ambient pressure

Commercial REBCO tape, with a superconducting-layer thickness of $t \approx 2$ μm and a diameter of $d \approx 4$ mm, and a piece of optimally doped BSCCO, with $d \approx 4$ mm and $t \approx 300$ μm, were chosen as specimens for testing the proposed RF transmission method. In the geometry of a high-frequency transformer with single-turn coils (inset in Fig. 3a) and a carrier frequency of $f_{carr}$ = 500 kHz, we observed a pronounced drop (step-like) in the intensity of the real part of the transmission voltage, Re $U$, at a temperature of 92-93 K. At the same time, a similar change is observable in the imaginary part, Im $U$ (Fig.3a). Exposing the sample to a periodically oscillating magnetic field, which is generated by a solenoid electromagnet, with a modulation frequency $f_{mod}$ = 33 Hz and a field amplitude $B_{max}$ = 4-5 mT, leads to the appearance of a peak in the second harmonic (66 Hz) in the region of the SC transition (Fig. 3b).

A BSCCO crystal (Bi-2122, $\bar{c}||\bar{H}$) installed between the same single-turn coils on printed circuit boards behaves similarly. At $f_{carr}$ ranging from 1 to 10 MHz, we observed a pronounced peak in the second harmonic of the low-frequency oscillating field ($f_{mod}$ = 19 Hz) (Fig. 3c). Finally, we placed a BSCCO microparticle, with $d \approx 100$ μm and $t \approx$ 5–10 μm, in a test high-pressure DAC with sputtered Lenz lenses. For $f_{carr}$ = 32.7 MHz, we were able to reliably detect the superconducting transition in the form of a step in both the real and imaginary part of the transmission voltage (Fig. 3d). Several additional experiments with another DAC and a 40 μm wide sample (from the same BSCCO sample) show good reproducibility of the signal position (Supplementary Fig. S6-S7).

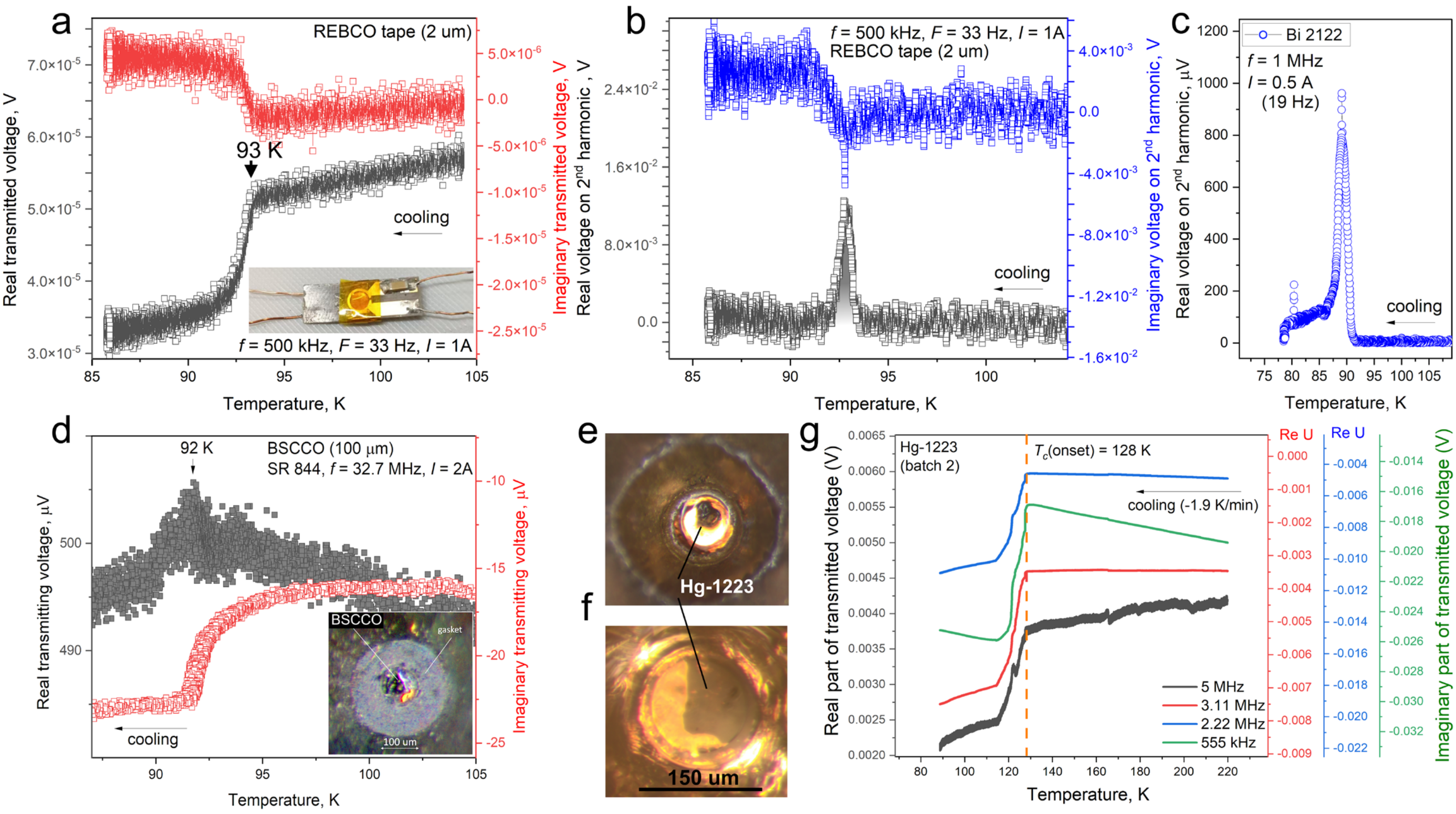

**Fig. 3 Temperature dependence of RF transmission signal in high-$T_c$ superconducting samples in weak AC magnetic field. a** REBCO tape (with a superconducting-layer thickness of $t \approx 2$ μm and a diameter of $d \approx 4$ mm) investigated in the configuration of a single-turn RF transformer. Real (Re) and imaginary (Im) parts of the transmission voltage, $U$ (measured at a carrier frequency, $f_{carr} = 500$ kHz). Inset: Image of the sample in the RF-transformer configuration. **b** Re $U$ and Im $U$ of the second harmonic of the modulation field ($f_{mod}$= 33 Hz) in the region of the SC transition. **c** Second harmonic for a BSCCO sample ($d \approx 4$ mm, $t \approx 300$ μm), measured in the same configuration as in **a** and **b**, at a carrier frequency of 1 MHz. **d** Re U and Im U($f_{carr} = 32.7$ MHz) measured for a BSCCO microparticle ($d \approx 100$ μm) in a test high-pressure DAC equipped with Lenz lenses at 0GPa. Inset: Image of the sample on the diamond anvil surrounded by the gasket. **e** Sample of Hg-1223 (batch 2) superconductor before compression. **f** Particle of Hg-1223 after compression in a test DAC. **g** Radio-frequency detection of superconducting transition in a Hg-1223 sample (batch 2) using four carrier frequencies (5, 3.11, 2.22 MHz and 555 kHz) in a cooling cycle (–1.9 K/min).

Moving on to tests of higher-$T_c$ superconductors, we investigated the mercury-containing cuprate superconductor Hg-1223 ($HgBa_2Ca_2Cu_3O_{8+\delta}$) in a test DAC. In the first experiment (batch 1, low quality sample), the initially compact crystal was broken into many particles after the diamond anvils were brought together and the DAC was closed. As a result, a large number of empty gaps were formed on the culet between individual crystallites and the efficiency of RF transmission measurements was significantly reduced (Supplementary Fig. S8b,c). Nevertheless, for $f_{carr} = 175$ MHz, $f_{mod} = 19$ Hz, and $B_{max} = 3$ mT we were able to resolve a peak-like feature around 141 K in the real and imaginary parts of both the transmission voltage as well as the second harmonic of the low-frequency modulation (Supplementary Fig. S8b,c). This value of the transition temperature corresponds to a small pressure buildup of about 2-3 GPa in the test DAC as it cools down[36]. The superconducting transition signal and signal-to-noise ratio was improved by using a second batch of Hg-1223 with higher superconducting volume fraction (batch 2, particle ~80 μm, Fig. 3e-g).

Thus, we hereby demonstrated that the detection of superconductivity in macroscopic samples and microparticles could reliably be achieved with the experimental setup, combining DACs and Lenz lenses with the RF transformer configuration described above.

### Feasibility tests II: Detection of SC and its field-dependence in BSCCO, NbTi and $MgB_2$

An anomaly in the RF signal transmission is not direct evidence for the superconducting nature of the underlying transition in the sample. A key signature of superconductivity is its response to an external magnetic field. To study the behavior of RF transmission anomalies and its evolution with magnetic field, $H$, we firstly chose a massive BSCCO sample ($d = 4$ mm, $t = 0.3$ mm) coupled to a single-turn transformer at ambient pressure in a 16 T magnet system. In addition, we conducted similar measurements on a thin (t = 0.2 mm) REBCO-tape sample (Supplementary Note 4, Supplementary Figs. S10), confirming the reliability of the preceding results. Experiments with millimeter-scale samples yield very good signal (Fig. 4) and serve as proof of concept for radio-frequency detection.

Next, as a comparison, we mounted a powder sample of $MgB_2$ ($d$ = 120 μm, Shanghai Aladdin Biochemical Technology Co., LTD) and a micron-sized crystalline sample of NbTi ($d \approx 30$ μm) to different test DACs, equipped with Lenz lenses, and studied their response in fields up to 2 and 6 T, respectively, under zero-pressure conditions (Supplementary Fig. S11 c, d).

For the massive BSCCO sample, we were able to detect a step-like feature in the RF transmission voltage at around 90 K that shifts to lower temperatures upon increasing the external magnetic field aligned perpendicular to the crystallographic $c$ direction (Fig. 4a). A clear feature is discernible in both the real and imaginary parts of the signal (Fig. 4a and b). The obtained value of the derivative $d\mu_0 H_{c2}/dT|_{T=Tc} \approx -2.8$ T/K is consistent with previous reports for $H \perp c$[21,37]. In these measurements, we did not include the low-frequency modulation and, hence, were not able to collect information about the second harmonic. From Figs. 4a, b it can be seen that the step in the RF transmission signal broadens as the magnetic field increases.

For the finely dispersed powder sample of $MgB_2$, we observe a step-like feature as well that we associate with the superconducting transition. This is confirmed by the observed shift towards lower temperatures upon increasing magnetic field (Figs. 4c, d). Anisotropic upper critical magnetic field in $MgB_2$ depends significantly on the morphology and impurities of the sample. Nevertheless, the field dependence matches well the previous results from transport measurements [38,39].

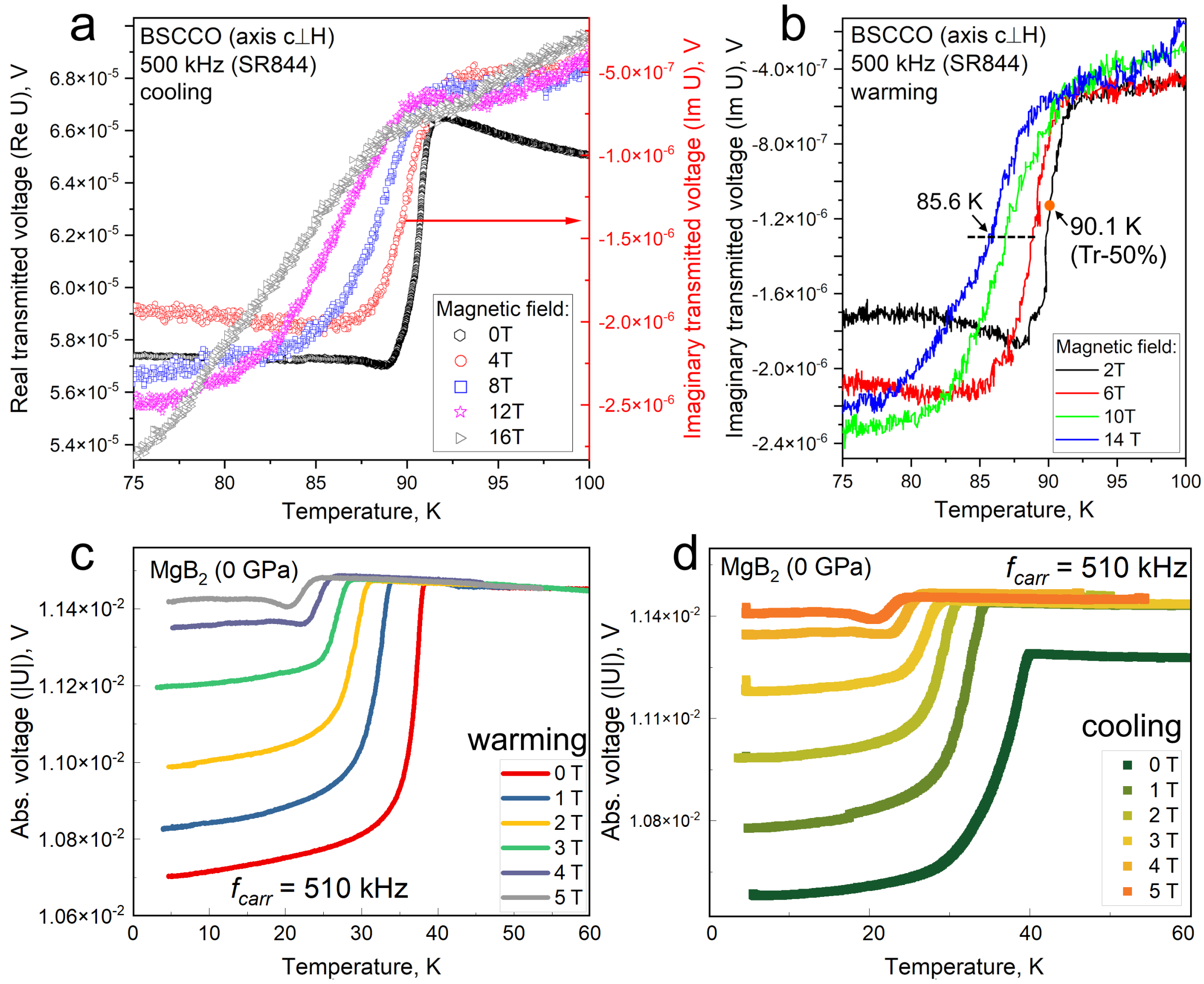

**Fig. 4 The effect of external magnetic fields on superconductivity studied by RF transmission detection.** **a** Temperature dependence of the transmitted signal (mainly the real part) at 500 kHz through a single-turn transformer with a BSCCO core ($d$ = 4 mm, $t$ = 0.3 mm) in magnetic fields of 0, 4, 8, 12, 16 T. Cooling cycle. **b** Same as in panel "a", but for the imaginary part of the signal in the heating cycle in magnetic fields of 2, 6, 10, and 14 T. The half-height (Tr-50%) of the step, indicated by a dotted line and an orange circle, was used to determine $B_{c2}(T)$. **c** Temperature

dependence of the transmitted signal (absolute value) at 510 kHz through a multi-turn transformer with a $MgB_2$ core ($d$ = 2 mm, $t$ = 2 mm) in magnetic fields of 0-5 T. Warming cycle. **d** Same as in panel "c" but for cooling cycle.

The NbTi sample was investigated at low temperatures in a test DAC, made from BeCu, with a diamond-anvil culet size of 30 μm. The sample of approximately the same diameter ($d \approx 30$ μm, $t$ = 1–2 μm) was placed in an ammonia borane ($NH_3BH_3$) medium at zero pressure (0 GPa). We used anvils with deposited Lenz lenses that were etched into the metal film with the help of a Ga focused-ion-beam system. BeCu DACs usually create some additional pressure during cooling, due to thermal contraction of the DAC's body material. This, in turn, may lead to some non-uniform stress distribution in the sample, potentially affecting its superconducting properties. As the measurement results show (Supplementary Fig. S11d), even such a small sample can be successfully investigated using closely adjacent Lenz lenses.

To confirm the superconducting nature of the observed transition, we performed a multi-frequency study of the sample's RF transmission in magnetic fields of 3 T and 6 T (Fig.4e). We used a step of 3T due to the high upper critical field $B_{c2}(0)$ of NbTi, exceeding 14 T. We found that the error of in determining the $T_c$ by the high-frequency method is about ±1 K (Supplementary Figs. S11b, S13). In this case, the most convenient for comparison is the step-like signal, which appeared in our system at different frequencies from 50 to 150 MHz. Second harmonic generation cannot be used here because of the large Ampere forces acting on the solenoid from the external magnetic field. Comparison with the transport measurement data shows that the studied sample has a slightly lower $T_c$ (Supplementary Figs. S11b). Nevertheless, the obtained points of the phase diagram are in agreement with the transport measurements.

It is important to note that the RF method is inconvenient for low-temperature measurements ($T$< 15 K). Because our studies used Ta-based Lenz lenses, and various tantalum carbides could form during deposition of Ta on diamond, we observed additional parasitic signals corresponding to superconducting transitions in various parts of diamond anvil cells, and possibly also in the Pb-based solder of wires. Such parasitic transitions significantly complicate the study at low temperatures. Nevertheless, the study of high-$T_c$ superconductivity is free from this drawback (see the following paragraphs).

**Feasibility tests III: Detection of magnetic and metal-to-insulator transitions**

By testing other types of phase transitions, we demonstrate the ability/disability of the RF transmission method to identify the difference in the behavior of magnetic, superconducting and phase transitions, which allows them to be distinguished using the proposed technique.

As a next target, we investigated the paramagnetic-to-ferromagnetic (PM → FM) transition in gadolinium. With the literature value $T_C$ = 293 K, this phase transition is convenient for our tests, in that it occurs near room temperature within a narrow temperature interval. Recalling the theory of transformers, we immediately notice that a ferromagnetic core should significantly improve the operation of a high-frequency transformer (see discussion in Supplementary Note 5). Indeed, for Gd we observed a sharp peak in the RF transmission voltage (Fig. 5a, b). We repeated this measurement for various frequencies ($f_{carr}$ = 25, 50, 200, and 300 kHz). The determined $T_C$ varies between 289 and 296 K depending on $f_{carr}$. Independent of $f_{carr}$, the detected feature around $T_C$ for Gd describes a sharp peak(s) in the Re U and Im U signals (see also Supplementary Fig. S15). The presence of two peaks in Fig. 4b may indicate the formation of a domain structure in ferromagnetic Gd below $T_C$, or non-uniform pressure distribution across the sample. It is important to note that there is also a pronounced signal at the second harmonic (*2F*) of the modulating field (inset in Fig. 5b).

We also investigated a small sample of Terbium ($d$ = 200 μm) mounted to a test DAC at near zero-pressure conditions. This material, exhibits multiple magnetic phase transitions: at the Néel temperature, $T_N$ = 230 K, it transitions from a PM to an antiferromagnetic (AFM) ground state, and below the Currie temperature of $T_C$ = 218 K it turns ferromagnetic (FM). We were able to detect these magnetic transitions with the RF transmission method (Fig. 5c, Supplementary Fig. S16). The detected AFM transition temperature value varied between 220 and 230 K indicating a possible increase in pressure in the test DAC during cooling due to thermal contraction of the DAC's material within a few GPa. Notably, features related to magnetic transitions mainly exhibit a peak and can also be picked up by the 2nd harmonic of the modulation field (Fig. 5d).

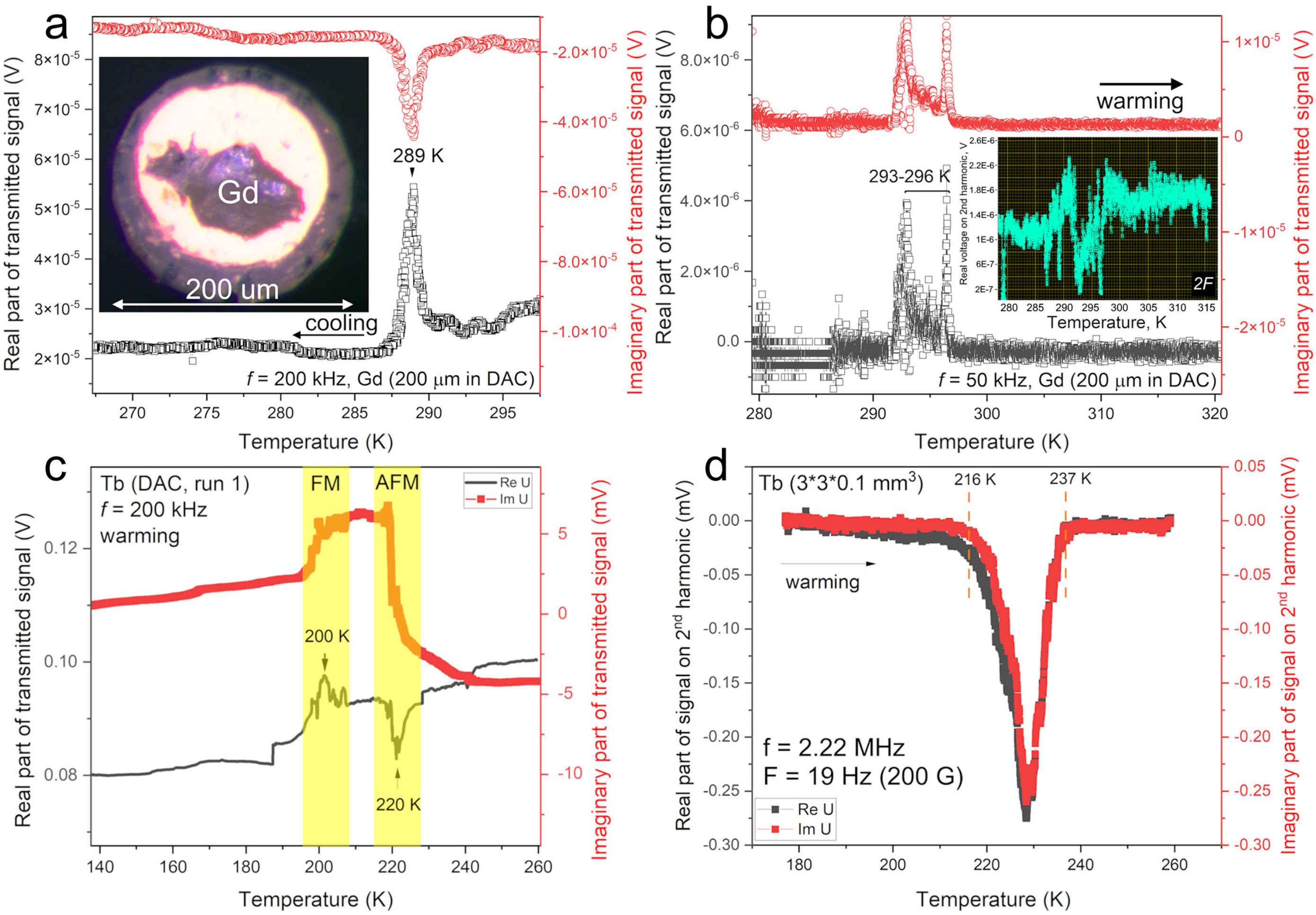

**Fig. 5 Detection of magnetic transitions in Gd and Tb by RF transmission method. a** Temperature dependent real and imaginary parts of the transmitted RF signal ($f$ = 200 kHz) measured for a Gd sample ($d \approx 200$ μm) in the cooling cycle (rate: 3K/min). Inset: photograph of the test DAC's chamber with the loaded Gd sample at 0 GPa. **b** Real and imaginary parts of RF signal at 50 kHz carrier frequency in a warming cycle for the same sample. Differences in transition temperature may be due to differences in the $T$-cycle type (cooling/warming). Inset: the second harmonic (2$F$) signal in its direct demodulation ($f_{mod}$ = 19 Hz). **c** Temperature dependent real and imaginary parts of the transmitted RF signal ($f$= 200 kHz) measured for a Tb sample ($d \approx 200$ μm) in a warming cycle. **d** Magnetic transitions in Tb investigated for a bulk millimeter-size piece of Tb (3×3×0.1 mm$^3$, see also Supplementary Fig. S16).

## RF detection of high-temperature SC in La and La-Ce superhydrides under megabar pressure

Finally, after demonstrating that we are fully capable of detecting superconducting transitions of very small samples in a high-pressure DAC via the RF transmission method, we can turn to the most topical high pressure superconducting materials. The goal of this work is to detect the superconducting transition in metallic superhydrides with a complementary RF transmission method. Some results of radio-frequency studies of superconductivity in hydrides at high pressure were presented by us in previous publications on the La-Sc-H system [25] at a pressure of 150-200 GPa, and phases of the composition $LaH_{\sim 12}$ [10]. As part of this study, we chose the well-studied cerium $CeH_{9\text{-}10}$ [40], lanthanum $LaH_{10\text{-}x}$ and lanthanum-cerium polyhydrides $(La,Ce)H_{9\text{-}10}$. The last compound is remarkable since it has the highest $T_c$ at the lowest possible pressure among superhydrides to date. Previous works provided electrotransport evidence for superconductivity with $T_c$(onset) ≈ 180–190 K at a pressure of 110–120 GPa [41,42].

Cerium superhydrides are stabilized at very low pressures of approximately 80-90 GPa [40,43,44], and in combination with lanthanum can be stabilized even at 70 GPa [45]. Unfortunately, the Cooper pair scattering by local magnetic moments of the Ce $f$-electrons and spin fluctuations suppress superconductivity and reduce $T_c$ in Ce-containing polyhydrides [46]. Studies of structure and properties of $CeH_9$ and $CeH_{10}$ have been conducted repeatedly, in particular, using four-probe transport measurements in magnetic fields [40,47], and NV centers magnetometry implanted in a diamond anvil [48]. However, cerium hydrides had never before been studied using noncontact techniques (low frequency AC susceptibility, SQUID magnetometry) to measure changes in surface conductivity and diamagnetic permeability. Here, we addressed this gap by employing the proposed radio-frequency transmission technique (Fig. 6).

We prepared two high-pressure DACs loaded with Ce/$NH_3BH_3$ and heated by an IR laser (1.064 μm) at 100 and 120 GPa. In both cases, we used a setup with single-stage Lenz lenses made of molybdenum, ≈100 μm in diameter, sputtered directly onto both diamond anvils and connected directly to an RF generator and an SR844 lock-in amplifier. The resistance of the single-turn coils was approximately 20 Ω at room temperature. RF test measurements in this configuration were performed for YBCO and showed a good signal-to-noise ratio of *S/N* ≈ 20 (Supplementary Fig. S17).

An investigation of the RF signal transmission at 75 MHz reveals a small anomaly at 91–92 K both in the high-frequency channel (Fig. 6b) and at the second harmonic (2×33 Hz) of the modulating magnetic field of approximately 20 Gauss (Fig. 6c). The low signal-to-noise ratio (S/N ~1) corresponds to a low filling factor of the single-turn cavity by the sample (≈ 9%). X-ray synchrotron diffraction study of the sample at 100 GPa shows the formation of pure *hcp*-$CeH_9$ (V = 33.6 Å$^3$/Ce).

A higher filling factor of the single-turn coil's cavity achieved within the second experiment at 120 GPa leads to a significantly more pronounced RF signal of superconducting transition in the high-frequency (116 MHz, Fig. 6d) channel, as well as at the 2$^{nd}$ harmonic signal (2×19 Hz, Fig. 6e). Unlike the experiment at 100 GPa, the sample apparently also contains $CeH_{10}$, which accounts for the anomaly at 106 K (Supplementary Fig. S18).

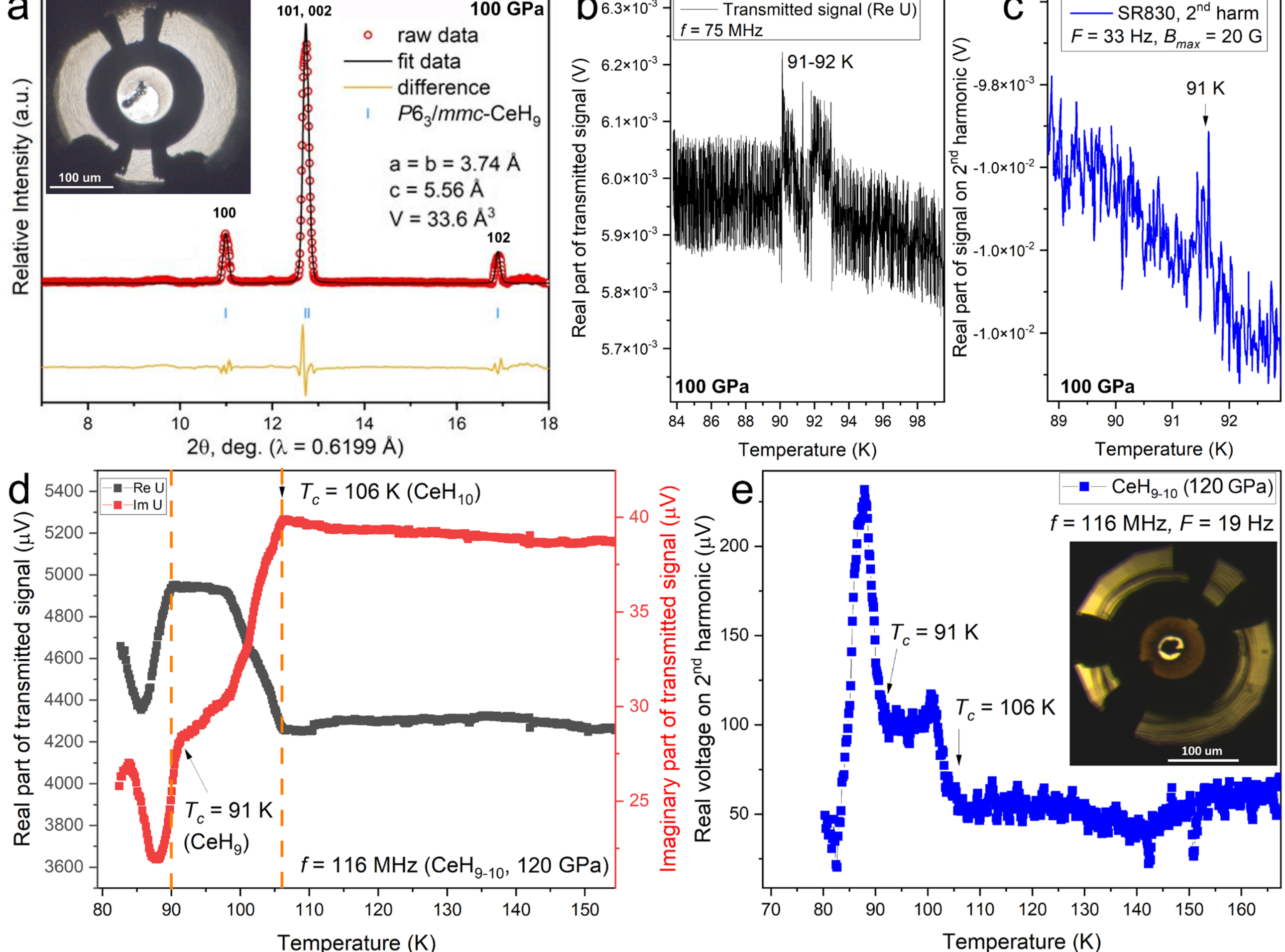

**Fig. 6. Radio-frequency characterization of cerium superhydrides under high pressure**. **a** X-ray diffraction pattern of the sample at 100 GPa (red circles) accompanied by a Le Bail refinement (black curve) of the *P*6$_3$/*mmc*-$CeH_9$ phase (V = 33.6 Å$^3$/Ce). The orange curve represents the difference profile. Inset: photo of the sample inside the chamber. **b** Temperature dependence of the real part of the transmitted radio-frequency signal at $f$ = 75 MHz under a pressure of 100 GPa, showing a distinct anomaly in the vicinity of 91–92 K. **c** Corresponding second-harmonic signal anomaly at modulation frequency $F$ = 33 Hz, and magnetic field amplitude $B_{max}$ = 20 G recorded at 91 K for the same sample. **d** Transmitted RF signal components: real (black curve) and imaginary parts (red curve), measured at $f$ = 116 MHz for a mixed-phase $CeH_9$–$CeH_{10}$ sample synthesized at 120 GPa. The steps indicate two superconducting transitions at $T_c$ =

91 K and $T_c$ = 106 K. **e** Corresponding real voltage part of the second-harmonic signal ($f$ = 116 MHz, $F$ = 19 Hz) for the $CeH_{9-10}$ mixture at 176 GPa, clearly resolving the two distinct transition temperatures. Inset: photo of the $CeH_9$ -$CeH_{10}$ sample inside the chamber.

Recently, several new superconducting compounds with a hexagonal structure have been discovered and studied in the La-H system under pressure: $hP$-$LaH_{9-10}$ [49,50] with $T_c$ = 160 K and 216 K, respectively, as well as $hP$-$LaH_{12}$ [51] which exhibits a pronounced transition in the radio-frequency transmission at 270-280 K, which may indicate superconductivity in this compound. In the first test DAC, we synthesized hexagonal (refined as $P6_3/mmc$) lanthanum superhydride with the unit cell volume of 33.2 Å$^3$/La (Fig. 7c) in a mixture with tetragonal hydride $LaH_{4+x}$ (V = 27.9 Å$^3$/La) and another uninterpreted compound (marked *) at 153 GPa starting from lanthanum metal and $NH_3BH_3$ using the standard procedure (see, for example, Ref. [52]). The resulting sample showed a pronounced signal in RF measurements at a carrier frequency of 90 MHz at 220-225 K in both the high-frequency (Fig. 7a) and low-frequency (2$^{nd}$ harmonic, Fig. 7b) channels. This shows that the high frequency signal from the sample is sensitive to the weak magnetic field and is suppressed in the region ≈ 5 K below $T_c$, where lower critical magnetic field $B_{c1}(T)$ < 60 Gauss. Extrapolation based on the Ginzburg-Landau theory [53,54] dependence $H_{c1}(T) \propto (1 - T/T_c)$ leads to the following estimation $B_{c1}(0)$ ~ 0.27 T.

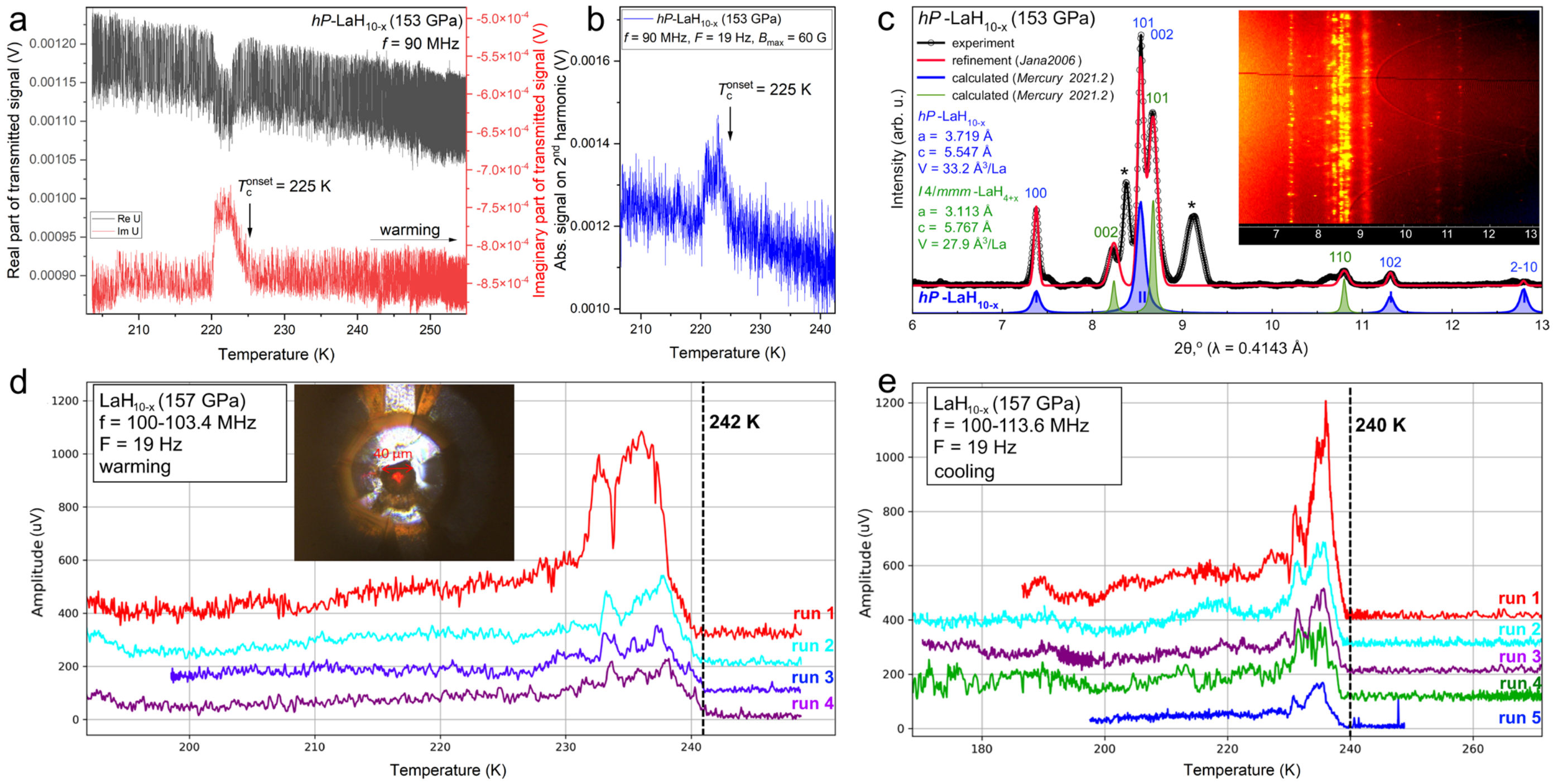

**Fig. 7. Superconducting transitions and X-ray diffraction of $LaH_{9-10}$ under pressure of about 1.5 Mbar**. **a** Real (black) and imaginary (red) parts of the RF transmission voltage ($f_{carr}$ = 90 MHz) measured for $LaH_{10-x}$ at 153 GPa in the warming cycle. An onset of the transition is at 225 K. **b** Signal at the second harmonic of the sinusoidal modulating magnetic field ($B_{max}$ = 60 Gauss, $F$ = 19 Hz) obtained from the $hP$-$LaH_{10-x}$ sample. **c** Typical powder X-ray diffraction pattern (λ = 0.4143 Å, beam size is 4×8 μm$^2$) from the $hP$-$LaH_{10-x}$ sample containing also the tetragonal phase $LaH_{4+x}$. Le Bail refinement is shown by the red curve, the experimental pattern is shown by the black one. Inset: X-ray diffraction image ("cake") used for the integration. **d, e** The absolute voltage at the second harmonic of the modulating magnetic field ($B_{max}$ = 40-110 Gauss, $F$ = 19 Hz) from the second test sample with $LaH_{10-x}$ and $T_c$(onset) = 240-242 K. The signal is well reproduced in the warming (d) and cooling (e) cycles (rate: 2-3 K/min). Inset: optical photograph of the sample at 157 GPa with laser spot indicating the point where the pressure was measured.

In the second experiment with lanthanum superhydrides, we used an identical synthesis procedure from La and $NH_3BH_3$ at a slightly higher pressure of 157 GPa. This resulted in a hydride similar in properties to $LaH_{10}$ with an onset critical temperature of approximately 240–242 K (Fig. 7d, e). The appearance of the second harmonic signal within the RF measurements ($f_{carr}$ = 100-113 MHz) was observed during numerous (9 experiments) warming and cooling cycles as part of reproducibility tests. During subsequent measurement cycles at different carrier frequencies, both DACs showed degradation of the useful signal (run 1-5) followed by cracking of the diamond anvils, indicating the possible destructive nature of the RF induced currents in superhydride samples.

Already in our previous work on the properties of (La,Ce)$H_{9-10}$[42], we noted that in the La-Ce-H system, in addition to the main phase with $T_c \approx 190$ K, there are other impurity hydride phases with higher $T_c > 205$ K. An independent synthesis using the $La_6Ce$ alloy led to a polyhydride with a pronounced resistive transition at $T_c$(onset) $\approx$ 202-205 K[17] (Supplementary Fig. S21). However, we can ask ourselves a question: is this the full picture for the La-Ce-H system? Could additional phases or properties remain undetected by four-electrode transport measurements, due to their isolation from the current path? Additional anomalies in the resistance detected above the main transition (Supplementary Fig. S21) may indicate presence of impurities with higher $T_c$'s in the sample and therefore may be overlooked by electrical transport measurements. To clarify this question, we performed a RF-transmission study of a La-Ce hydride sample.

For the RF study, we synthesized a sample of (La,Ce)$H_x$ using standard procedure, $La_6Ce$ alloy and $NH_3BH_3$ as starting materials. Our X-ray characterization revealed a new, likely hexagonal, polyhydride with chemical formula of (La,Ce)$H_{12}$ (see further details in Supplementary Note 8). We identified cubic $Fm\bar{3}m$-(La,Ce)$H_3$, and possibly tetragonal hydride (La,Ce)$H_{4-x}$ as impurities (Supplementary Figs. S25, S26, Table S3). A new, previously unknown La-Ce polyhydride with small-angle diffraction reflections (at 6.6°) is also present in the mixture (Supplementary Fig. S28). Taking into account multiphase composition of this sample and previous XRD studies, we will further use the chemical formula of (La,Ce)$H_{10-12}$ or (La,Ce)$H_{10+x}$ for this sample.

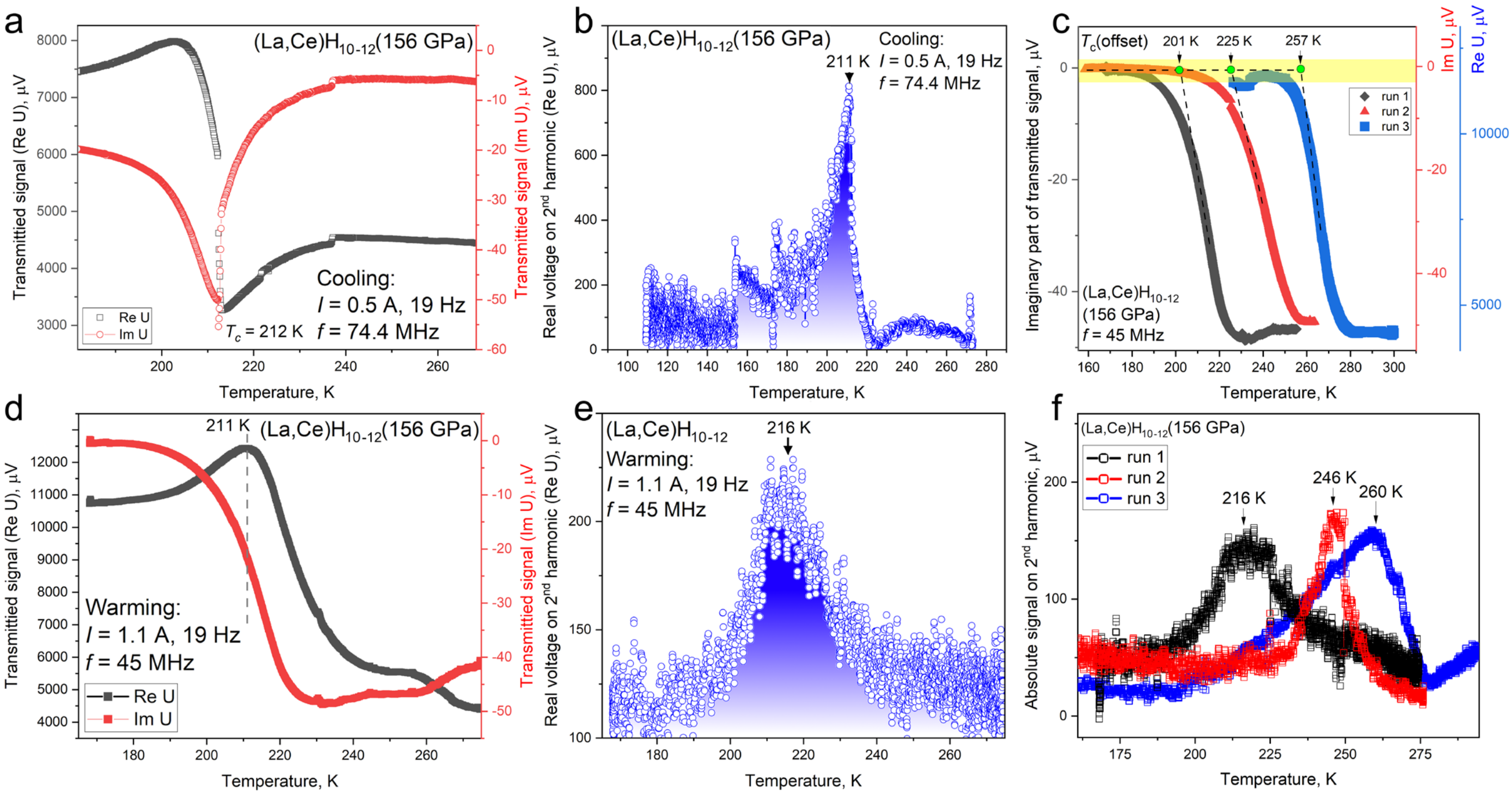

**Fig. 8 Superconducting transitions in (La,Ce)$H_{10-12}$ under pressure**. **a, d** Real and imaginary parts of the RF transmission voltage ($f_{carr}$ = 74.4 and 45 MHz) measured for pressurized (La,Ce)$H_{12}$ at 156 GPa. A middle point of the transition can be observed at about 212 K during a cooling cycle. **b, e** Two different experiments detecting the second harmonic of the modulation at $f_{mod}$ = 19 Hz, revealing the transition middle point at ≈211 K and 216 K, respectively. **c, f** Time evolution of the observed anomaly in the RF signal and in the second harmonic during three warming and cooling cycles (runs 1-3). **c** Real (Re $U \sim mV$) and imaginary (Im $U \sim \mu V$) components of the transmitted RF signal as a function of temperature at 74.4 MHz (run 1), 45 MHz (runs 2, 3). **f** Temperature dependence of the amplitude of the signal at the second harmonic of the modulating field in runs 1-3.

We conducted various measurements at 14.6, 45, 55, 74.4 MHz, and at many other frequencies (Fig. 8, and Supplementary Figs. S22-S24) confirming that the obtained La-Ce polyhydride sample exhibits a very pronounced feature in the RF transmission signal, centered around 211-215 K (Fig. 8a-d). Along with the clear feature in the RF channel, we observed a characteristic broad peak in the second harmonic of the low-frequency ($f_{mod}$ = 19 Hz) modulated signal (Fig. 8b, d). The determined transition temperature is even higher than the one reported for a similar (La,Ce)$H_{10+x}$ sample at 148 GPa from transport measurements [17]. However, it was also reported that additional anomalies in the resistance were resolved above $T_c$ that are suggestive of superconducting transitions in parts of the sample, likely associated with different superconducting phases [17].

Notably, the onset of the detected feature in the second harmonic (Figs. 8b, d) reaches close to a temperature of 220-240 K, or even higher, indicating that our sample may contain grains with significantly enhanced $T_c$. The observed critical temperature is comparable to the $T_c$ reported for $LaH_{10}$. One explanation for such an early onset of the superconducting transition in the La-Ce polyhydride may be the presence of a new phase modification, namely hexagonal *hP*-(La,Ce)$H_{12}$. Our X-ray characterization confirmed the presence of this phase (see diffraction pattern in Supplementary Fig. S26). The observed diffraction pattern from the sample is very similar to that reported earlier in the work on $LaH_{12}$ [10]. This is not surprising, given the high lanthanum content in the initial alloy ($La_6Ce$) and the great chemical similarity of Ce and La. Indeed, for example, in the La-Ce-Th series, all three elements form identical structures of $XH_{10}$ and $XH_9$ compositions.

Changing the pressure in the La-Ce DAC leads to a regular change in the unit cell volume of this hydride (Supplementary Fig. S27 and Table S3). After a pressure drop to 18 GPa due to cracking of anvils, the X-ray diffraction signal disappeared. Estimating the critical temperature in $LaH_{12}$ as $\approx 280$ K[10], and the negative influence of *f*-electrons of cerium on $T_c$ of Ce-containing superhydrides via simple linear formulas

$$T_c(A_xB_{1-x}H_n) \approx xT_c(AH_n) + (1-x)T_c(BH_n), \qquad (1a)$$

$$T_c(AH_n)/T_c(AH_{n+1}) \approx T_c(BH_n)/T_c(BH_{n+1}), \qquad (1b)$$

one can obtain $T_c \approx 130$ K for $CeH_{12}$ and $T_c \approx 258$ K for (La,Ce)$H_{12}$ with x = 0.86. In formulas 1a, b, the "A" is heavy atom No. 1, "B" is heavy atom No. 2 (the properties of atoms A and B must be close), "H" is a hydrogen atom, $n$ is the atomic content of hydrogen in the hydride, $x$ is the concentration of element A in the alloy $A_xB_{1-x}$. This may explain why in this experiment (Fig. 8c,f) the onset of the RF transmission feature is so high in temperature.

Indeed, during the reproducibility tests (at least 20 warming/cooling cycles were performed with the La-Ce DAC), we noticed that the probable superconducting transition temperature changes with time, likely due to changes in the morphology and composition of the sample, caused by heating from induced RF currents, migration of hydride ions in an electric field, as well as from the thermal cycling of the DAC. Indeed, COMSOL simulation estimates of the induction heating power (Fig. 2e, f) yield a density of $\sim 10^4$ W/m$^2$. This dissipation power for isolated microsamples of ~100 μm would result in their rapid heating by hundreds and thousands of Kelvins. However, if there is good thermal contact with the diamond surface, the sample temperature change is < 1 K.

Intriguingly, for the (La,Ce)$H_x$ sample, we observe an increase in the transition temperature from 214-230 K (Figs. 8b-e) up to roughly 260 K (Figs. 8c, f) and even higher (Supplementary Figs. S22-S24). Note that the anomaly in the RF transmission through the (La,Ce)$H_{10-12}$ sample is clearly visible both in the high-frequency channel (Figs. 8a,d, and c) and in the second harmonic of the modulating AC field (Figs. 8b,e, f).

## Discussion

The important capability of the RF transmission method with Lenz lenses is its sensitivity to superconducting transitions in heterogeneous samples under extreme pressure. This approach can detect contributions from impurity-hydride phases that are non-uniformly distributed over the volume of the sample, or located in isolated grains, separated by non-superconducting grains of molecular or lower hydrides. By contrast, the detection based on electrical resistance usually only probes the superconducting transition of the dominant phase and, in particular below $T_c$, becomes insensitive to additional phases (Supplementary Fig. S3). The RF transmission method enables the identification of superconducting transitions in such inhomogeneous samples.

The advantages of the proposed method with Lenz lenses with respect to conventional AC-susceptibility (see, for example, Ref. [55]) or four-contact transport probes are:

- A high sensitivity to micron-size samples (signal-to-noise ratio > 50 at 1.5 Mbar);

- The stability of Lenz lenses with respect to significant surface damage;
- The possibility of detecting “hidden” phases in very inhomogeneous samples;
- A low-effort manufacturing process and rather simple use of Lenz lenses.

Moreover, the RF transmission method offers the possibility of studying not only superconducting, but also other phase transitions, such as magnetic or metal-insulator transitions.

Even though this method is superior to electrical transport, there are also specific drawbacks: a very low noise level of the cryogenic system and the experimental setup is required (e.g., stable gas flow, stable basement, remote control of the system); For a proper confirmation and identification of transitions it is essential to conduct multiple repeated scans at different frequencies and their harmonics in both cooling- and warming-mode operations; The overall structure of features in the RF transmission signal is complex and may change for each measurement cycle. The latter effect is related to potential drifts in the inductance and capacitance of connecting coax wires during cooling/warming cycles; Measurements in magnetic field are complicated by a broadening of RF features.

In general, the RF transmission is only sensitive to local changes of the sample that require a complementary method in order to identify the physical nature of the observed features. Consequently, RF transmission provides only additional information, this is not a first-choice method. However, combined with the second-harmonic detection at low frequency, it could provide a solid proof of Meissner-type response of the samples, as suggested by Timofeev et al.[22]

Supplementary Table S1 presents a summary on the typically observed RF anomalies in the case of superconducting, magnetic, or other phase transitions in the sample. Experimental practice shows that it is possible to distinguish between these transitions. Superconducting and magnetic transitions usually induce a narrow signature in the RF channel and a strong second harmonic of the transmission signal, induced by a low-frequency modulation of the magnetic field. On the contrary, a broad RF signature accompanied with a large temperature hysteresis indicates first-order phase transitions. Notably, the magnetic transition signal is well pronounced at frequencies of hundreds of kilohertz. However, in most cases it subsides for frequencies above 10 MHz caused by the inertia of spins that have not enough time to respond to such a rapid change in the magnetic field.

## Conclusions

We have demonstrated the detection of superconducting, magnetic, and other phase transitions under extreme pressure by a novel radio-frequency (RF) transmission method employing Lenz lenses. Using examples of Bi-2212, YBCO, REBCO, Hg-1223, $MgB_2$, NbTi, $CeH_{9-10}$, $LaH_{9-10}$ and $(La,Ce)H_{10-12}$, we have shown that RF transmission combined with the detection of the second harmonic signal of the modulating AC magnetic field is an effective complementary method for detecting superconductivity and investigating the phase diagram of small-volume samples in high magnetic fields, at least up to 10 T. Furthermore, using examples of Gd and Tb, we have demonstrated that this method is highly effective for detecting magnetic phase transitions (ferromagnetic and, in some cases, antiferromagnetic), as well as metal-insulator transitions.

Due to its high sensitivity, the RF transmission method can be used cautiously as a complementary technique alongside classical transport measurements. The method provides useful information and compatible with powder and single-crystal XRD studies. It can potentially be combined with electrodes, which enables the application of multiple complementary experimental techniques to a sample in one and the same DAC: NMR, RF, electrical transport, and single-crystal or powder XRD studies can be performed all in one diamond anvil cell. This would significantly reduce the experimental effort and time for comprehensive high-pressure studies. This approach further qualifies to be combined with, e.g., Nitrogen-vacancy detection or optical detection methods, such as IR transmission or femtosecond spectroscopy, for the study of phase transformations of matter under high pressure.

Lenz lenses, unlike electrodes, exhibit high stability against damage and compression. They can be prepared using the simplest metal deposition techniques in combination with lithography assisted by laser cutting, focused ion beam, or other etching approaches. In our opinion, the most important application of the developed RF method is the detection of superconducting phases with high $T_c$’s in non-uniform superhydride samples. A detection of multiple phases is challenging for resistivity measurements. Hence, the RF-method offers a sensitive complementary tool.

## Methods

### Radio-frequency transmission method

The SR844 lock-in amplifier (Stanford Research Systems) was used as a source and analyzer of the radio-frequency signal. Specific frequencies for study were selected based on the output signal level of several millivolts or higher. To create an external modulated magnetic field, a signal from the built-in generator of the lock-in amplifier SR830 was used, amplified by the Yamaha PX3 audio frequency power amplifier. The wires connecting the sample and the lock-in amplifiers were selected to be of minimal length and were fixed at the maximum distance from each other to minimize mutual induction. To reduce vibration in the system, all wires were fixed using glue, plastic ties, and blue tack. All data are presented as obtained; no background signals were subtracted.

When it is possible to determine, the $T_c$(onset) corresponds to the onset of the RF transmission anomaly or the onset of the second harmonic signal. However, for inhomogeneous multiphase samples, this is not always possible due to the very smooth onset of this anomaly, indistinguishable from the smooth change in the background signal. In such cases, we determined the transition $T_c$ based on the position of the RF transmission signal extremum or midpoint (criterion Tr-50%).

### Samples and sample preparation

For the RF study, we synthesized a sample of (La,Ce)$H_x$ based on the $La_6Ce$ alloy prepared by arc melting of La and Ce in an argon atmosphere[17,42]. The high-pressure study was carried out in a BeCu DAC (BX-90) with diamond anvils culet size of 100 μm. Lenz lenses were prepared by magnetron sputtering of silver with subsequent lithographic process. The insulating gasket was made of Ta-W (10 at. %) alloy, pre-compressed, laser-drilled, and oxidized at a temperature of 1000-1500 °C in air as described in Ref. [35]. The hydrogen source was ammonia borane, laser heating of which at 153-156 GPa using an IR laser (1.06 μm) led to the formation of various superhydrides (Supplementary Note 8). For the synthesis of lanthanum superhydrides, we used anvils with culet diameter of 50 μm and an insulating gasket made of $CaF_2$/epoxy.

For the test experiments at 0 GPa, DACs with a diamond anvil culet diameter of 100-250 μm were used. In these cases, we sputtered two-layer Ta/Au lenses and shaped them with a Ga focused ion beam (Ga FIB, Thermo Fisher Scientific). To reduce the resistance of Ga-doped diamond after using the FIB, surface treatment with argon ions (3.5 keV, 1 min) or oxygen plasma was used. The temperature sensor (Cernox or Pt-100) was located inside DACs in most experiments, reducing the possible temperature hysteresis between cooling and warming cycles to 3-5 K. During cyclic measurements, the typical cooling or warming time was 1-1.5 hours; after 3-5 minutes of preparation, the next stage of the experiment was started (e.g., cd1 → wu1→cd2→wu2→…). The number of cooling and warming cycles for each sample did not exceed 10-12.

### X-ray diffraction

The structural characterization of $La_6Ce$ alloy and (La,Ce)$H_{10-12}$ hydrides was carried out at European Synchrotron Research Facility (ESRF, beamlines ID27, $\lambda$ = 0.3738 Å, and ID31, $\lambda$ = 0.1652 Å) and Shanghai Synchrotron Radiation Facility (SSRF, beamline BL15U, $\lambda$ = 0.6199 Å). The structural characterization of $LaH_{10-x}$ sample was performed at SPring-8 synchrotron research facility (beamline, BL10XU, $\lambda$ = 0.4143 Å). The X-ray beam size at ESRF (ID27) is less than 1 μm in diameter, around 4×8 $\mu m^2$ at SPring-8, and about 7×15 $\mu m^2$ at SSRF. The obtained two dimensional XRD images are converted to one dimensional diffraction patterns with the help of Dioptas 0.6.1[56]. The XRD patterns were analyzed using the Jana2006[57] software. For the refinement of the unit cell parameters, we used Le Bail method[58]. Further details regarding the diffraction experiments can be found in the Supplementary Information.

## Data availability

All the data supporting the findings of this study are available within the article and its Supplementary Information. Source data are provided upon reasonable request by authors.

## Acknowledgments

This work was supported by the National Key Research and Development Program of China (grant 2023YFA1608900, subproject 2023YFA1608903, and grant 2022YFA1402301). V.V.S. acknowledges the financial support from Shanghai Science and Technology Committee, China (No. 22JC1410300) and Shanghai Key Laboratory of Materials Frontier Research in Extreme Environments, China (No. 22dz2260800). D. Z. thanks the Fundamental Research Funds for the Central Universities for support of this research. I. A. T. acknowledges the financial support of RSF grant 26-12-00447. We also express our gratitude to the teams of the high-pressure stations ID27 (ESRF), BL10XU (SPring-8), and BL15U1 (SSRF) for their help in the XRD investigations. This work was supported by HLD-HZDR, member of the European Magnetic Field Laboratory (EMFL). We further acknowledge support under the European Union´s Horizon 2020 research and innovation program through the ISABEL project (No. 871106). We thank Drs. Jinyu Zhao and Shu Cai, as well as Prof. Liling Sun (HPSTAR, Beijing), for providing the magnet time. We also would like to thank Prof. Xiaoli Huang and Dr. Jianning Guo for their assistance in the preparation of cerium hydride samples and the synchrotron XRD studies. We express our gratitude to the teams of the high-pressure stations ID27 (ESRF), BL10XU (SPring-8), and BL15U1 (SSRF) for their help in the XRD investigations. We acknowledge the European Synchrotron Radiation Facility (ESRF) for provision of synchrotronradiation facilities and Momentum Transfer for facilitating the measurementsof the La-Ce alloy. Jakub Drnec is thanked for assistance and support in using beamline ESRF ID31. The measurement setupwas developed with funding from the European Union's Horizon 2020 research and innovationprogram under the STREAMLINE project (grant agreement ID 870313). Measurements performedas part of the MatScatNet project were supported by OSCARS through the EuropeanCommission's Horizon Europe Research and Innovation programme under grant agreement No. 101129751.

## Author contributions

D.V.S., D.Z., W. C., I. A. T. and V.V.S. designed the experiments, prepared the DACs, Lenz lenses and electrodes, D.V.S. did the COMSOL modelling, performed the RF transmission measurements, transport, powder X-ray diffraction studies, and analyzed the XRD data. J. Z. prepared samples and equipment for measurements. D.V.S., D.Z., and T.H. performed measurements in magnetic fields. T.H. supervised the work at HZDR. V.V.S., H.K.M., Y. D, and T.H, supervised the work at HPSTAR and HLD HZDR. All authors contributed to the discussion of the results and to the preparation of the manuscript.

# Supplementary Information

## Radio-Frequency Method for Detecting Superconductivity Under High Pressure

Dmitrii V. Semenok[1,†,*], Di Zhou[1,†,*], Jianbo Zhang[1,†],Toni Helm[2,3], Wuhao Chen[4], Yang Ding[1] , Ho-kwang Mao[5,6,] Ivan A. Troyan[6] and Viktor V. Struzhkin[5,6,*]

[1] *Center for High Pressure Science and Technology Advanced Research (HPSTAR), Beijing 100193, P. R. China*

[2] *Dresden High Magnetic Field Laboratory (HLD-EMFL) and Würzburg-Dresden Cluster of Excellencect.qmat, Helmholtz-Zentrum Dresden-Rossendorf, 01328 Dresden, Germany*

[3] *Max Planck Institute for Chemical Physics of Solids, 01187 Dresden, Germany*

[4] *Quantum Science Center of Guangdong–Hong Kong–Macao Greater Bay Area (Guangdong), Shenzhen 518000, China*

[5] *Center for High Pressure Science and Technology Advanced Research (HPSTAR), Shanghai 201203, P. R. China*

[6] *Shanghai Key Laboratory of Material Frontiers Research in Extreme Environments (MFree), Shanghai Advanced Research in Physical Sciences (SHARPS), Pudong, Shanghai 201203, China*

[*]Corresponding authors, emails: dmitrii.semenok@hpstar.ac.cn (Dmitrii V. Semenok), di.zhou@hpstar.ac.cn (Di Zhou), viktor.struzhkin@hpstar.ac.cn (Viktor V. Struzhkin).

[†]These authors contributed equally to this work

# 1. Methods

## *1.1 Details of the RF approach*

In our experiments, the sample was cooled and warmed at an average rate of about 3-6 K/min. The thermometer (Cernox or Si diode, LakeShore 335) was mounted inside high-pressure cells, which were placed in a large solenoid powered by low-frequency alternating current (AC). In experiments without high-pressure DACs (e.g., with REBCO tape), the thermometer was fixed on the top of a sample. The SR830 (time constant 300 ms) served as a low-frequency signal generator, and the Yamaha PX-3 served as an amplifier. Before starting measurements, in most cases, we performed auto phasing of the SR844 lock-in to zero the imaginary component of the transmitted signal. Given the uncertainty in the capacitive-inductive characteristics of the supply wires at different temperatures, such auto phasing can lead to an arbitrary distribution of the useful signal between its imaginary and real components at the transition temperature. The choice of the frequency (e.g., 17, 19 Hz or 33 Hz) of external AC magnetic field does not affect the measurement results and manifestation of the signal at the second harmonic. In the warming cycle, we usually used a higher solenoid current to speed up the heating. A NiCr coil or high-power ceramic resistor was used as a heater, placed far from the sample and shielded by an aluminum holder of the high-pressure cell (Figure S1).

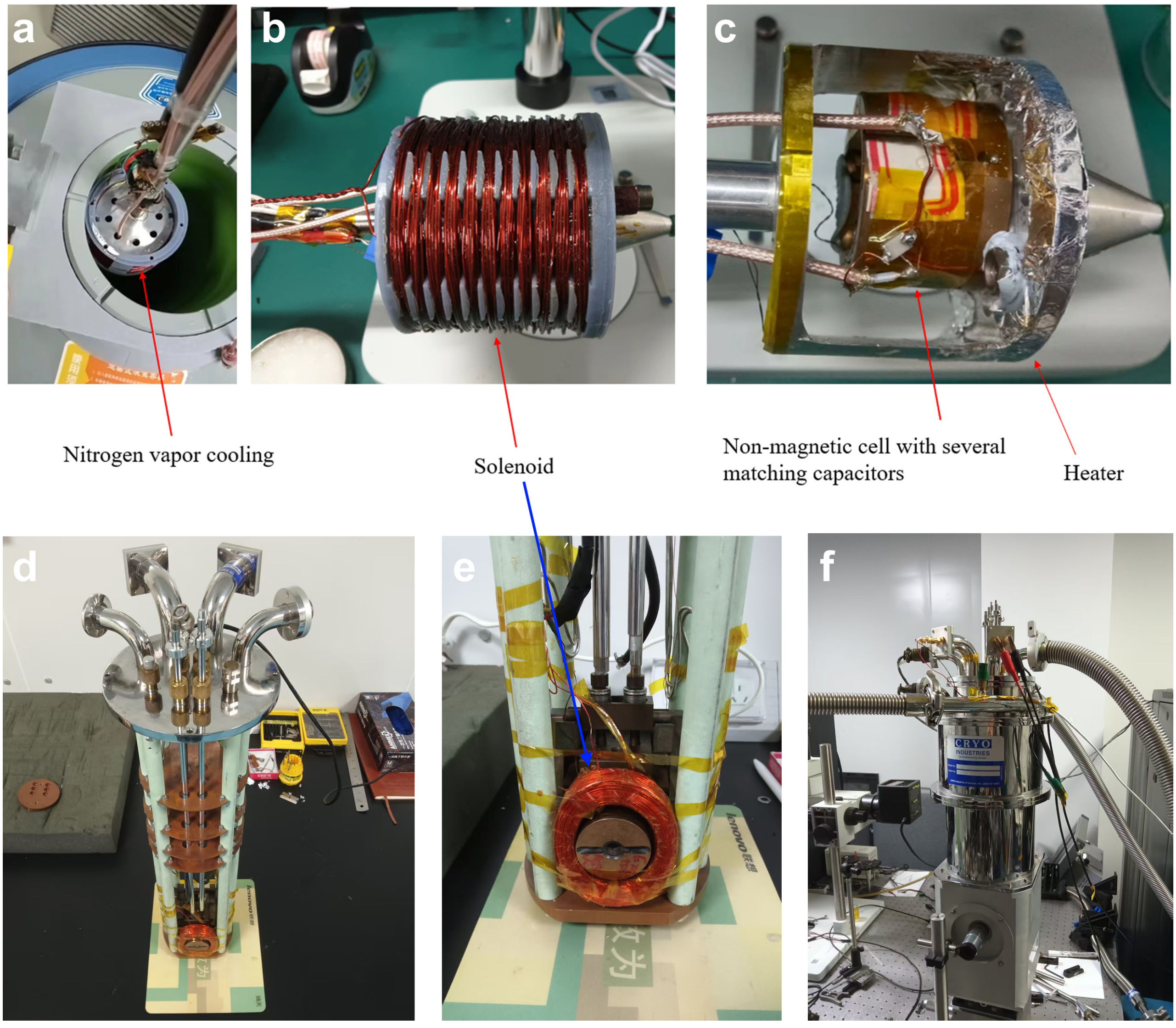

**Figure S1.** Cryogenic inserts, solenoids and DACs used in the RF transmission experiments. (a) Experiment in nitrogen vapor cooling mode. It is easy to perform and it gives low noise level in case of well-done grounding of all elements. (b) External view of a hand-made solenoid that creates AC modulating field of $B_{max}$ = 30-60 Gauss. (c) High-pressure diamond anvil cell (BX-90) fixed in an aluminum holder with RF coaxial leads. (d) Cryogenic insert of the cryostat. (e) High-pressure DAC and the solenoid for creating external AC magnetic field. (f) General view of the optical cryostat for performing the RF experiments. The survival of a DAC during the experiment can be monitored using optical imaging and Raman spectroscopy.

Since the superconducting samples were clamped in high-pressure cells, the cooling of which usually leads to an increase in pressure, the critical temperature measurements led to some scatter of about 5% in $T_c$ due to its pressure dependence.

It is important to point out that superhydride samples are often inhomogeneous. The presence of phases (e.g., *fcc*-$XH_{10}$ and *hcp*-$XH_9$ ) with different $T_c$ leads to the appearance of many RF signals. These are expressed to different degrees at various frequencies due to the fact that the penetration depth of electromagnetic field depends on the frequency, as well as due to variations in the quality factor of the micro resonators for different phases and grains. In our work, we focused on the most pronounced signals of superconducting transitions with the highest $T_c$. The advantage of the radio-frequency method over the standard practice of resistive four-probe measurements is the possibility of detecting superconducting transitions in impurity hydride phases or isolated grains, which are separated from the current paths by a non-superconducting or normal shell of molecular or lower hydrides, and which are very difficult to detect by the four-probe transport technique. An example is the Figure S2 where the transition to superconducting state (SC) in the center of shown grid of resistors, due to symmetry factors, is not accompanied by any change in the four-contact resistance.

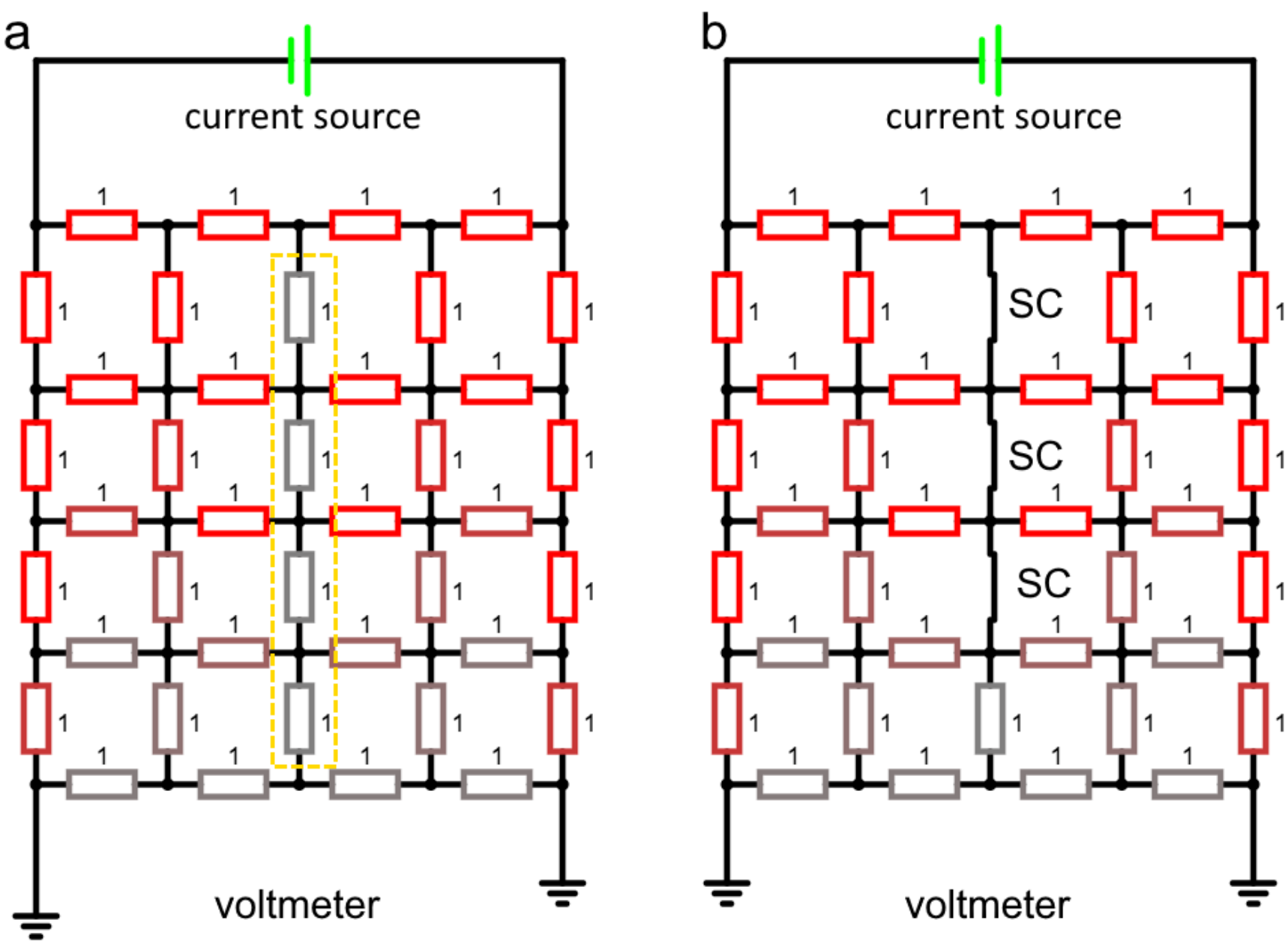

**Figure S2.** Illustration of superconductor inclusion in a van der Pauw. (a) A schematic representation of a standard van der Pauw configuration, consisting of a 2D network of resistors (1 Ω) with uniform resistance. A current source is connected to two opposing terminals, and a voltmeter measures the voltage across the other two terminals. The yellow dashed square highlights a subset of resistors within the network, which are subsequently replaced by superconductors (SC) in panel (b). (b) The same van der Pauw configuration as in (a), but with the resistors within the yellow dashed square region now replaced by superconductors. Despite the presence of these zero-resistance elements, the voltage measured by the voltmeter remains unchanged compared to the configuration in (a). This demonstrates that the introduction of superconductors at specific locations within the network does not necessarily result in a detectable voltage change using the standard van der Pauw measurement technique (modeled by www.falstad.com/).

### *1.2 Shape of radio-frequency transmission anomalies*

The target signal in the radio-frequency method most often has the form of a step or a peak. A step is a theoretically expected form for bulk samples and is more convenient when comparing experiments conducted under different conditions (for example, in a magnetic field). In the simplest case, the output signal ($\mathcal{E}_{\text{out}}$) has a contribution from the diamagnetic permeability ($\mu_{\text{s}}$), surface resistance ($R_{\text{s}}$) and their first derivatives:

$$\mathcal{E}_{\text{out}} = -\frac{d\Phi}{dt} \propto \frac{d(\mu_{\text{s}} I_{\text{ind}})}{dt} = I_{\text{ind}} \frac{d\mu_{\text{s}}}{dT}\left({}^{dT}\!/_{dt}\right) + \mu_{\text{s}} \frac{dI_{\text{ind}}}{dt} =$$

$$= (\mathcal{E}_{\text{ind}}/R_{\text{s}}) \sin(\omega t) \frac{d\mu_{\text{s}}}{dT}\left({}^{dT}\!/_{dt}\right) + \omega \mathcal{E}_{\text{ind}} \left(\frac{\mu_{\text{s}}}{R_{\text{s}}}\right) \cos(\omega t) - \mathcal{E}_{\text{ind}} \sin(\omega t) \frac{\mu_{\text{s}} dR_{\text{s}}}{R_{\text{s}}^{2} dT}\left({}^{dT}\!/_{dt}\right), \qquad (S1)$$

where $T$ is the temperature, $t$ – is time, $\omega = 2\pi f$ – is the angular frequency, $dT/dt$ is the cooling/warming rate (3-6 K/min in our experiments), index "s" relates to the sample, while index "ind" relates to the current induced on the sample's edge ($I_{\text{ind}}$), see Fig. 2e.

At the moment of transition, due to the large value of $\frac{dR_s}{dT}$ or/and $\frac{d\mu_s}{dT}$, the first term in eq. S1 can be significantly larger than the contribution from other terms. This term gives a peak and disappears when $R_s$ or $\mu_s$ take a constant value.

The second term of this equation, proportional to $\left({}^{\mu_s}/_{R_s}\right)$, takes on different values before and after the transition and gives a step-like signal. Finally, the contribution $\frac{\mu_s\, dR_s}{{R_s}^2 dT}$ in the third term also takes on peak values at the moment of superconducting transition (when $R_s \rightarrow 0$) causes the appearance of a peak in the RF signal.

An undesirable phenomenon in the measurement process is breakdown and leakage current in the region of the longitudinal section of the Lenz lens. Moreover, since the inhomogeneous structure, sample itself can contain geometries of the Lenz lens type formed by a random distribution of conductive grains, such breakdown and leakage of current can also be observed inside the sample at high induced current (Figure S3). This leads to the appearance of additional splashes and steps in the temperature dependence of the transmitted signal.

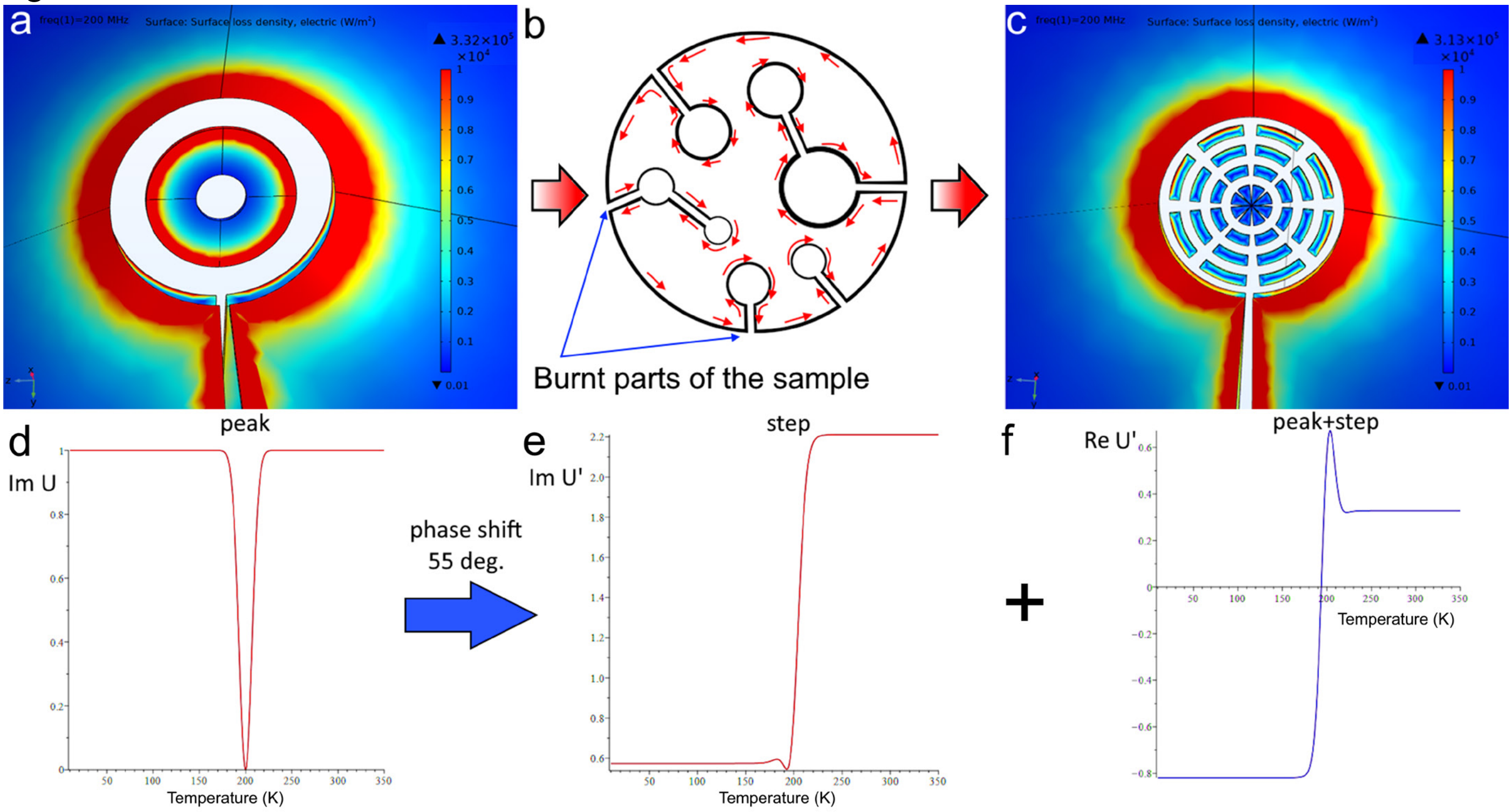

**Figure S3.** Possible mechanism of sample fragmentation and degradation in the RF transmission experiments, and dependence of the RF signal shape on the phase shift. (a) COMSOL simulation of the initial sample configuration, a disk, showing the distribution of inductive surface power losses density (SLD, colormap). The color gradient represents the SLD, with higher SLD indicated by red and lower density by blue. (b) Schematic illustration of the sample damage process. Arrows indicate the redistribution of induced currents due to the formation of gaps at the disk's edge. "Burnt parts of the sample" highlights the regions where damage is likely to occur due to localized high current densities. (c) COMSOL simulation demonstrating the final stage of the damage, where the sample undergoes significant defragmentation. (d-f) Results of phase change for the real and imaginary components of the complex voltage: (d) initial Im U vs T; (e) phase shift transformation Im(U′)=Re(U)·sin(ϕ) + Im(U)·cos(ϕ) versus T; (f) phase shift transformation Re(U′) =Re(U)·cos(ϕ) – Im(U)·sin(ϕ) versus T. In this example, $T_c$ = 200 K.

As stated above, the most typical sign of superconductivity is the change of the signal level on the receiving RF coil and the simultaneous occurrence of a wide or narrow peak when detecting the 2nd harmonic (2$F$) of the external AC magnetic field ($F$). In addition, due to the nonlinear magnetic properties of the sample's medium in the superconducting transition region, the high-frequency second harmonic (2$f$) often occurs in the RF channel simultaneously with many higher even harmonics in the low-frequency channel (*2F*, *4F*, *6F*, *8F*, …) of the external magnetic field (see Figure S4). In this case, a pronounced (and very useful) second harmonic signal is observed due to sample's non-linear magnetization *M(H)*

$$M(H) \propto \chi H + \beta H^2 + \ldots, = \chi H_0 \cos(2\pi F t) + \left(\frac{\beta}{2}\right) H_0^2 (1 + \cos(4\pi F t)) \ldots, \qquad (S2)$$

where $\beta$ – is the nonlinearity coefficient in the vicinity of $T_C$ or $T_N$.

*1.3 Measurements of the background signal of an empty DAC*

An empty DAC BX-90 mini (BeCu) with a sputtered Lenz lens on a diamond anvil (culet is 100 μm) was investigated using the built-in microcoil as an emitter and MFLI (Zurich Instruments) as a lock-in amplifier and radio-frequency signal generator. At the same time, the sample was placed in a weak periodic magnetic field (33 Hz, $B_{max} \approx$ 50 Gauss). The RF signal modulation was investigated using the SR830 lock-in. In all investigated cases, the transmitted RF signal has no pronounced features in the range of 160-300 K at 2.22, 3.11 and 5 MHz (Figure S5a-c). The signal at the 2$^{nd}$ harmonic is completely featureless (Figure 5d-e). It is important to underline that due to the deformation and displacement of the connecting coaxial wires during the cooling process, it is almost impossible to obtain a completely "flat" background of the RF signal in a wide temperature range.

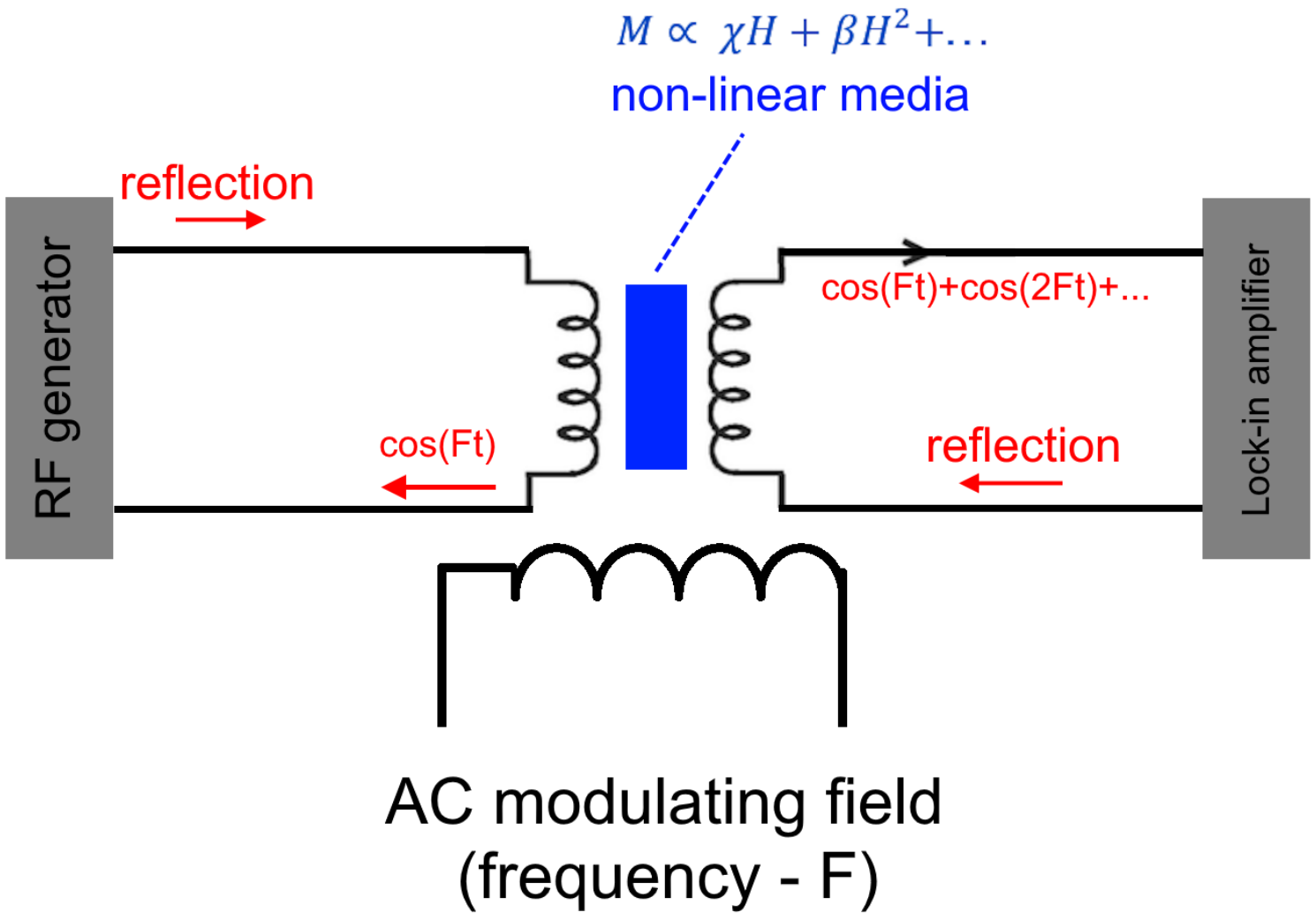

**Figure S4.** Scheme explaining the appearance of higher low-frequency harmonics (*2F*, *4F*, *6F*, *8F*, …) due to multiple passage and re-reflection of the signal induced by the modulating solenoid ("L"). The passage of the induced signal through the nonlinear medium of the sample leads to the appearance of many (mainly even) low-frequency harmonics in the spectrum.

*1.4 General remarks on reproducibility*

What is reproducible and what is not in the RF transmission method (microscopic samples):

- Because the LC circuit of the coax cables changes its parameters due to thermal contraction and expansion, as well as changes in the relative position of the wires, the phase of the RF signal changes during the measurement process, and the signal shape from microscopic samples is not constant and not always reproducible (see Fig. S3d-f).

- Cooling and warming cycles differ due to vortex dynamics in a superconductor. The radio-frequency field disrupts vortices from pinning centers and causes them to move during the warming cycle. Therefore, the radio-frequency signal during warming cycles generally differs from that during cooling cycles. Furthermore, cooling and warming cycles differ in the deformation of the supply coax wires.

- The amplitude of the RF anomaly depends on the degree of fragmentation of the microsample (e.g., see Figure S3a-c), its shape, its position relative to the Lenz lens, the carrier frequency (field penetration depth), as well as the condition of the lens itself. The amplitude of the useful signal for a microsample generally varies from experiment to experiment. A good illustration of this is Fig. 6d, e, which shows a decrease in the amplitude of the SC transition signal with repeated cycling.

- The position of the anomaly in RF transmission on the temperature scale is *well reproducible*. This is especially true when performing simultaneous scanning at multiple carrier frequencies or using a noise generator covering a wide frequency range. In such cases, the transition width is also *well reproducible*.

- Measurements of millimeter samples of $MgB_2$, REBCO, BSCCO (Figs. 3-4) are *well reproducible*.

Therefore, at present, the radio-frequency method can be characterized as a method for detecting the position of superconducting (or phase) transition for micro-samples under high pressure, but it is not able to directly provide any other characteristics of the superconducting sample (e.g., superconducting volume fraction, penetration depth, critical magnetic field, except maybe rough estimation of $H_{c1}$).

In the same way, transport measurements at a relatively high current can also lead to degradation of polyhydride samples, diffusion of hydrogen under the action of an electric field. Heterogeneous samples may contain geometries close to the geometry of Lenz lenses, and the radial cut or "bridge" connecting the lens to the edge of the sample may be subject to RF breakdown during the experiment. This leads to the appearance of additional noise, and artifacts (spikes) of the RF signal. Measurements should be performed taking into account the degradation and change in the frequency response of the system during the experiment.

Another reason for the partial irreversibility of the measurements is a random distribution of Abrikosov vortices in a sample and their motion [1,2], as well as the memory effects of the sample [3], which explains the different shape of the transmission curves in the warming and cooling cycles. During warming from the SC state, the sample contains a trapped magnetic flux, which affects the detected signal, while during cooling, the trapped magnetic flux is zero.

**Table S1.** The basic properties of radio-frequency transmission anomalies in the vicinity of the first-order phase, magnetic and superconducting transitions.

| Transition type | RF signal | 2nd harmonic of the AC magnetic field | Additional criteria |
|---|---|---|---|
| Superconducting | Step, peak or kink | Strong signal | Usually, sharp transitions<br><br>There is a shift in strong magnetic fields |
| Magnetic | Step or peak | Strong signal | Usually signal disappears at high frequency (> 10 MHz) |
| Phase transition (first-order) * | Broad step/bell-shaped peak | Weak or no signal | Large temperature hysteresis |

*See the example of $VO_2$ demonstrating the 1st order phase transition (Supplementary Figure S17-S18).

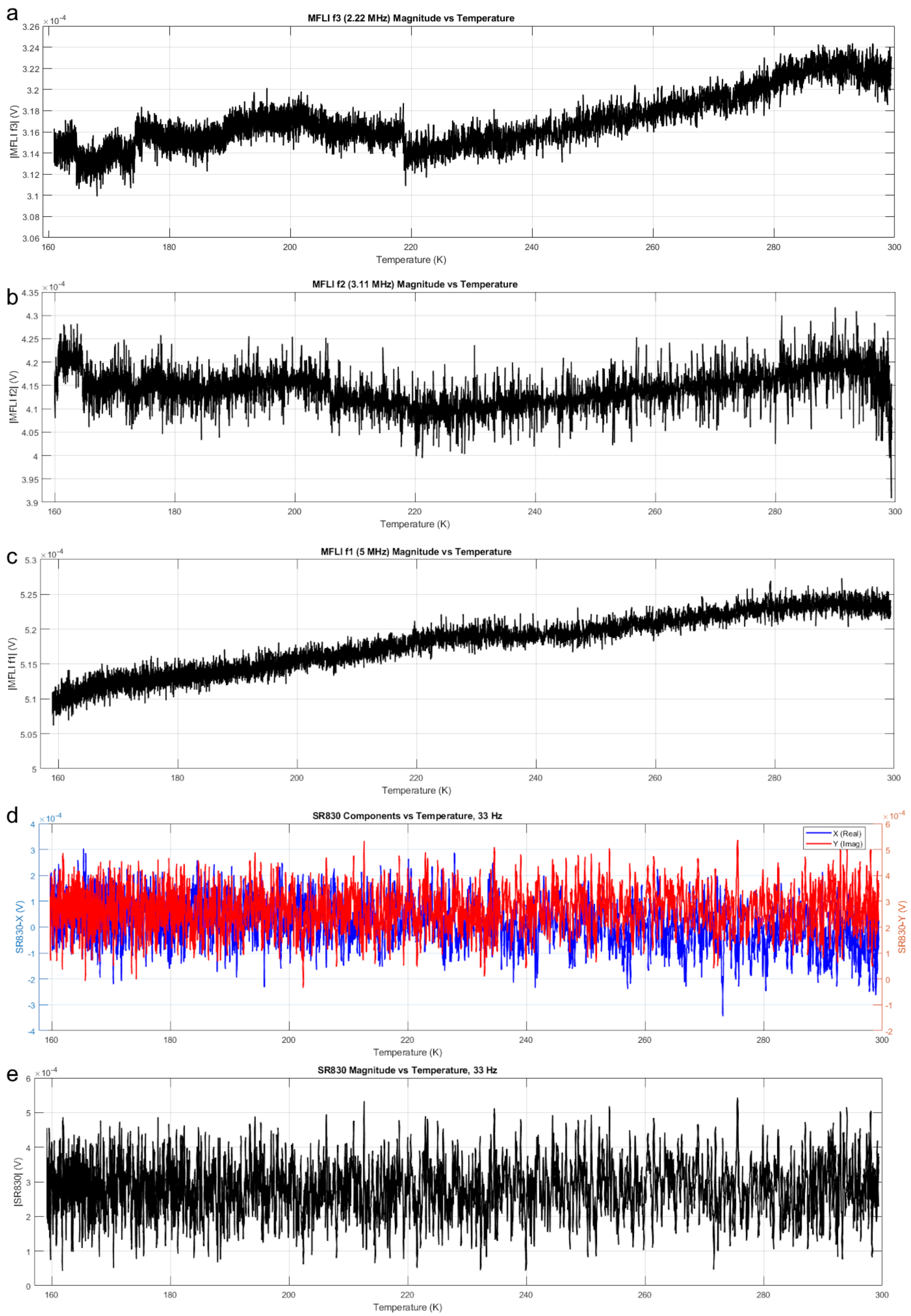

**Figure S5.** Absolute value of the radio-frequency signal transmitted through an empty BX-90 mini DAC at about ≈0-5 GPa equipped with a Lenz lens (culet diameter is 100 um) in the temperature range from 160 K to 300 K (cooling cycle). (a) $f_{carr}$ = 2.22 MHz, (b) $f_{carr}$ = 3.11 MHz, (c) 5 MHz. (d) Real (blue) and imaginary (red) components of the signal at the second harmonic *2F* of the modulating magnetic field ($F$ = 33 Hz, $B_{max}$ ≈ 50 G). (e) The absolute value of the signal at the second harmonic *2F*.

## 2. BSCCO, REBCO and Hg-1223 radio-frequency transmission tests

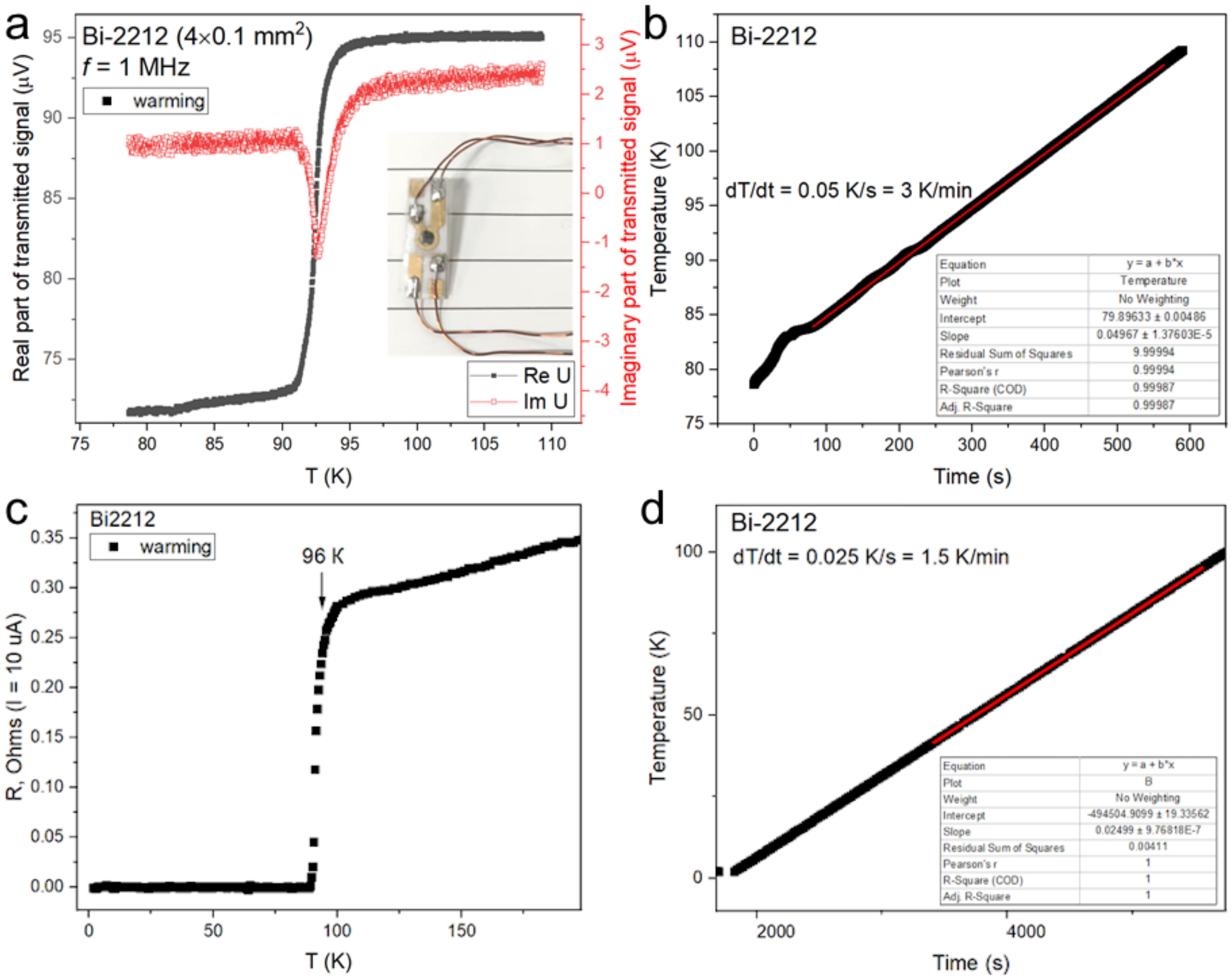

**Figure S6.** Comparison of superconducting transition detection methods for optimally doped Bi-2212 (BSCCO) sample under varying warming rates. (a) Temperature dependence of the real (Re U) and imaginary (Im U) components of the transmitted radio-frequency signal at $f_{carr}$ = 1 MHz, with the inset illustrating the experimental setup. (b) Temperature-time profile establishing a linear heating rate of 3 K/min (0.05 K/s) used during the data acquisition. (c) Resistive transition of the same Bi-2212 sample measured via the standard four-probe method ($I_{DC}$ = 10 μA), showing a characteristic transition onset at approximately 96 K. (d) Temperature-time profile for a warming rate of 1.5 K/min (0.025 K/s) used in elctrotransport experiment, demonstrating the absence of significant temperature hysteresis in the position of the superconducting transition in transport and radio-frequency experiments.

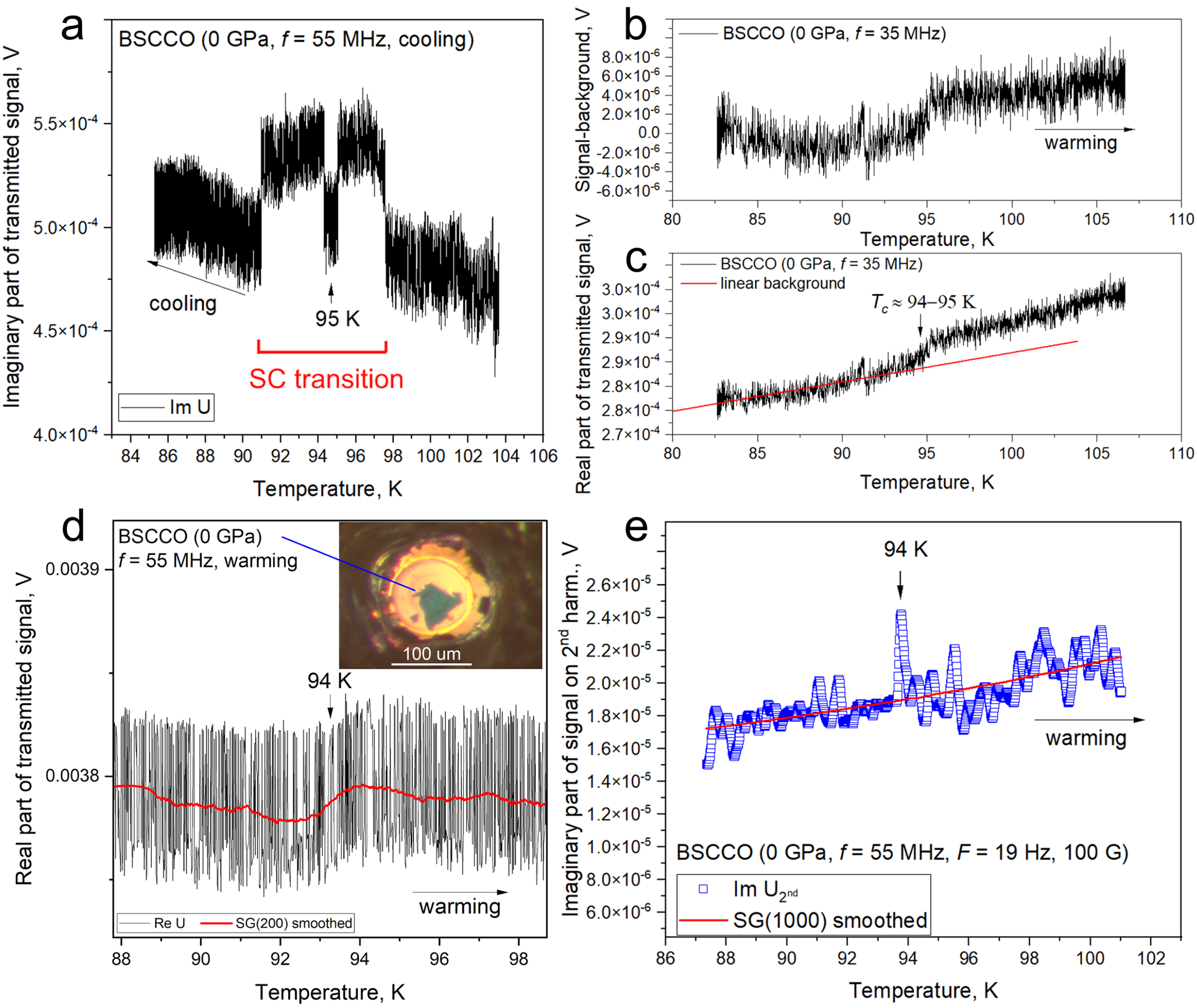

**Figure S7.** Radio-frequency transmission measurements of a test BSCCO sample at ambient pressure in the RF DAC. (a) Imaginary component of the transmitted radio-frequency signal (Im $U$) as a function of temperature during a cooling cycle, measured at 55 MHz. (b) Real component of the transmitted RF signal as a function of temperature during a warming cycle measured at 35 MHz. A linear background has been subtracted from the original data to highlight the transition. (c) The original (unprocessed) real component of the transmitted RF signal, corresponding to the data presented in panel (b). The transition temperature ($T_c$) is estimated to be within the 94-95 K range. (d) Real component of the transmitted RF signal (Re $U$) as a function of temperature during a warming cycle, measured at 55 MHz. A Savitzky-Golay (SG) smoothing filter (200 points) is applied to the data, represented by the red line, revealing a transition at about 94 K. The inset shows an optical microscope image of the BSCCO sample inside of the test DAC. (e) Imaginary component of the 2$^{nd}$ harmonic signal (Im $U_{2nd}$) in a low-frequency (19 Hz) modulating magnetic field of about $B_{max}$ = 100 G, as a function of temperature during a warming cycle. A Savitzky-Golay smoothing filter (1000 points) is applied to show the straight background line.

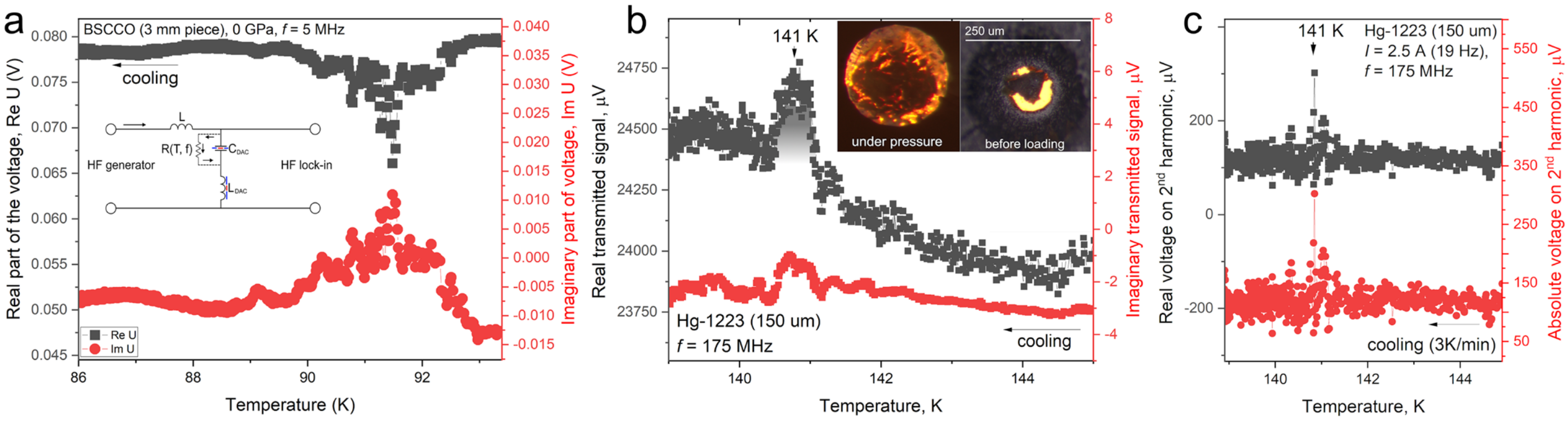

**Figure S8.** (a) Detection of the superconducting transition in a large BSCCO sample at 0 GPa (diameter 3 mm, thickness about 0.1-0.2 mm) in a scheme with one Lenz lens (the scheme is shown in the inset). In this case, we used a frequency of 5 MHz and detected a change in the LC characteristics of the circuit, as well as a jump in the absorption of electromagnetic waves in the BSCCO sample. (b) Re U and Im U ($f_{carr}$ = 175 MHz) measured for a Hg-1223 sample inside of a closed test DAC. Inset: Images of the Hg-1223 sample (batch 1) on the diamond-anvil culet before and after closing the DAC. The brittle single crystal broke into multiple pieces during compression. (c) Simultaneously measured second harmonic ($f_{mod}$ = 19 Hz, $B_{max}$ = 3 mT) shows a small peak at 141 K.

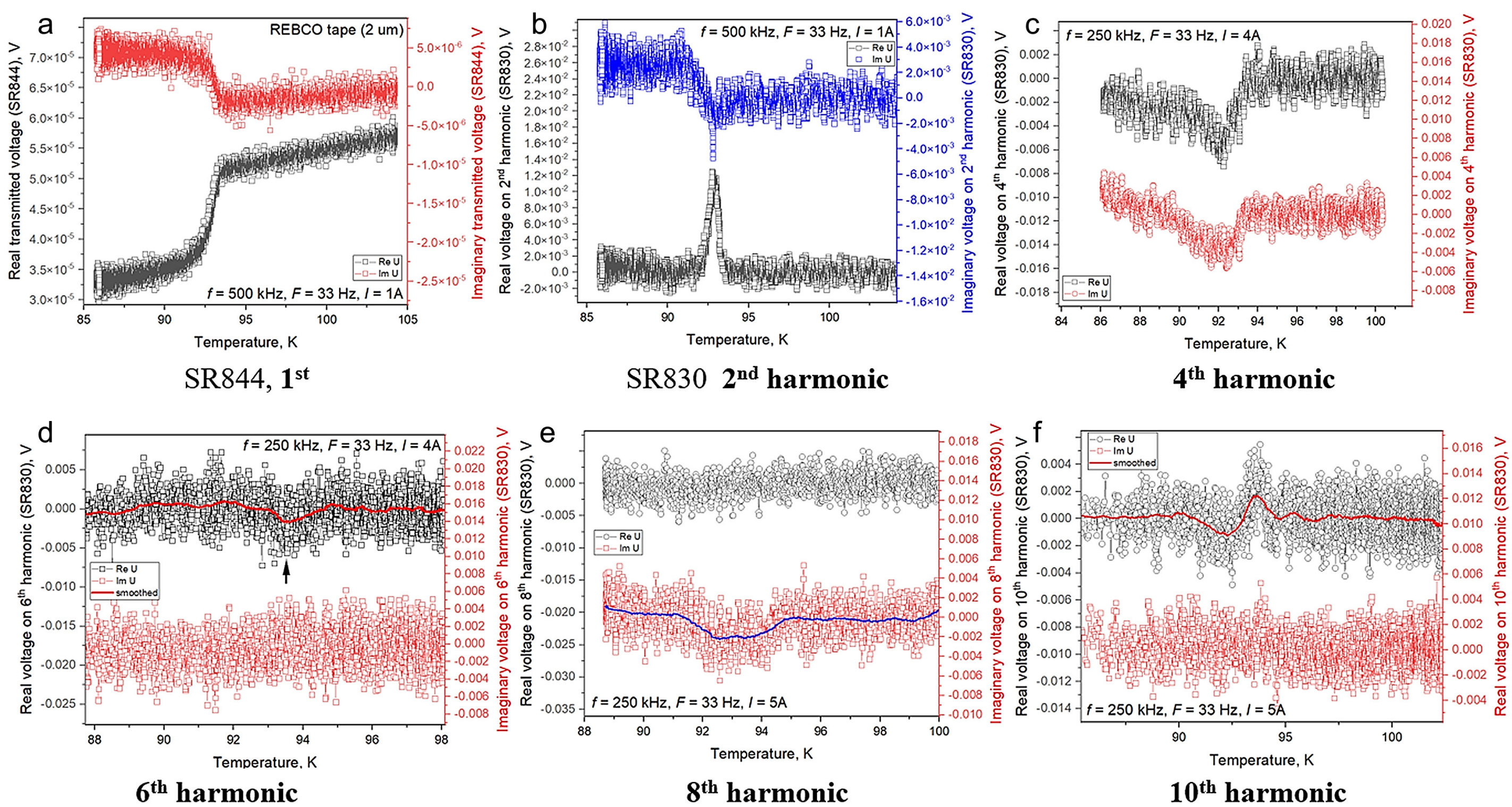

**Figure S9.** Higher harmonic generation and anomalies in the vicinity of the superconducting transition in a piece ($4\times4\times0.2$ $mm^3$) of REBCO tape placed in a single-turn transformer. Carrier frequency $f$ = 500 kHz, modulating field frequency $F$ = 33 Hz, amplitude – from 10 to 40 Gauss (4 A). We see that in the vicinity of the superconducting transition, pronounced signal anomalies are observed at all even harmonics of the modulating field (see also Figure S4).

## 3. Radio-frequency transmission measurements of $MgB_2$, NbTi, and REBCO tape in magnetic fields

The purpose of the experiment with a macroscopic sample of REBCO tape (4×4×0.2 $mm^3$) placed in a single-turn transformer was to establish a qualitative shift of the radio-frequency transmission anomaly in a magnetic field. However, the precise orientation of the film with respect to the magnetic field direction was maintained only approximately.

In case of $c \parallel H$ REBCO tape orientation (Figure S11a), we were unable to unambiguously determine Tc(onset) and needed to scale the RF transmission anomalies for different magnetic fields to the same start and end points of the superconducting transition and use the middle of the transition interval to estimate the $T_c$ (Tr-50% criterion, "Tr" – is transmission).

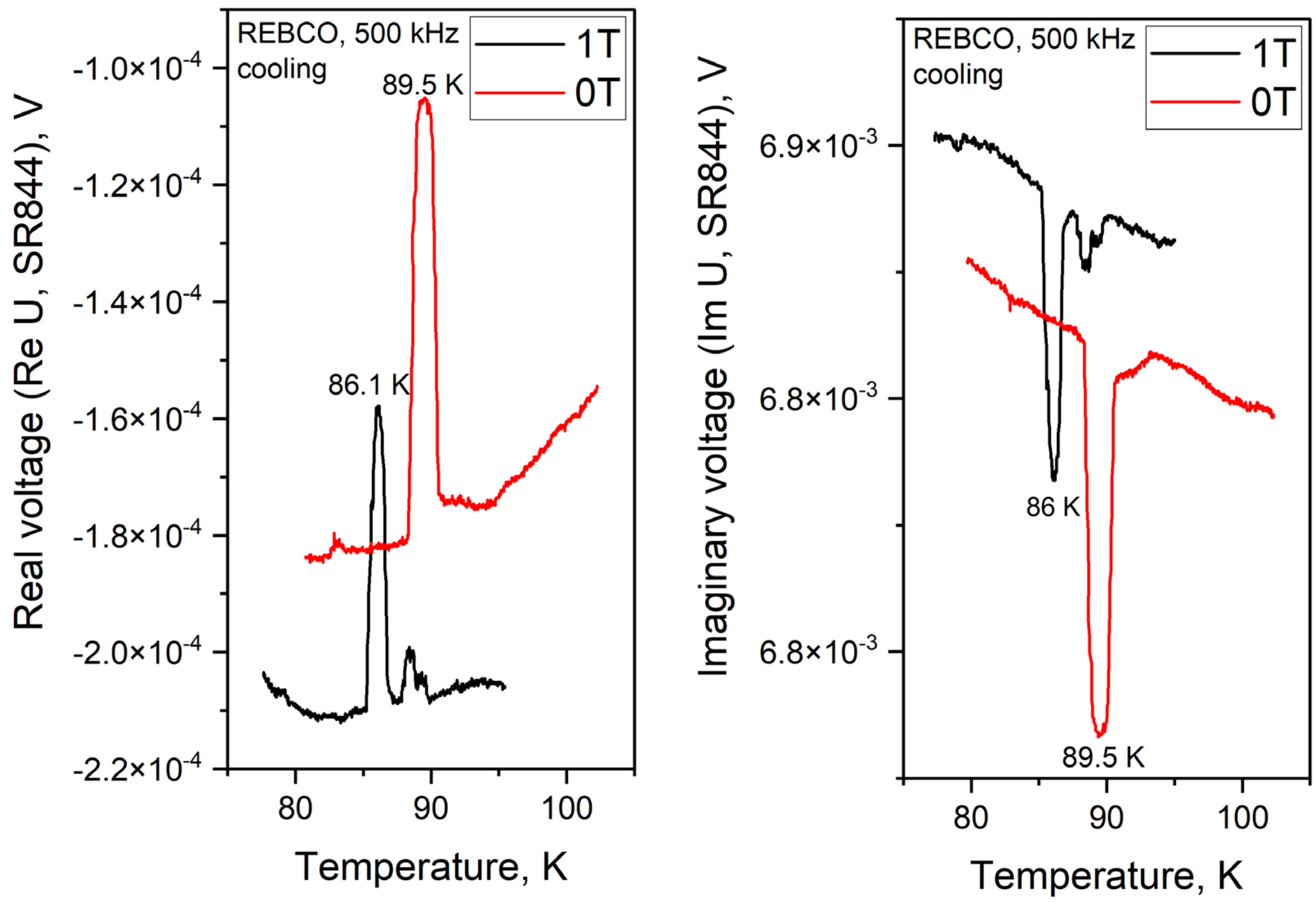

**Figure S10.** Temperature dependence of the radio-frequency transmittance trough the REBCO film on temperature at a frequency of 500 kHz in the absence of a magnetic field (0 T), and in a magnetic field of 1 T in a single-turn high-frequency transformer within a cooling cycle. REBCO tape orientation is near $c \perp H$. SC transition manifests as a peak. The presence of a small additional signal likely corresponds to the presence of another compound or a different orientation of REBCO microcrystals in the superconducting tape with a different $B_{c2}(T)$ dependence. As a result, this small fraction of the sample at 1 T may have higher $T_c$ and $B_{c2}$ than the main superconducting phase.

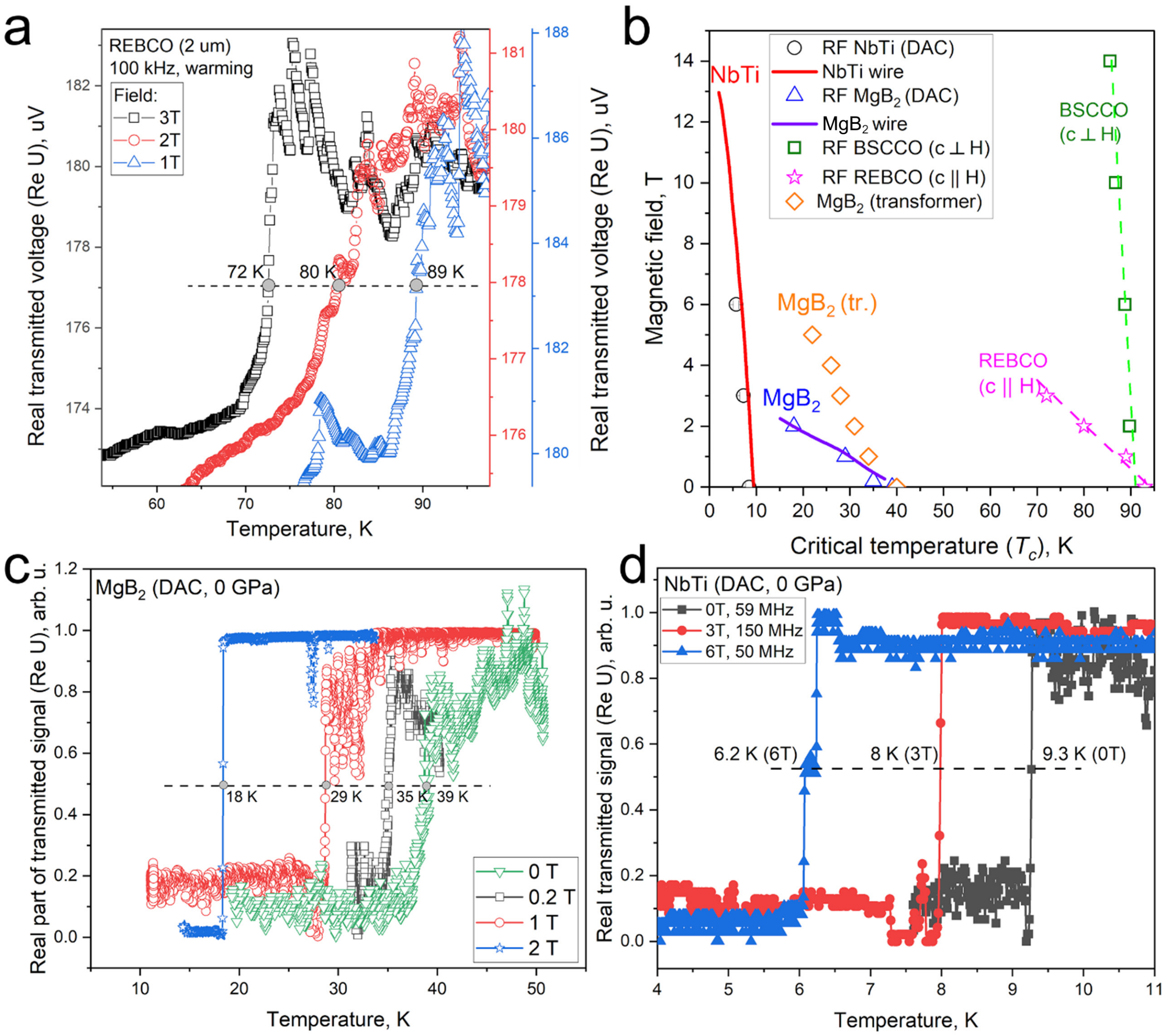

**Figure S11.** Temperature dependence of RF transmittance trough REBCO film, $MgB_2$ and NbTi. Dashed lines correspond to 50% of transmission step criteria (Tr-50%) used for $B_{c2}$ estimation. (a) REBCO film transmittance at 100 kHz in the warming cycle in magnetic field near the transition temperature. REBCO tape orientation axis is c || H. The differences in the $T_c$ value at 1 T may be due to the fact that the measurements were carried out in different systems (Oxford instruments 16T magnet, and QD MPMS) with different thermometer positions and in different cycles (cooling and warming). (b) Superconducting phase diagram for $MgB_2$, NbTi, REBCO ($c \parallel H$) and BSCCO (axis $c \perp H$), with data points obtained by the RF method (circles, triangles, stars and squares) and plotted based on literature data (solid lines). (d) Temperature dependence of the real part of the transmitted signal in the test DAC with $MgB_2$ ($d \approx 120$ μm) at 0 GPa in a magnetic field of 0, 0.2, 1, and 2 T. For clarity, the comparison is made at frequencies where the change in the intensity of the transmitted signal was a step-like. (d) Temperature dependence of the level of the transmitted signal in the test DAC with a NbTi particle (≈ 30 μm) in a magnetic field of 0, 3, and 6 T. For clarity, the comparison is made at frequencies of 50, 59, and 150 MHz, where the change in the intensity of the transmitted signal was a step-like.

## 4. Reproducibility of radio-frequency transmission measurements

The purpose of this section is to show that multiple scans of NbTi and $MgB_2$ samples at 0 GPa and with different carrier frequencies yield pronounced RF transmission anomalies in a narrow temperature band in the vicinity of $T_c$.

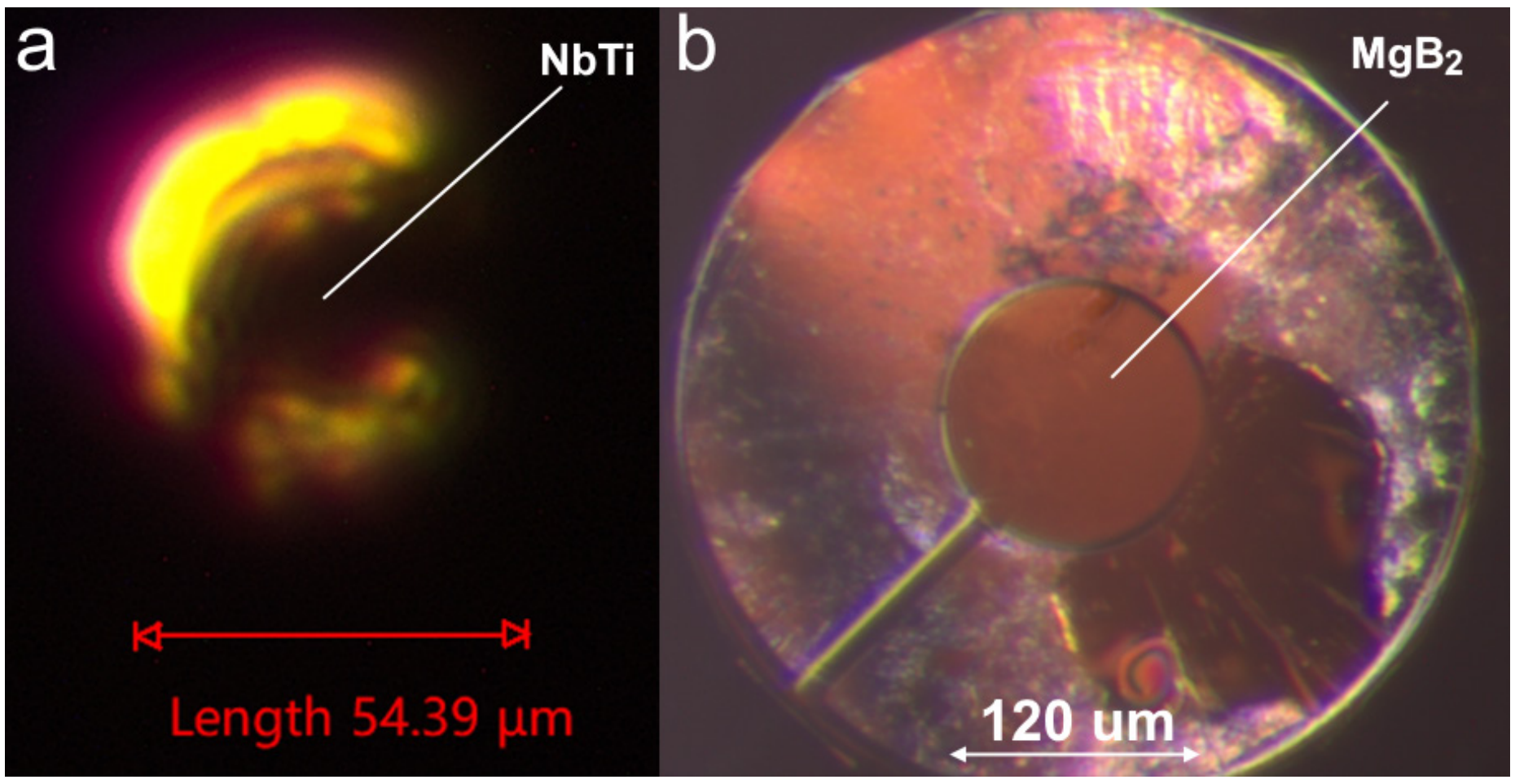

**Figure S12.** Photographs of test chamber cells with loaded samples at 0 GPa: (a) NbTi particle with diameter of about 30 µm in transmitted light; (b) pressed $MgB_2$ powder in reflected light. The Lenz lens is clearly visible.

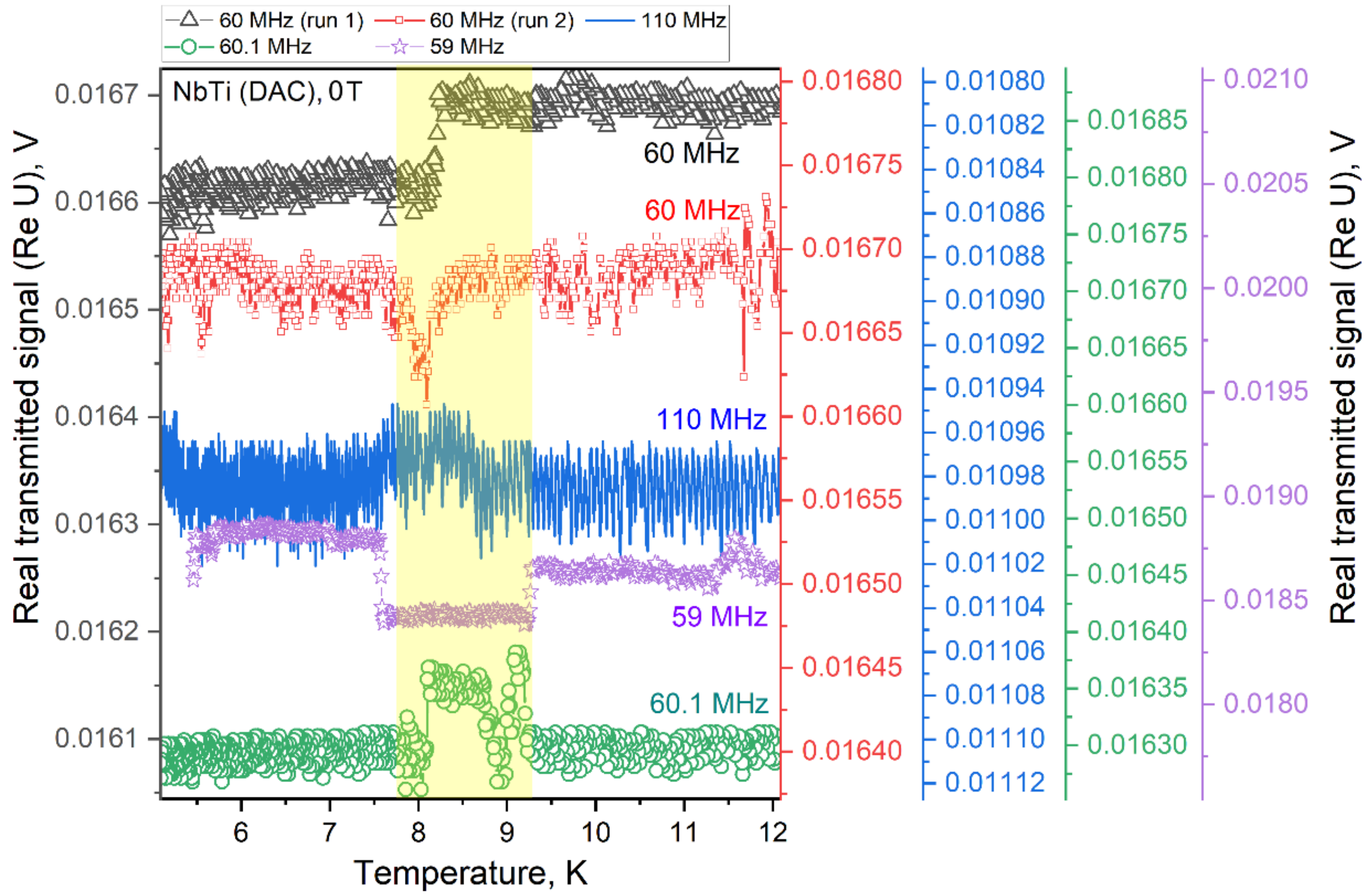

**Figure S13.** Change in the level of the transmitted radio-frequency signal (Re U) at different frequencies from 59 MHz to 110 MHz in the vicinity of the superconducting transition in 30-µm particle of NbTi at 0 GPa (in DAC). There is some variation of $T_c$ = 8-9 K depending on the carrier frequency, which may be due to the non-uniform distribution of stresses in the sample. As can be seen, the reproducibility of the transition is at the level of ± 1 K.

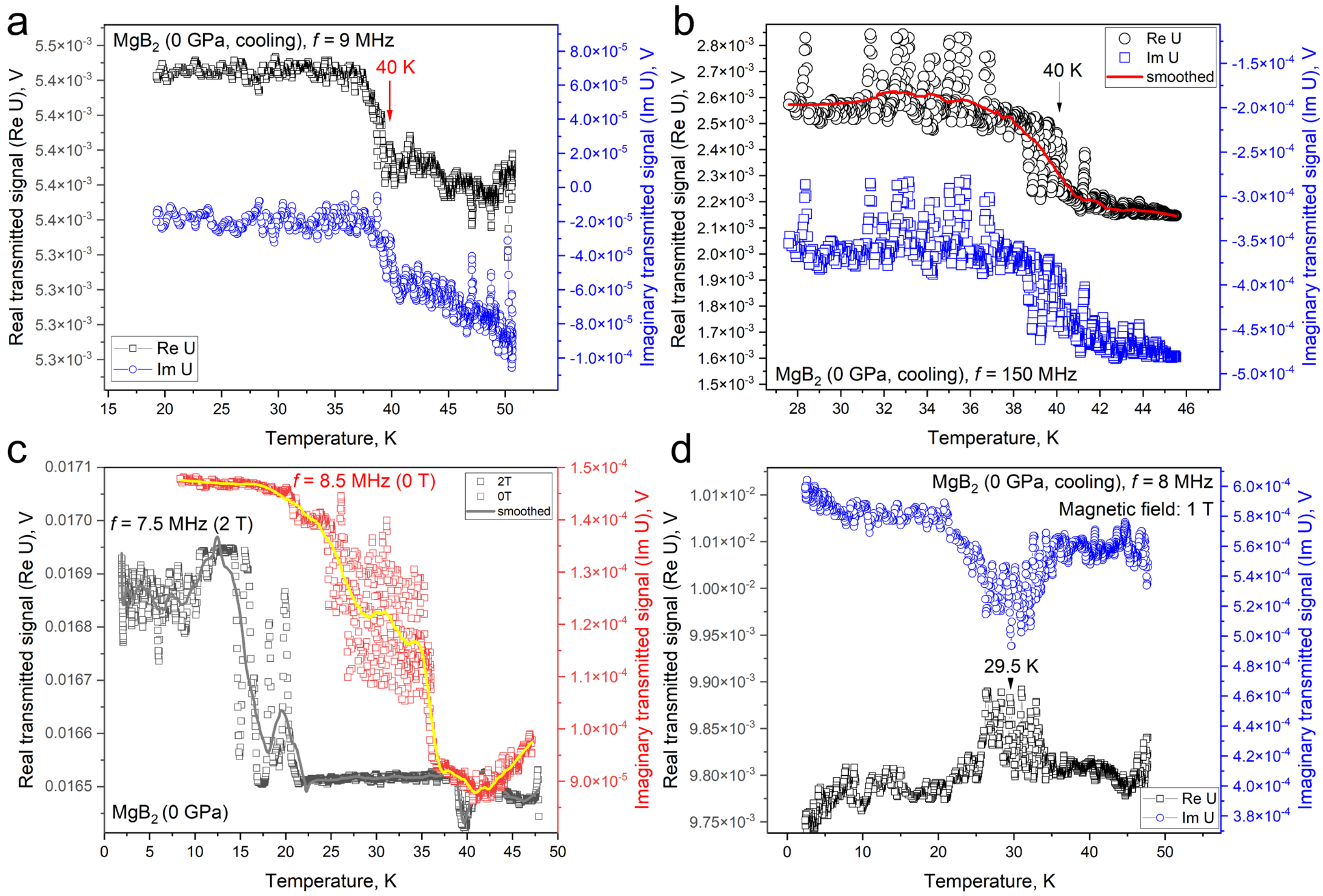

**Figure S14.** Radio-frequency transmission anomalies for a $MgB_2$ sample (about 1-2 mm diameter, 0.2 mm thick) in a single-turn transformer circuit at ambient pressure and different magnetic fields in the vicinity of the superconducting transition. (a) 9 MHz carrier frequency, 0 T cooling cycle. (b) 150 MHz carrier frequency, 0T cooling cycle. (c) 7.5 MHz and 8.5 MHz carrier frequency, zero (0 T) and 2 T cooling cycles. (d) 8 MHz carrier frequency, 1T field, cooling cycle. The superconducting transition is expressed as a peak.

## 5. Detection of ferromagnetic transitions in gadolinium and terbium

Experiments with magnetic metals were conducted to demonstrate the possibility of detecting changes in diamagnetic permeability during ferromagnetic and antiferromagnetic transitions if they occur in a sufficiently narrow temperature range. We investigated the ferromagnetic (PM → FM) transition in Gadolinium (Figure S15), and FM and AFM transitions in Terbium (Figure S16). In the case of Gd, direct demodulation of the second harmonics (*2F*) at a low-frequency modulating AC field (33 Hz) was used, which is possible according to Supplementary Figure S4. Magnetic transition in Gd is convenient in that it occurs near room temperature within a narrow temperature range. Recalling the theory of transformers, we immediately notice that a ferromagnetic core should significantly improve the operation of a high-frequency transformer. Indeed, for Gadolinium we observed a sharp peak in the level of the transmitted signal (Fig. 5a), which, however, returns almost to its original value upon subsequent cooling. The presence of two peaks in Fig. S15b may indicate the formation of a domain structure in ferromagnetic Gd below $T_C$, or non-uniform pressure distribution across the sample. All signals detected for Gd have a peak shape.

With a further decrease in temperature, the domain structure of ferromagnetic Gd stabilizes, the nonlinearity decreases, and the signal at the second harmonic disappears. The key difference between magnetic and superconducting transitions is the disappearance of the magnetic transition signal at a high frequency above ≈10 MHz due to the impossibility of rapid reorientation of the magnetic moments of atoms following the RF field. Using this criterion, it is possible to distinguish between magnetic and superconducting transitions.

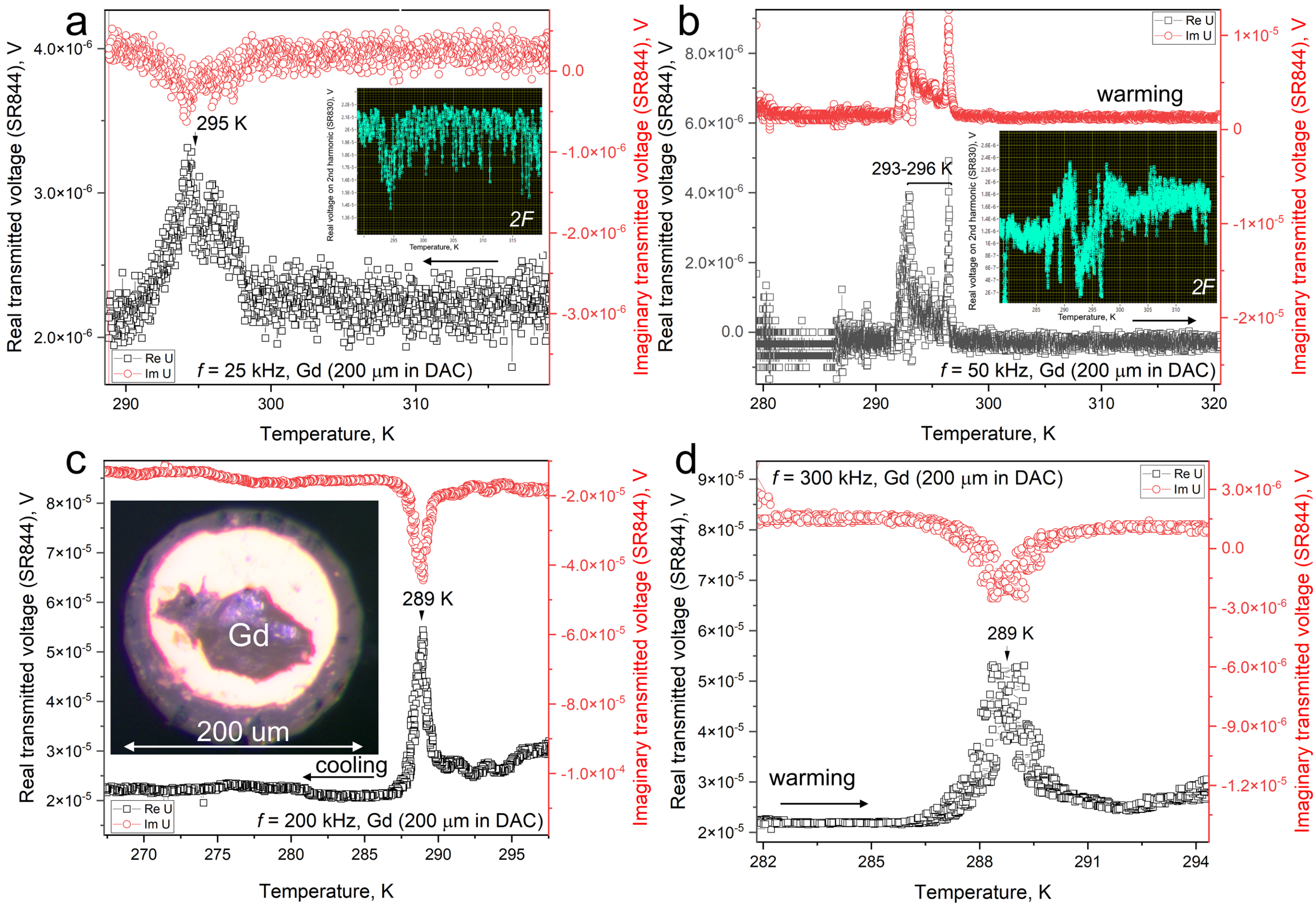

**Figure S15.** Investigation of the ferromagnetic transition in a gadolinium microparticle (~200 μm, in DAC) using the radio-frequency transmission method. (a) Real and imaginary parts of the transmitted RF signal, carrier frequency of 25 kHz. Inset: signal at the second harmonic of the oscillating AC magnetic field (33 Hz). There is a feature at 295 K which corresponds to the ferromagnetic transition in Gd. (b) The same during measurements in the warming cycle at the carrier frequency of 50 kHz. Inset: signal at the second harmonic of the AC modulating field. (c) The same at the carrier frequency of 200 kHz. Inset: photograph of the test DAC's chamber with the loaded Gd sample at 0 GPa. (d) The same at the carrier frequency of 300 kHz.

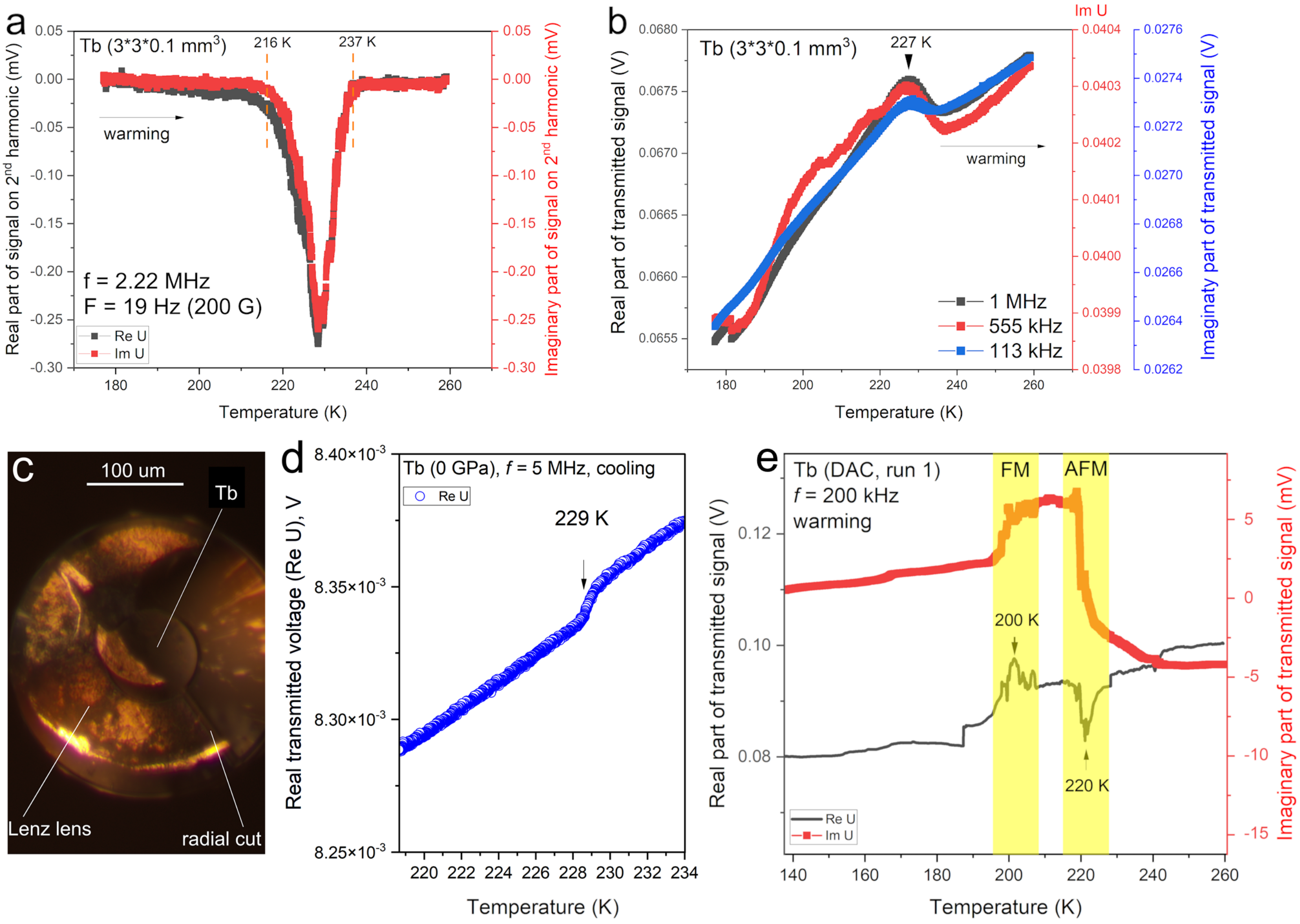

**Figure S16.** Magnetic transitions in Tb investigated for a bulk millimeter-size piece of Tb (a, b) and a particle of about 200 μm in a test DAC (c, d, e). Since the test DAC was used multiple times, the Lenz lens was partially damaged, but the Tb particle is still located between the Lenz lenses (see inset in panel "c"). The measurements were performed at carrier frequencies up to 5 MHz (panel "d"). The best signal in DAC was obtained at 200 kHz (panel "e"), where both ferromagnetic and antiferromagnetic transitions in the sample can be observed [4-6]. The second harmonic signal (a) is also observed between 216 K and 237 K.

## 6. Additional radio-frequency and XRD measurements for $CeH_{9-10}$

Before experiments with cerium hydrides, we firstly tested a standard yttrium-barium-copper-oxide (YBCO) calibration sample with a diameter of approximately 30 μm in a test DAC and a pressure close to 1 atm. The frequencies of the excitation single-turn and modulation coils were $f$ = 40 MHz and $F$ = 19 Hz, respectively, and the total resistances of the single-turn coils and twisted pairs were 25 Ω and 26 Ω, respectively (Supplementary Fig. S17). As shown in Figure 2, during the warming cycle, we used a modulation method to measure a significant redistribution of the magnetic field within the sample cavity near 90 K via signal at the 2$^{nd}$ harmonic ($2F$). S/N ratio was about 20 in this experiment.

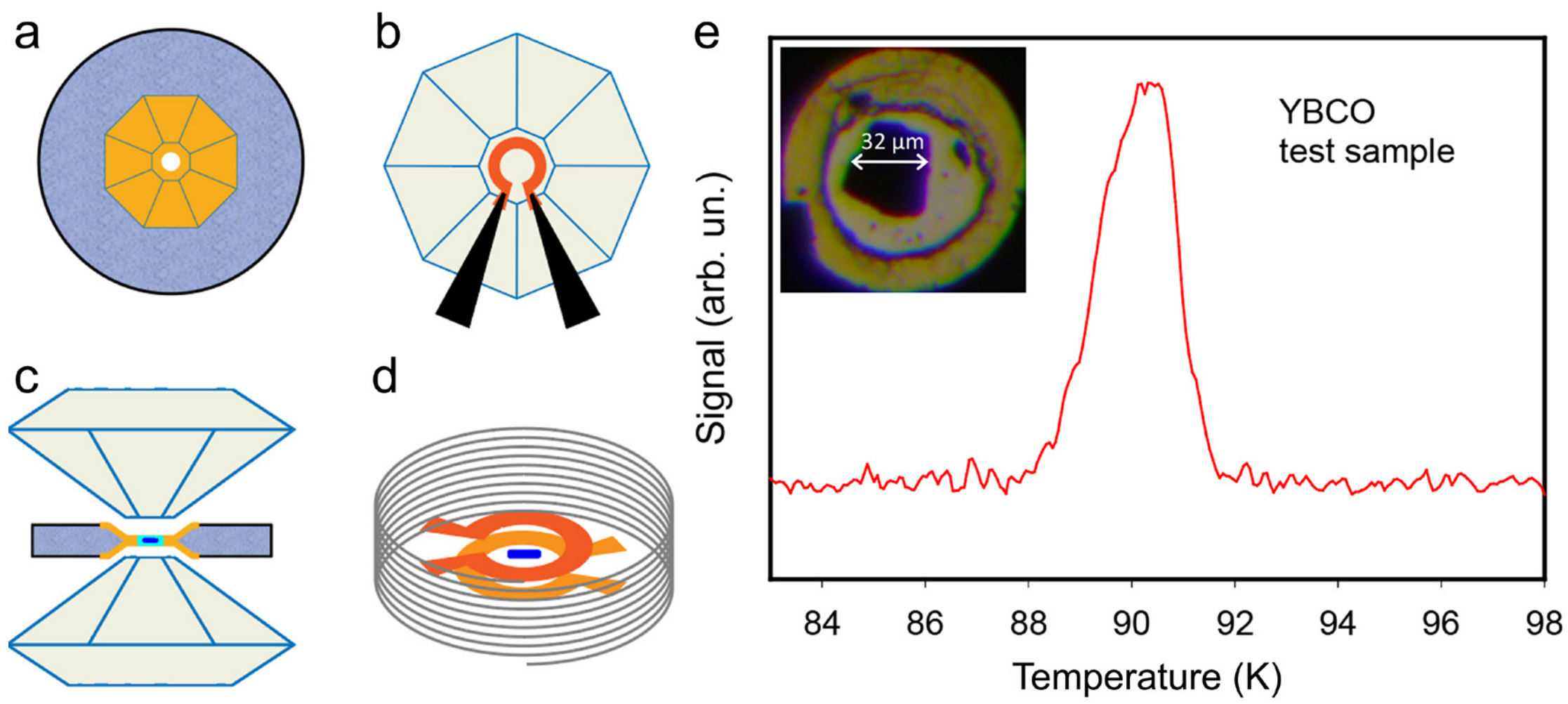

**Figure S17.** Preparation process and test measurements of YBCO in a DAC. (a) Prepare a double-sided insulating gasket. (b) Preparation of excitation and induction coil using lithography. (c) Assembled DAC with anvils and the insulating gasket. (d) Single-turn excitation and receiving coil, as well as the solenoid to create a modulating AC field. (e) Second-harmonic signal ($2F$ = 38 Hz) obtained for a 30-μm YBCO sample during a reheat cycle at 40 MHz with a modulating field of approximately 20 Gauss.

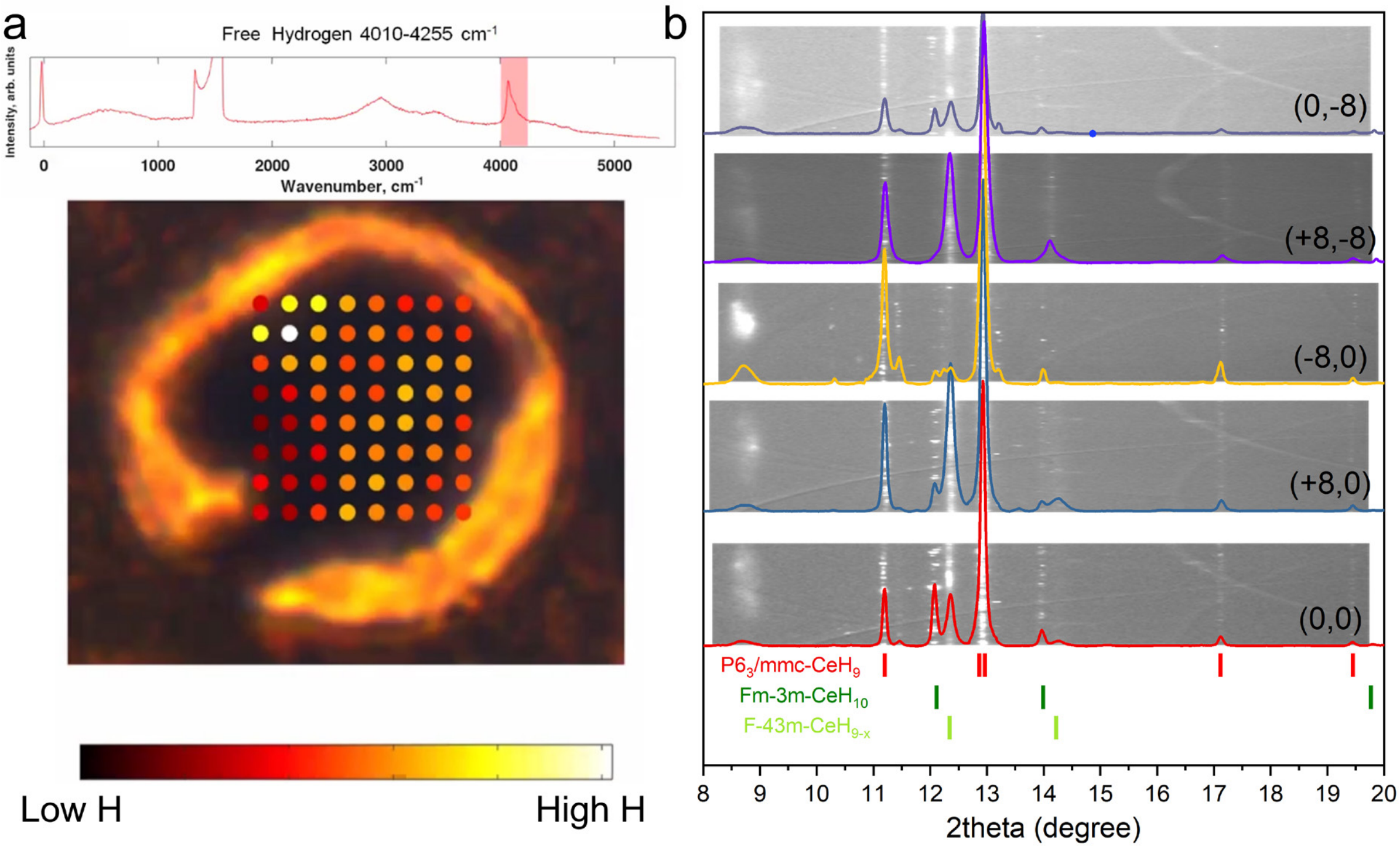

**Figure S18.** Raman spectroscopy and X-ray diffraction of a $CeH_{9-10}$ sample at 120 GPa. (a) Typical Raman spectrum and color map of free hydrogen distribution in the $CeH_x$ sample after laser heating. (b) Raw X-ray diffraction patterns and integrated spectra at different points of the sample at 120 GPa. Diffraction peaks from not only $CeH_9$ and $CeH_{10}$ but also unknown phases at smaller diffraction angles can be seen.

## 7. Additional radio-frequency measurements for (La, Ce)$H_x$

The aim of experiments in this section was to demonstrate the occurrence of anomalies at the second harmonic in the RF channel (*2f*) in the vicinity of the superconducting transition in the (La,Ce)$H_x$. We also found that properties of the synthesized sample do not remain constant and undergo transformations over time in the conditions of the radio-frequency transmission experiment. In particular, in a series of about 20 warming and cooling cycles, the transition temperature and the corresponding anomalies in RF transmission moved from 210-220 K (corresponds to the (La,Ce)$H_{9-10}$ phase) at the beginning of the experiment, to 260-270 K (corresponds to (La,Ce)$H_{10-12}$ phase) at the end of the experiment (Figure 7 and S23-S24). When repeating the study, a week later, we found that the RF transmission anomalies had almost disappeared.

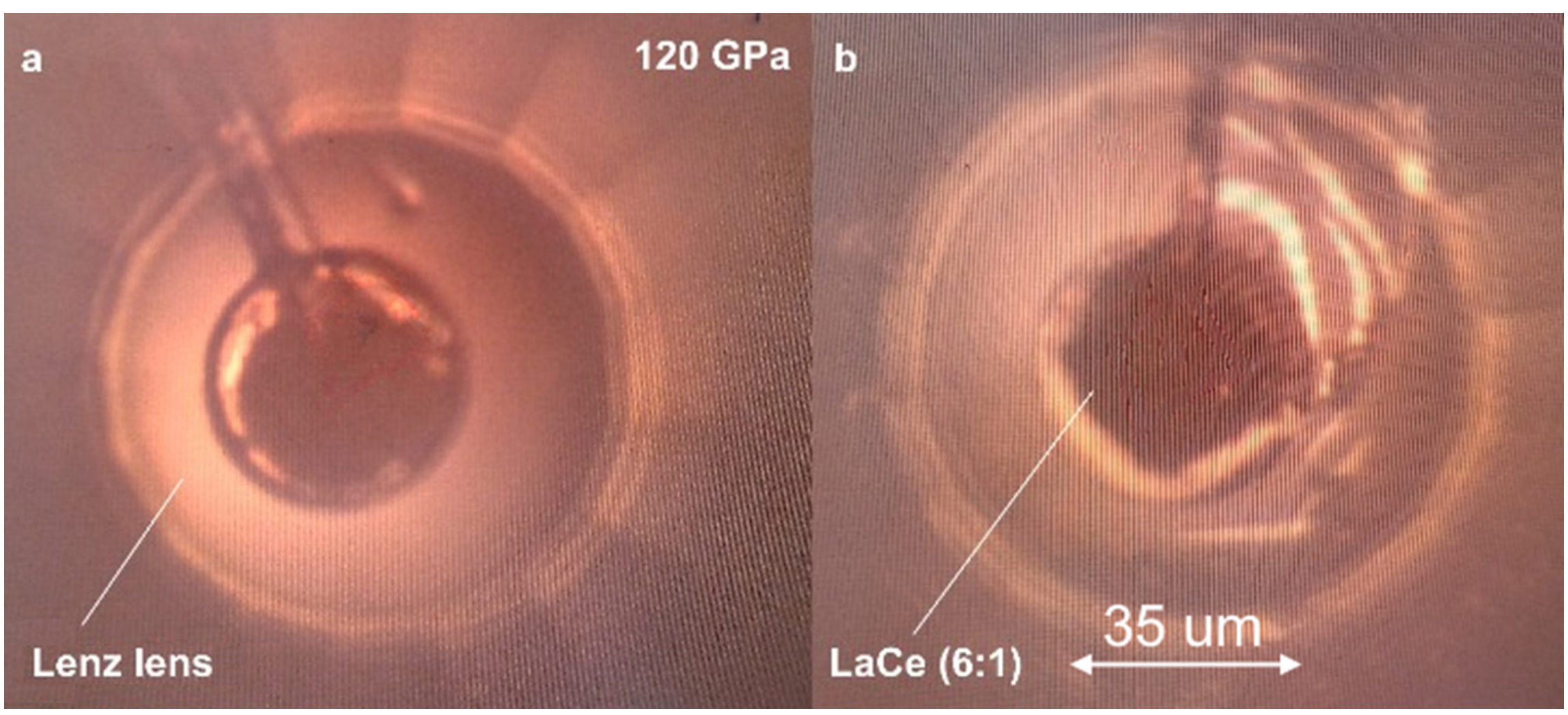

**Figure S19.** Photographs of DAC cell with Lenz lenses and (La,Ce)$H_x$ sample at about 120 GPa in reflected light: (a) piston side, (b) cylinder side. The pressure in this DAC was subsequently increased after laser heating and transportation to the ESRF from 120 GPa to 156 GPa.

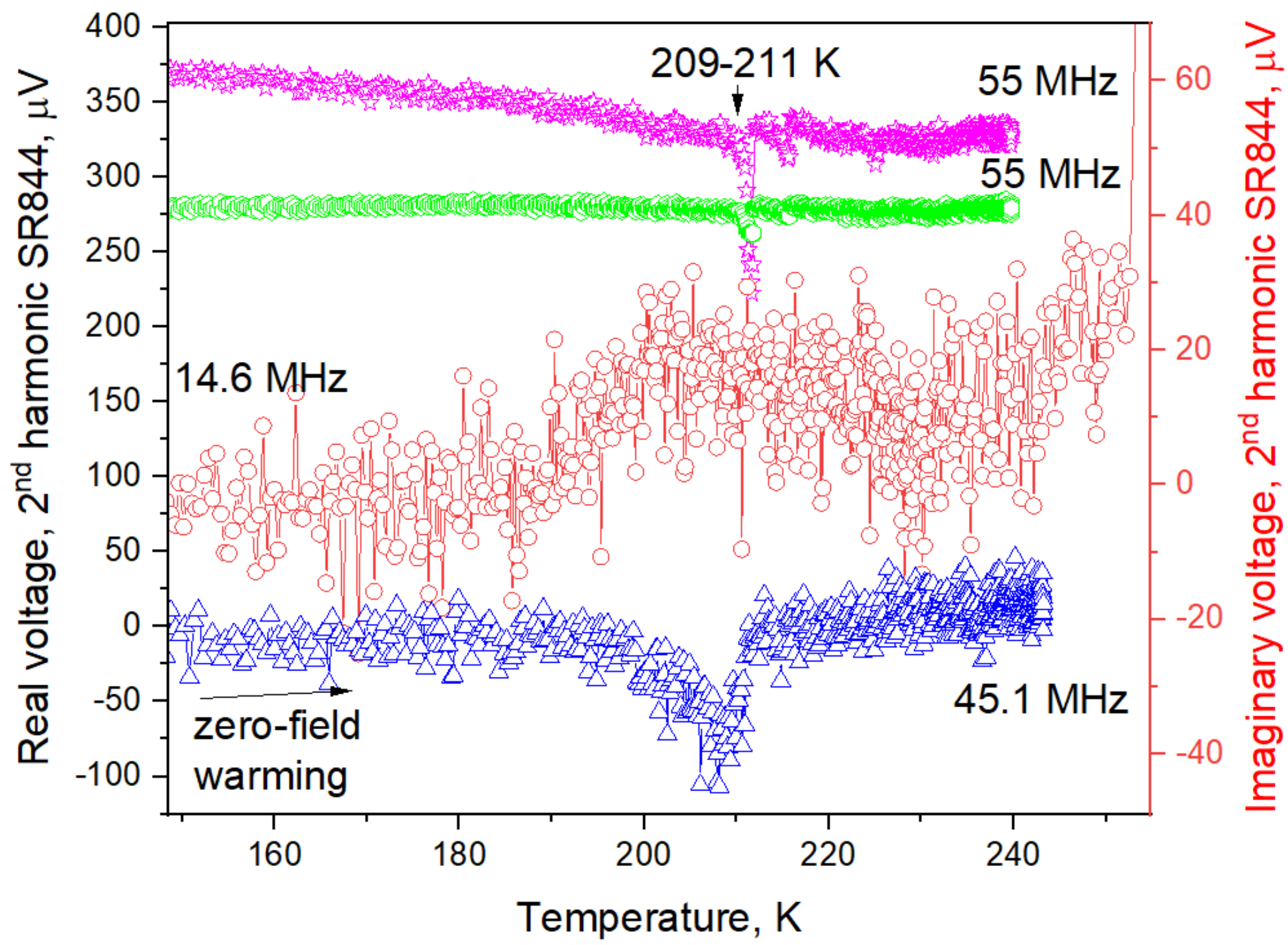

**Figure S20.** Anomalies of the 2nd harmonic of radio-frequency (*2f*) transmission signal as a function of temperature for the (La,Ce)$H_x$ (156 GPa) sample at several carrier frequencies: 55 MHz (run 1 – pink, run 2 - green), 14.6 MHz (red), and 45.1 MHz (blue). Modulating AC field is absent.

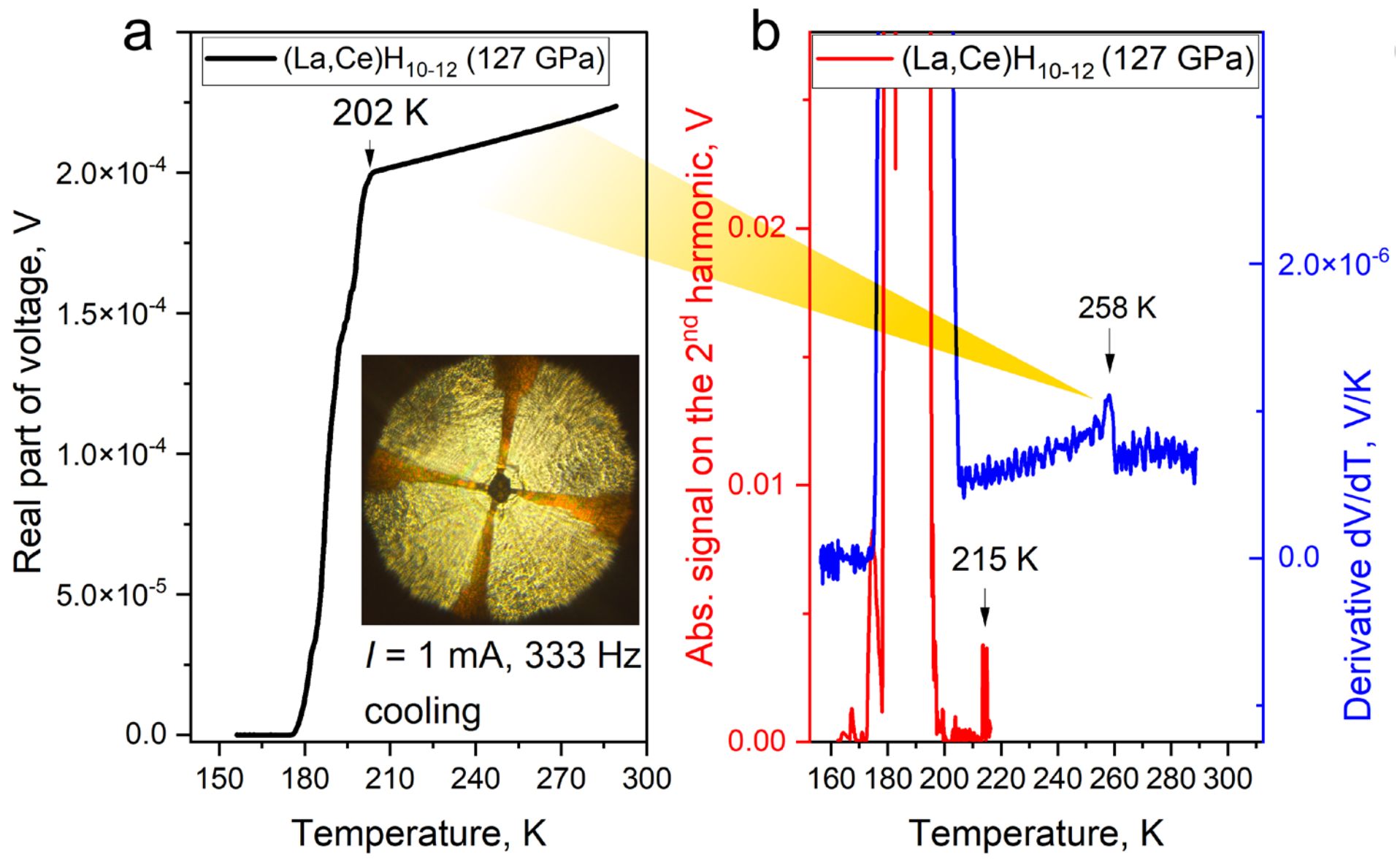

**Figure S21.** Superconducting transition in (La,Ce)$H_x$ under pressure. (a) Electrical resistance as a function of temperature *R(T)* for the (La,Ce)$H_x$ sample synthesized at 127 GPa, see Ref. [7]. Graph shows a clear superconducting transition starting from 202 K. Inset: optical image of the sample. (b) Anomalies detected in the 1st derivative of the measured voltage ($dV/dT$) and in the 2nd harmonic of low-frequency AC signal (2×333 Hz), indicating unknown peaks at 215 K and 258 K in the same sample.

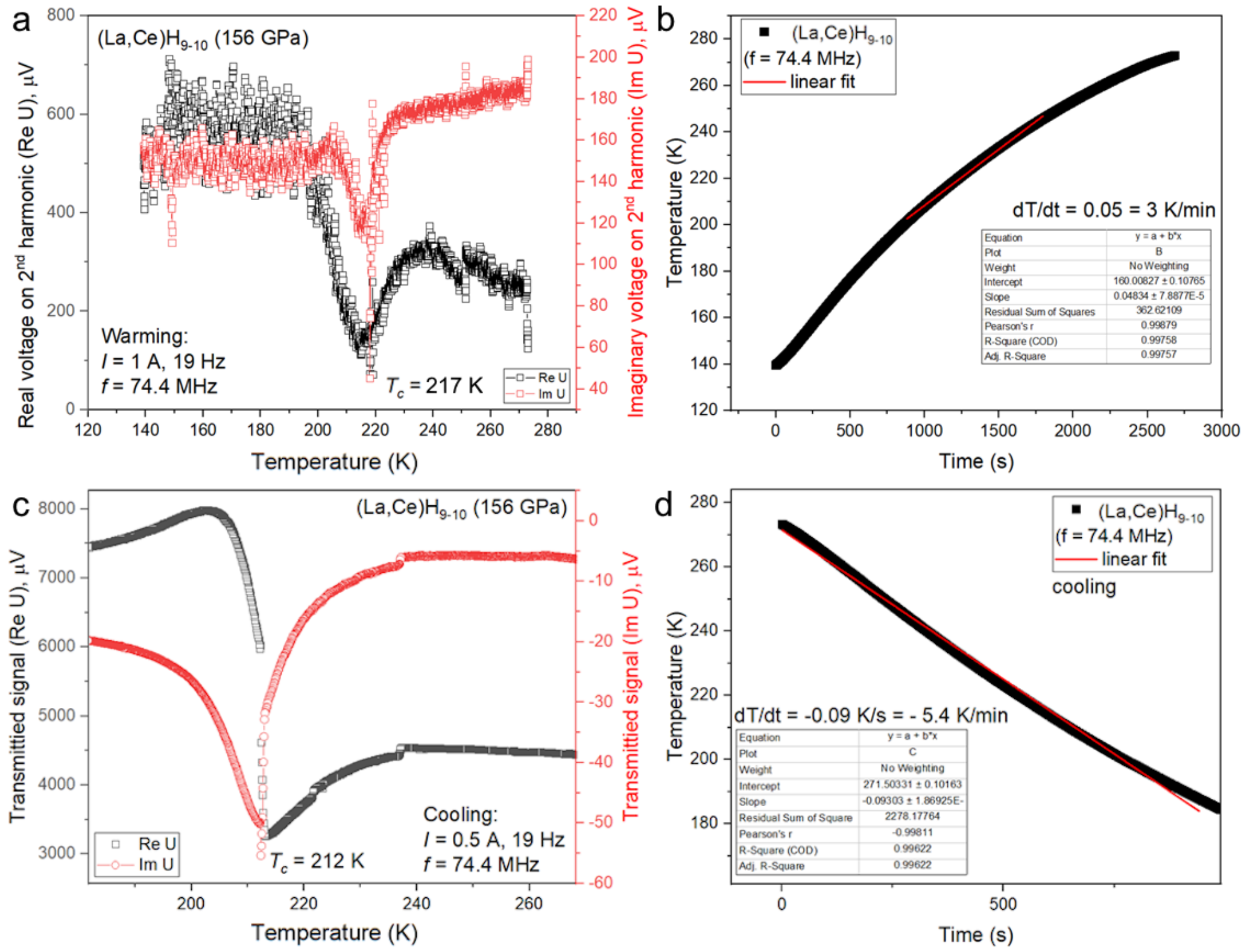

**Figure S22.** Position of the radio-frequency anomaly and hysteresis between warming and cooling cycles for the (La,Ce)$H_x$ sample at 156 GPa at the initial stages of the radio-frequency experiment. (a) Signal at the second harmonic of the modulating magnetic field ($B_{max} \approx$ 100 Gauss, $F$ = 19 Hz), warming cycle. (b) Temperature as a function of time in the warming cycle, the rate at the SC transition region is +3 K/min. (c) Signal of the high-frequency channel at $f_{carr}$ = 74.4 MHz in the cooling cycle. The observed hysteresis of $T_c$ is about 5-6 K. (d) Temperature as a function of time in the cooling cycle. The cooling rate was -5.4 K/min.

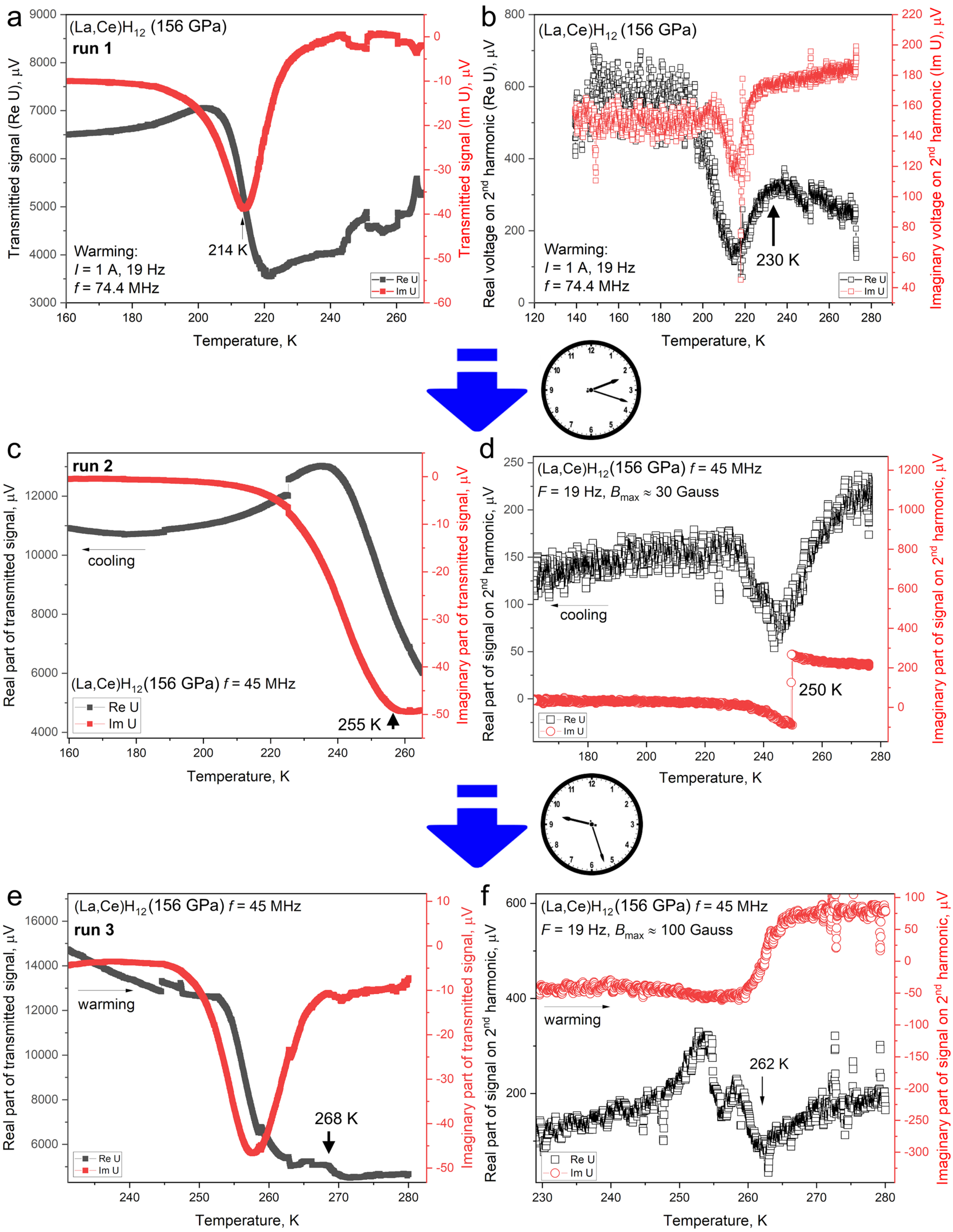

**Figure S23.** An example of the time evolution of real and imaginary components of the radio-frequency signal at the first (RF, panels “a, c, e”) and second harmonics (modulating filed, 19 Hz, panels “b, d, f”) of the $(La,Ce)H_{10-12}$ sample, observed during cooling and warming cycles.

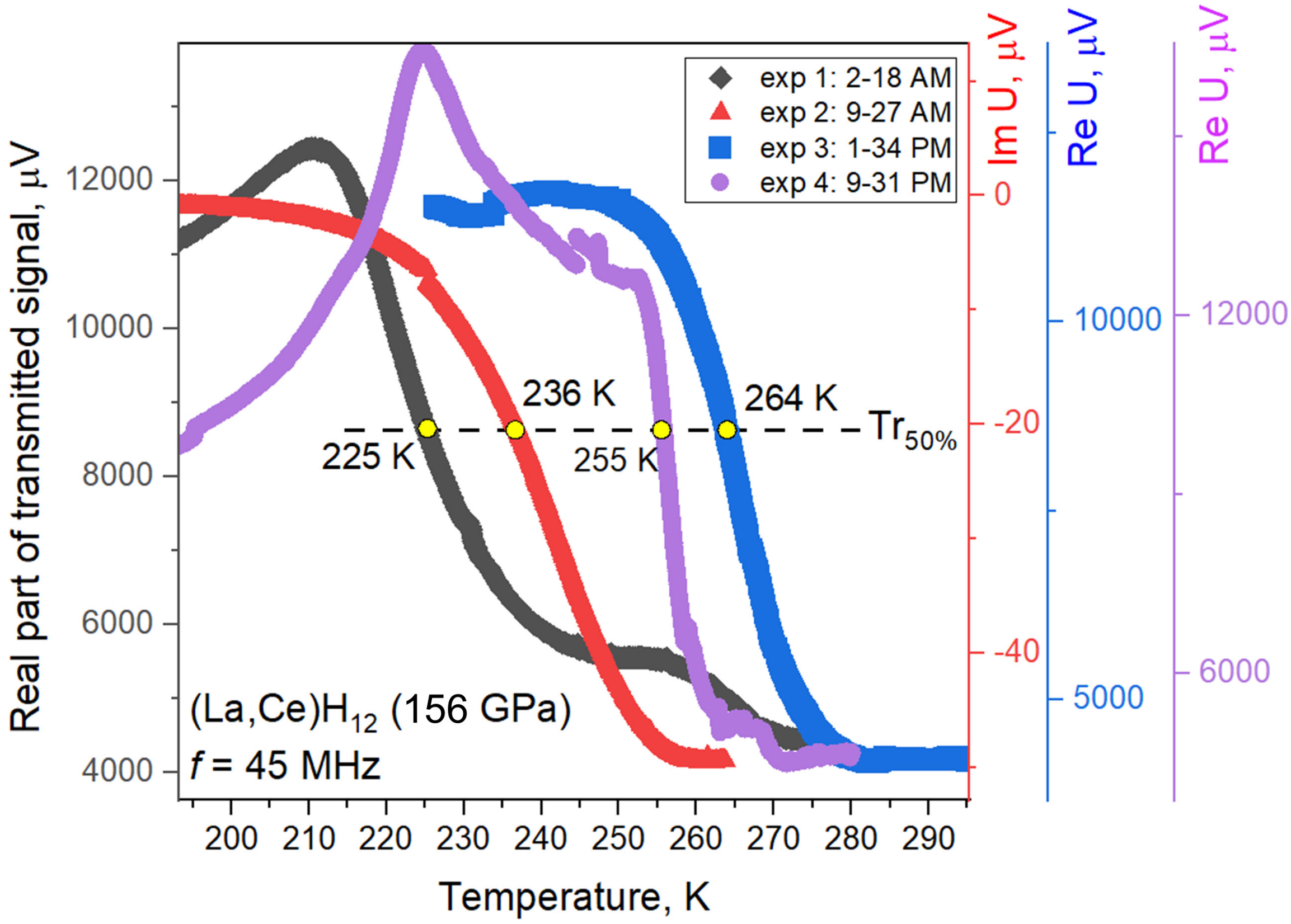

**Figure S24.** Example of RF transmission signal evolution for $(La,Ce)H_{12}$ (156 GPa) sample during multiple repetitions of RF experiment within warming/cooling cycles (rate was about 2-4 K/min). Evolution of transition and $T_c$ growth were observed for less than 24 hours from 2-18 AM (exp 1) to 9-31 PM (exp 4) on June 28, 2024, using constant scan parameters (SR844: time constant 100 μs, 18 dB) and carrier frequency of 45 MHz. Only the amplitude of modulating low-frequency magnetic field was changed.

## 8. Powder X-ray diffraction studies

At 0 GPa, powder diffraction analysis of the $La_6Ce$ alloy, used for the synthesis of La-Ce polyhydrides, shows the presence of three cubic (*fcc*) phases and one *dhcp* phase, of which the main ones are *dhcp*-(La,Ce) and *fcc*-(La,Ce) with lattice parameters close to those of pure lanthanum at ambient conditions (Figure S25, Table S2).

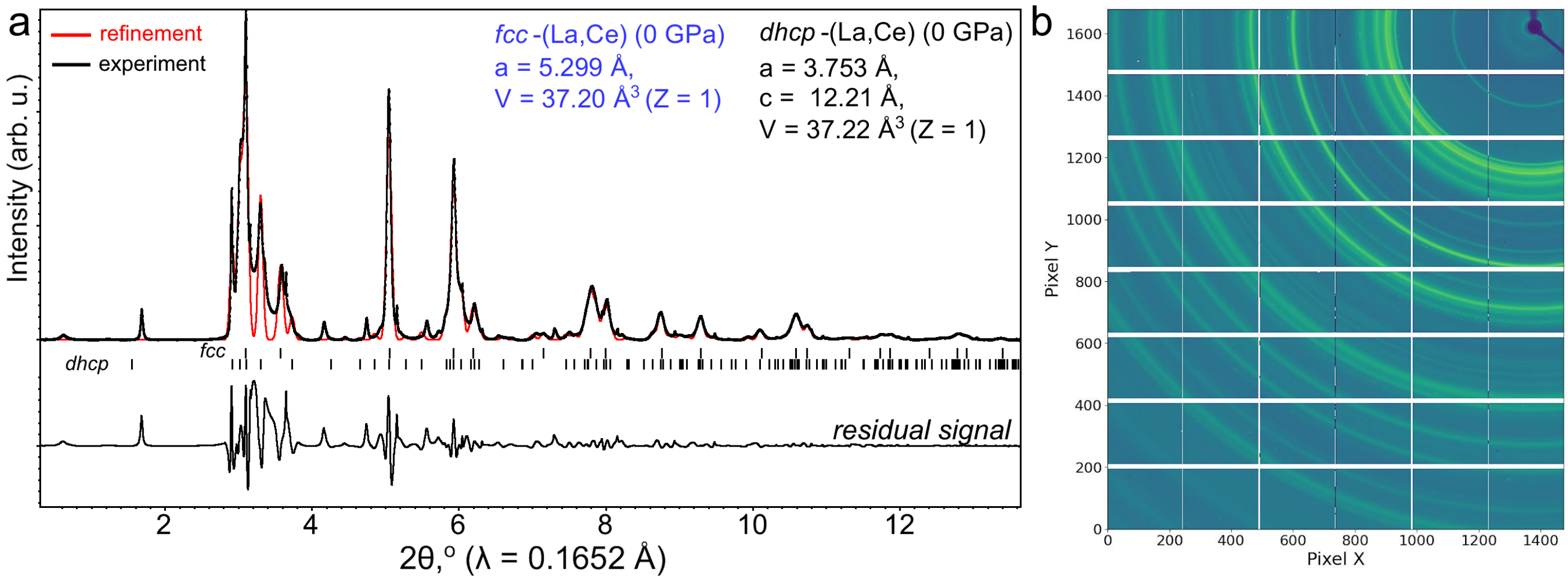

**Figure S25.** X-ray diffraction pattern and Le Bail refinement of the unit cell parameters of (La,Ce) alloy $La_6Ce$ at 0 GPa. (a) The experimental diffraction pattern (black line) is shown with the Le Bail refinement fit (red line). The tick marks below indicate the calculated peak positions for the face-centered cubic (*fcc*) and double hexagonal close-packed (*dhcp*) phases of (La,Ce). The residual signal (difference between experiment and refinement) is presented at the bottom. The refined lattice parameters and unit cell volumes for both *fcc*-(La,Ce) and *dhcp*-(La,Ce) phases are listed. The measurement was performed with X-ray beam of wavelength λ = 0.1652 Å. (b) 2D raw diffraction image, showing the distinct diffraction rings from the polycrystalline sample.

**Table S2.** Lattice parameters and unit cell volumes for various phases of the lanthanum-cerium system at ambient conditions.

| Compound | a, Å | c, Å | V, $Å^3$/atom |
|---|---|---|---|
| α-La | *dhcp:* 3.7742<br>*fcc*: 5.304 | *dhcp:* 12.171 | *dhcp:* 37.41<br>*fcc*: 37.30 |
| β- and γ-Ce | *dhcp:* 3.6811<br>*fcc*: 5.1612 | *dhcp:* 11.587 | *dhcp:* 34.75<br>*fcc*: 34.37 |
| $Ia\bar{3}$-$La_2O_3$ | 11.36 | - | 45.81 |
| *dhcp*-(La,Ce) * | 3.753 | 12.21 | 37.22 |
| *fcc*-(La,Ce) * | 5.299 | - | 37.20 |
| *fcc-2:* γ-(Ce,La) * | 5.196 | - | 35.07 |
| *fcc*-3: $(La,Ce)_2O_3$ * | *fcc:* 5.646<br>$Ia\bar{3}$: 11.288 | - | 44.99<br>44.94 |

*This work

Direct calculation of unit cell volumes shows that the cerium content in the *fcc* and *dhcp* phases is in the range of 3-7 at.%. In addition, in the XRD spectrum of the La-Ce alloy, there are reflections of almost pure cerium (phase *fcc*-2), which can explain the deviation of the calculated composition from the results of the energy dispersive X-ray (EDX) analysis. Moreover, diffraction signals of the product of oxidation of the alloy, cubic $Ia\bar{3}$-$(La,Ce)_2O_3$ (phase *fcc*-3), can also be seen.

An X-ray structural study of $(La,Ce)H_{10-12}$ sample within the decompression run was performed at the Shanghai Synchrotron Research Facility (SSRF). An attempt to decompress the DAC and determine the stability limits of the various La-Ce hydrides was unsuccessful: at the 3rd step of decompression the anvils were broken. In addition to the main product *hP*-$(La,Ce)H_{12}$ with very broad diffraction lines and, accordingly, very small crystallite size, we found traces of lower hydrides $Fm\bar{3}m$-$(La,Ce)H_3$, and $I4/mmm$-$(La,Ce)H_4$ (see

Supplementary Table S3), as well as pronounced diffraction reflections from an unknown compound (marked with asterisk in Supplementary Figure S28).

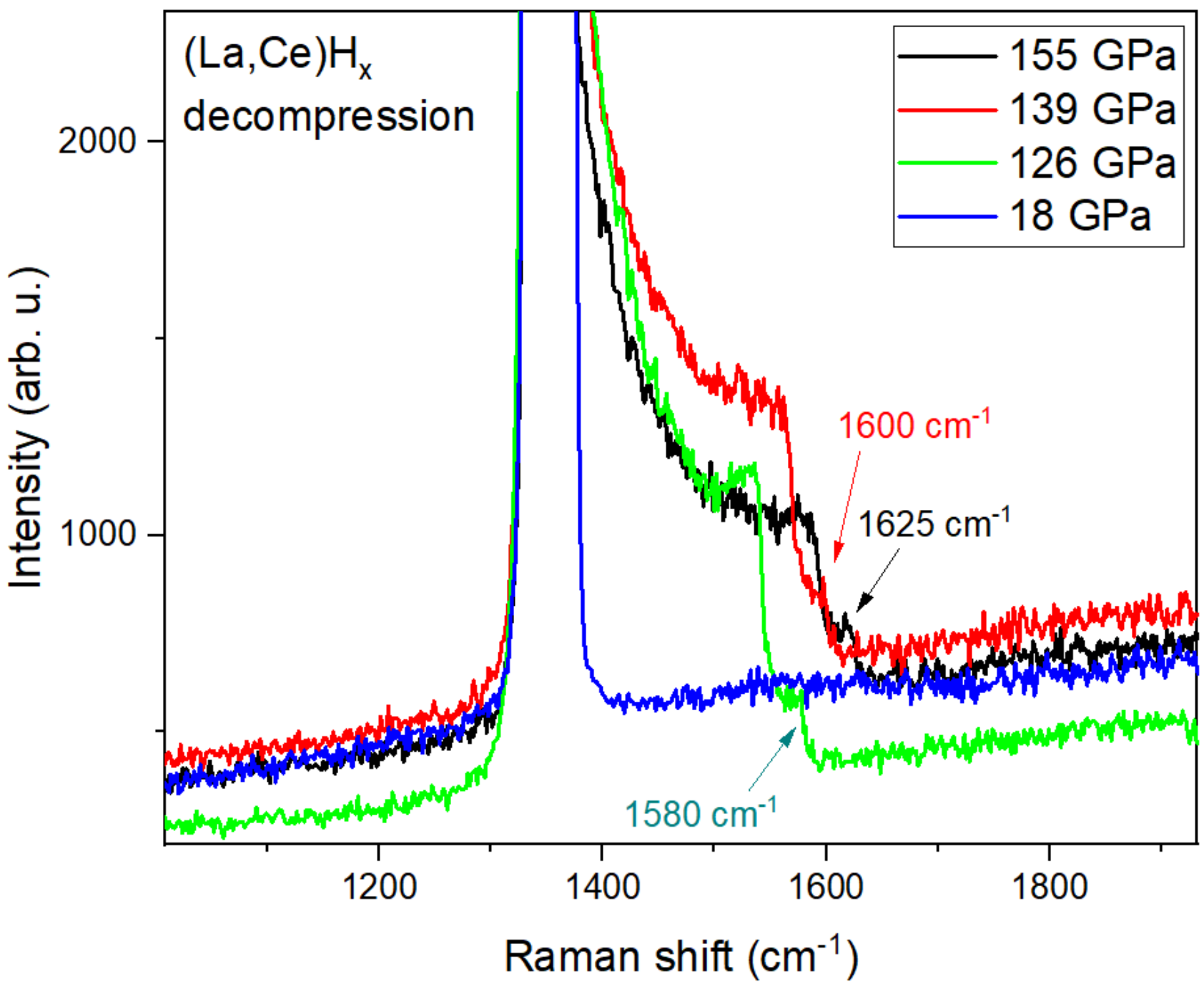

**Figure S26.** Raman spectra of (La,Ce)$H_x$ DAC during decompression from ≈155 GPa to 18 GPa (broken anvil). The Raman spectra show the evolution of the vibrational mode of diamond with decreasing pressure [8]. It is important that the diamond anvil has an incorrect (not [100]) orientation. That is why there are two steps at the edge of the diamond Raman signal. Therefore, an unambiguous determination of the pressure from the Raman spectra is not possible.

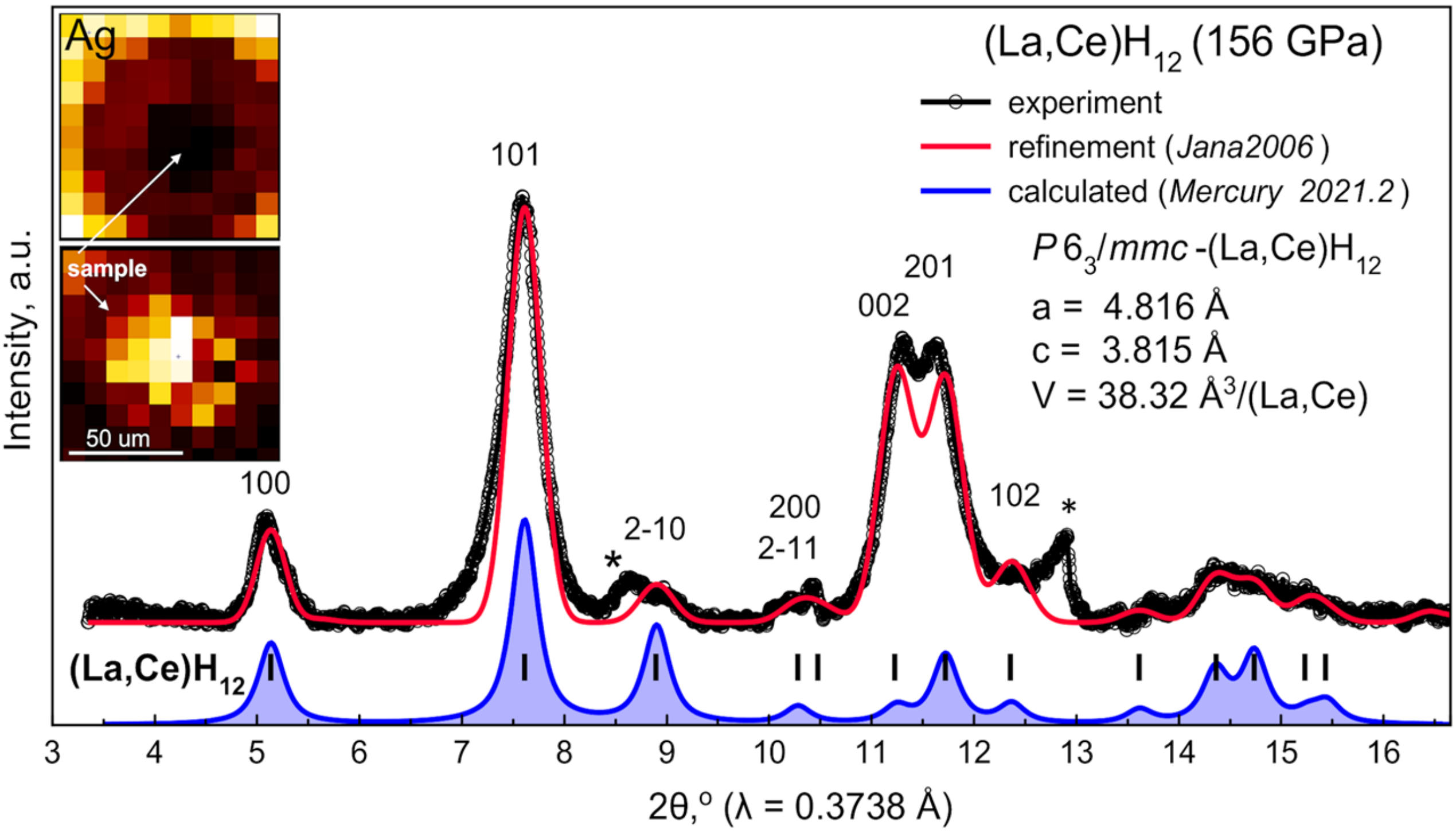

**Figure S27.** Powder X-ray diffraction analysis and the Le Bail refinement of the *hP*-(La,Ce)$H_{12}$ (refined as *P*$6_3$/*mmc*) unit cell parameters at 156 GPa. Black circles – are experimental points, red line – is the Jana2006 refinement, and the blue line – is the calculated XRD pattern for *hP*-(La,Ce)$H_{12}$ (Mercury 2021.2). Insets: XRD intensity maps of the sample, and silver (Ag) Lenz lens.

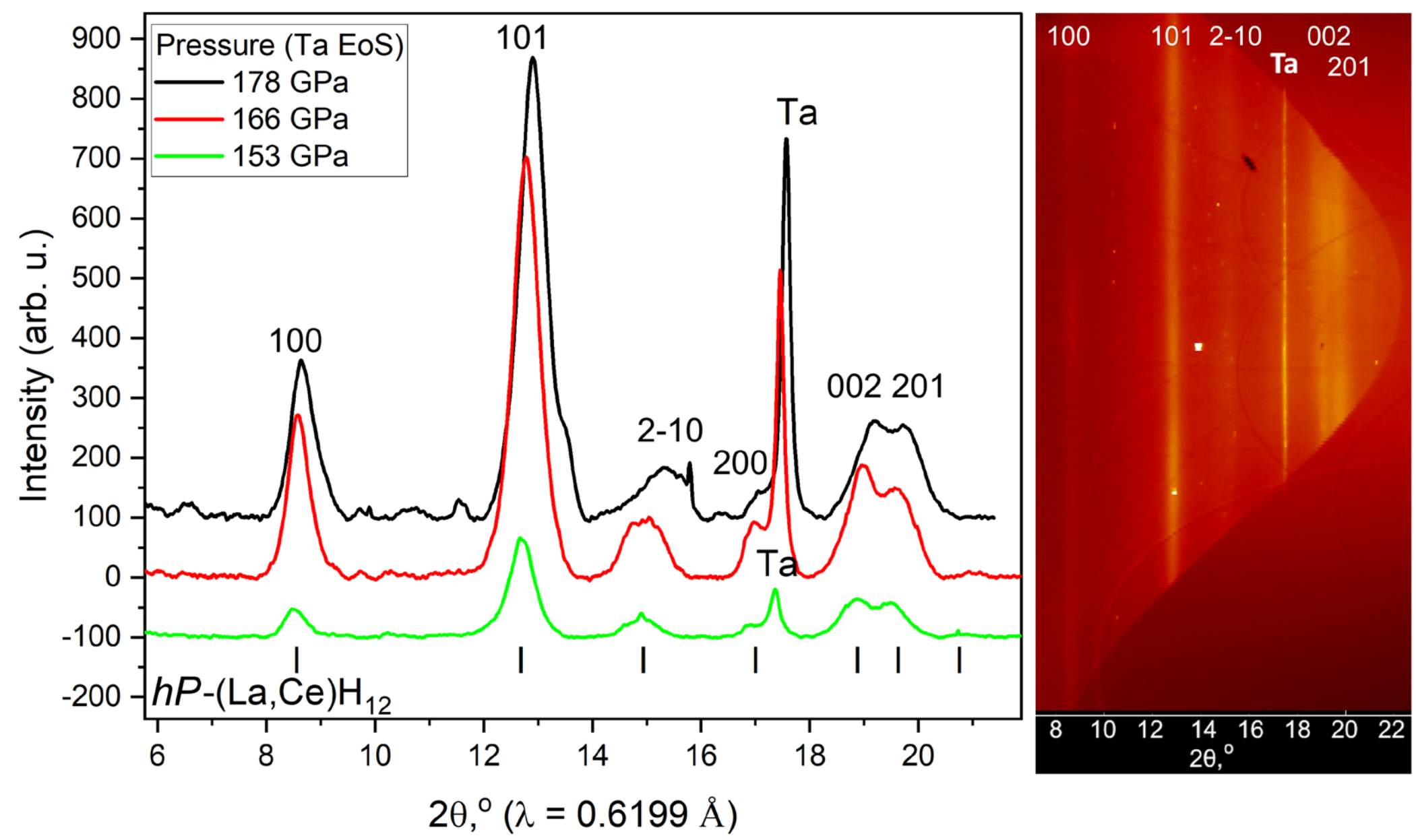

**Figure S28.** X-ray diffraction (XRD) patterns of *hP*-(La,Ce)$H_{12}$ under high pressure refined as *P*6$_3$/*mmc*. (Left) Integrated XRD patterns at various pressures measured in the decompression run (178 GPa, 166 GPa, 153 GPa according to the Ta gasket equation of state). Peaks corresponding to the *hP*-(La,Ce)$H_{12}$ phase are indexed, along with peaks from the tantalum gasket. (Right) A representative 2D raw diffraction image ("cake") at 166 GPa showing the diffraction lines from the sample and the Ta gasket, from which the 1D integrated patterns were obtained.

**Table S3.** Unit cell parameters of some phases found during X-ray diffraction studies of La-Ce hydrides. In the diamond cell, an uneven pressure distribution is observed, reaching a maximum on the Ta/W gasket. When the pressure dropped below 153 GPa, the DAC collapsed.

| Pressure, GPa | ***Fm*$\bar{3}$*m*-(La,Ce)$H_3$** | ***I*4/*mmm*-(La,Ce)$H_{4-x}$** | ***P*6$_3$/*mmc*-(La,Ce)$H_{12}$** |
|---|---|---|---|
| 155 (Raman)<br>178 (Ta EoS)* | a = 4.594 Å<br>V = 24.24 Å$^3$/(La,Ce) | a = 2.788 Å<br>c = 5.830 Å<br>V = 22.65 Å$^3$/(La,Ce) | a = 4.729 Å<br>c = 3.731 Å<br>V = 36.13 Å$^3$/(La,Ce) |
| 139 (Raman)<br>166 (Ta EoS) | a = 4.614 Å<br>V = 24.56 Å$^3$/(La,Ce) | - | a = 4.745 Å<br>c = 3.743 Å<br>V = 36.49 Å$^3$/(La,Ce) |
| 126 (Raman)<br>153 (Ta EoS) | - | - | a = 4.792 Å<br>c = 3.780 Å<br>V = 37.58 Å$^3$/(La,Ce |

* Due to the incorrect orientation of the diamond anvil plane, unambiguous determination of pressure is difficult and in addition to the Raman pressure scale [8], we used the Ta/W gasket XRD signal.

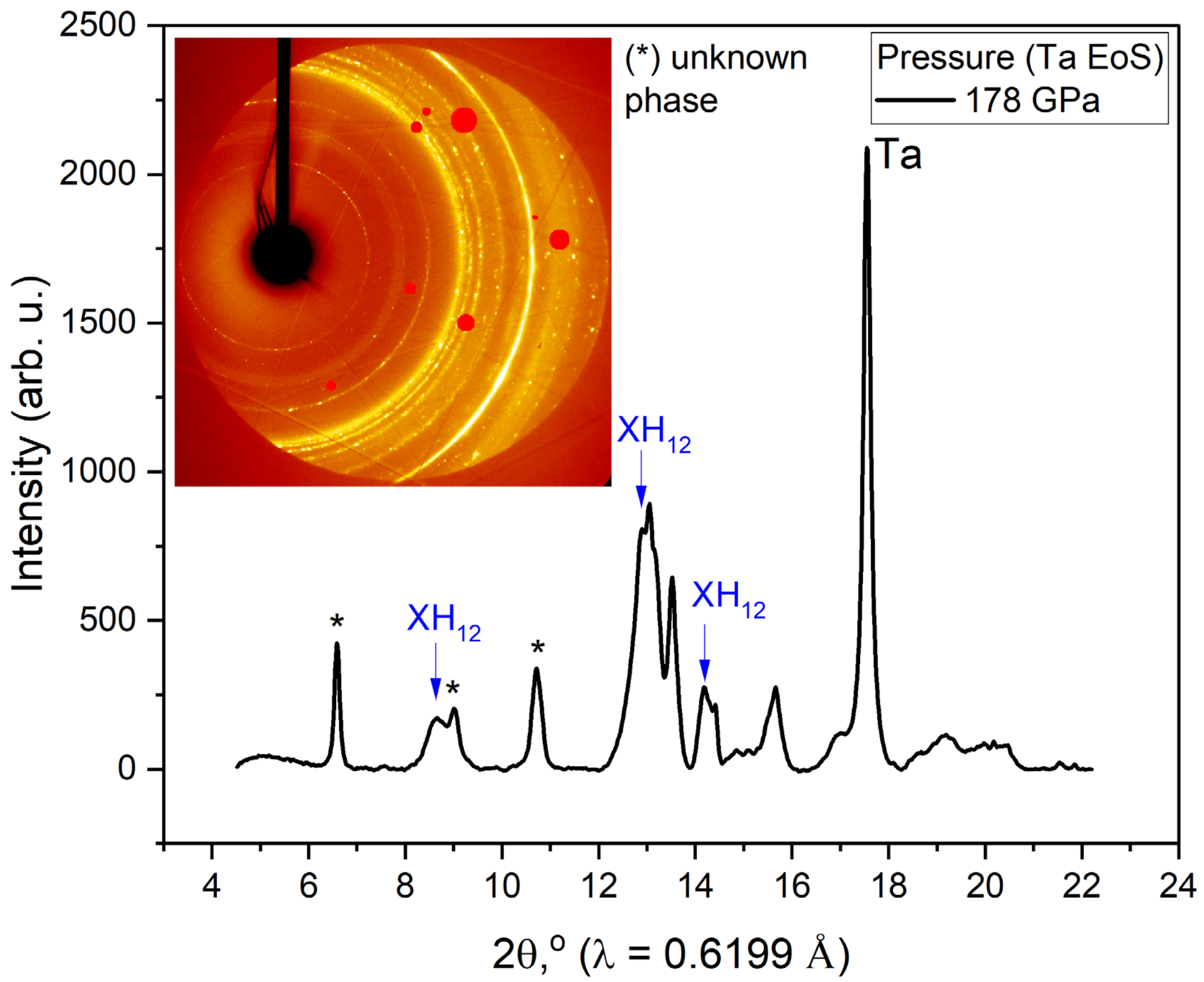

**Figure S29.** X-ray diffraction (XRD) pattern of La-Ce hydrides with unidentified phases at 178 GPa. The main panel shows the integrated XRD intensity. Peaks corresponding to the *hP*-$(La,Ce)H_{12}$ are indicated by blue arrows. The peak from the Ta gasket was used for the pressure determination. XRD peaks marked with an asterisk (*) correspond to an unknown La-Ce-H phase. The inset shows the raw 2D diffraction image, from which the 1D pattern was obtained.